\documentclass{article}

\usepackage{times}
\usepackage{graphicx}

\newcommand{\dir}{Figs}

\newcommand{\ie}{i.e.}
\newcommand{\eg}{e.g.}
\newcommand{\refcite}{\cite}

\begin{document}

{\LARGE \bf Coarse-grained models of complex fluids at equilibrium
and unter shear}


\begin{center}
{Friederike Schmid}

\noindent
Fakult\"at f\"ur Physik, Universit\"at Bielefeld,\\
Universit\"atsstra\ss e 25, D-33615 Bielefeld \\
E-mail: schmid\@physik.uni-bielefeld.de
\end{center}

\begin{quotation}
Complex fluids exhibit structure on a wide range of length and time scales,
and hierarchical approaches are necessary to investigate all facets of their 
often unusual properties. The study of idealized coarse-grained models at 
different levels of coarse-graining can provide insight into generic structures
and basic dynamical processes at equilibrium and non-equilibrium.

In the first part of this lecture, some popular coarse-grained models for 
membranes and membrane systems are reviewed. Special focus is given to 
bead-spring models with different solvent representations, and to 
random-interface models. Selected examples of simulations at the molecular 
and the mesoscopic level are presented, and it is shown how simulations
of molecular coarse-grained models can bridge between different levels.

The second part addresses simulation methods for complex fluids under shear.
After a brief introduction into the phenomenology (in particular for liquid
crystals), different non-equilibrium molecular dynamics (NEMD) methods are
introduced and compared to one another. Application examples include the
behavior of liquid crystal interfaces and lamellar surfactant phases under 
shear. Finally, mesoscopic simulation approaches for liquid crystals under
shear are briefly discussed.
\end{quotation}

\section{Introduction}     

The term ``Complex fluid'' or ``soft condensed matter'' -- these two are often
used interchangeably -- refers to a broad class of materials, which are usually
made of large organic molecules and have a number of common 
features~\cite{daoud,larson_book}: 
They display structure on a nanoscopic or mesoscopic scale; characteristic energies 
are of the order of $k_B T$ at room temperature, hence the properties are to a large 
extent dominated by entropic effects, and the materials respond strongly to weak 
external forces~\cite{israelachvili}; the characteristic response times span one 
or several orders of magnitude, and the rheological properties of the fluids are 
typically non-Newtonian~\cite{larson_book,kroeger}. Some examples are polymer melts 
and solutions~\cite{flory,degennes1,doi1,doi2}, emulsions, 
colloids~\cite{russel,hunter,borowko}, amphiphilic systems~\cite{safran,schick},
and liquid crystals~\cite{degennes2,chandrasekhar,chaikin}.

Computer simulations of complex fluids are particularly challenging, due
to the hierarchy of length and time scales that contributes to the material
properties. With current computer resources, it is practically impossible
to describe all aspects of such a material within one single theoretical
model. Therefore, one commonly uses different models for different length
and time scales. The idea is to eliminate the small-scale degrees of
freedom in the large-scale models, and to incorporate the details of the
small-scale properties in the parameters of a new, simpler, model. 
This is called coarse-graining.

There are two aspects to coarse-graining. First, systematic coarse-graining 
procedures must be developed and applied in order to study materials 
quantitatively on several length and time scales. This is the challenge of 
multiscale modeling, one major growth area in materials science. It is, 
however, not the subject of the present lecture. Some recent reviews can be 
found in Refs.~\refcite{mplathe1,nielsen}

Second, coarse-grained idealized models are used to study generic properties of 
materials on a given scale, and physical phenomena which are characteristic for 
a whole class of materials. Here, microscopic details are disregarded because 
they are deemed irrelevant for the physical properties under consideration.
This approach to materials science has a long tradition, going back to Ising's
famous lattice model for magnetism.

In the present lecture, we shall concentrate on coarse-grained models of the 
second type. We shall discuss how such models can be constructed, how they can 
be studied by computer simulations, and how such studies help to improve our 
understanding of equilibrium and nonequilibrium properties of complex materials.
We will not attempt to present an overview -- this would require a separate book 
-- but rather focus on selected case studies. In the first part, we will consider 
coarse-grained model systems for amphiphilic membranes and their use for studies of
equilibrium properties. In the second part, we will turn to complex fluids under 
shear, \ie, nonequilibrium systems, and discuss coarse-grained approaches for 
membrane systems and liquid crystals.

\section{Coarse-grained models for surfactant layers\label{sec:membranes}} 

In this section, we shall discuss coarse-grained simulation models and methods for 
amphiphilic membranes and membrane stacks. After a brief 
introduction into the phenomenology and some general remarks on coarse-graining, 
we will focus on two particular types of coarse-grained models: Particle-based 
bead-spring models and mesoscopic membrane models. We shall present some typical 
examples, and illustrate their use with applications from the literature.

\subsection{Introduction}  

\begin{figure}[th] 
\centerline{\includegraphics[width=2.2in,angle=-90]{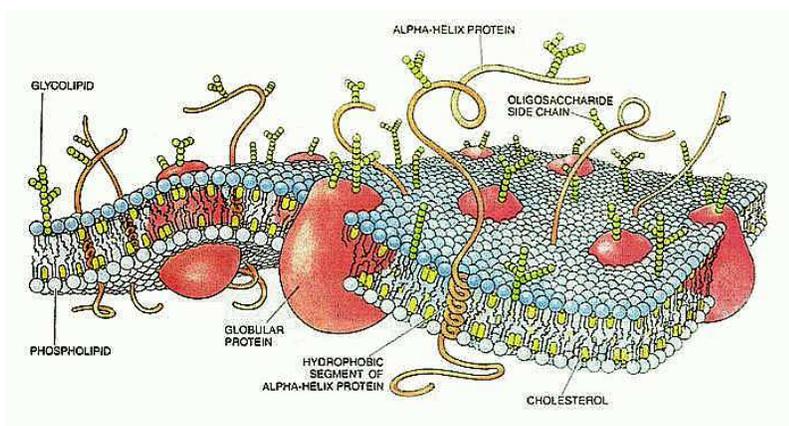}}
\vspace*{8pt}
\caption{
Artist's view of a biomembrane. 
Source: NIST
\label{fig:biomembrane}
}
\end{figure}

Amphiphilic membranes are a particular important class of soft material structures, 
because of their significance for biology~\cite{gennis,lipowsky,mouritsen}. 
Biomembranes are omnipresent in all living beings, they delimit cells and 
create compartments, and participate in almost all biological functions. 
Figure \ref{fig:biomembrane} shows an artist's view on such a biomembrane. 
It is mainly made of two coupled layers of lipids, which serve as a matrix 
for embedded proteins and other molecules.

\begin{figure}[b] 
\centerline{\includegraphics[width=1.5in]{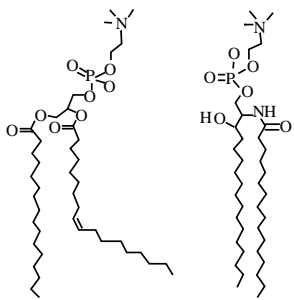}}
\vspace*{8pt}
\caption{
Examples of membrane lipids.
Left: Phosphatidylcholine (unsaturated);
Right: Sphingomyelin.
\label{fig:lipids}
}
\end{figure}

This structure rests on a propensity of lipids to self-assemble into bilayers when 
exposed to water. The crucial property of lipids which is responsible for this 
behavior is their amphiphilic character, \ie, the fact that they contain both 
hydrophilic (water loving) and hydrophobic (water hating) parts. Most lipid
molecules have two non polar (hydrophobic) hydrocarbon chain (tails), which are 
attached to one polar (hydrophilic) head group.  Figure~\ref{fig:lipids}) shows 
some typical examples. The tails vary in their length, and in the number and 
position of double bonds between carbon atoms. Whereas single bonds in a hydrocarbon 
chain are highly flexible, in the sense that chains can rotate easily around such 
a bond, double bonds are stiff and may enforce kinks in the chain. Chains with no 
double bonds are called saturated. The variety of polar head groups is even larger. 
Overviews can be found in~\cite{gennis}.

\begin{figure}[t] 
\centerline{\includegraphics[width=3.5in]{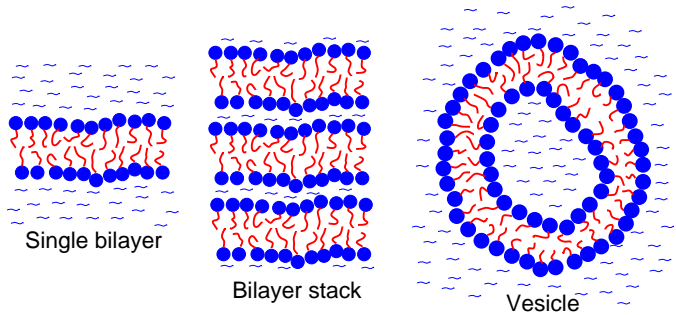}}
\vspace*{8pt}
\caption{
Self-assembled bilayer structures.
\label{fig:bilayers}
}
\end{figure}

By assembling into bilayers, the lipids can arrange themselves such that the head 
groups shield the tails from the water environment. Therefore, lipids in lipid-water 
mixtures often tend to form lamellar stacks of membranes separated by thin water 
layers (Fig.~\ref{fig:bilayers}). Such stacks can typically hold up to 20 \% water. 
In the more dilute regime, membranes may close up to form vesicles or other more
disordered structures. Vesicles play an important role in biological systems, since 
they provide closed compartments which can be used to store or transport substances.  
Single, isolated membranes are metastable with respect to dissociation for entropic 
reasons, but they have very long lifetimes.

Depending on the particular choice of lipids and the properties of the surrounding 
aqueous fluid (pH, salt content etc.), several other structures can also be observed 
in lipid-water systems -- spherical and cylindrical micelles, ordered structures
with hexagonal or cubic symmetry etc.~\cite{schubert}. In this lecture, we shall 
only regard bilayer structures.

\begin{figure}[b] 
\centerline{\includegraphics[width=3.7in]{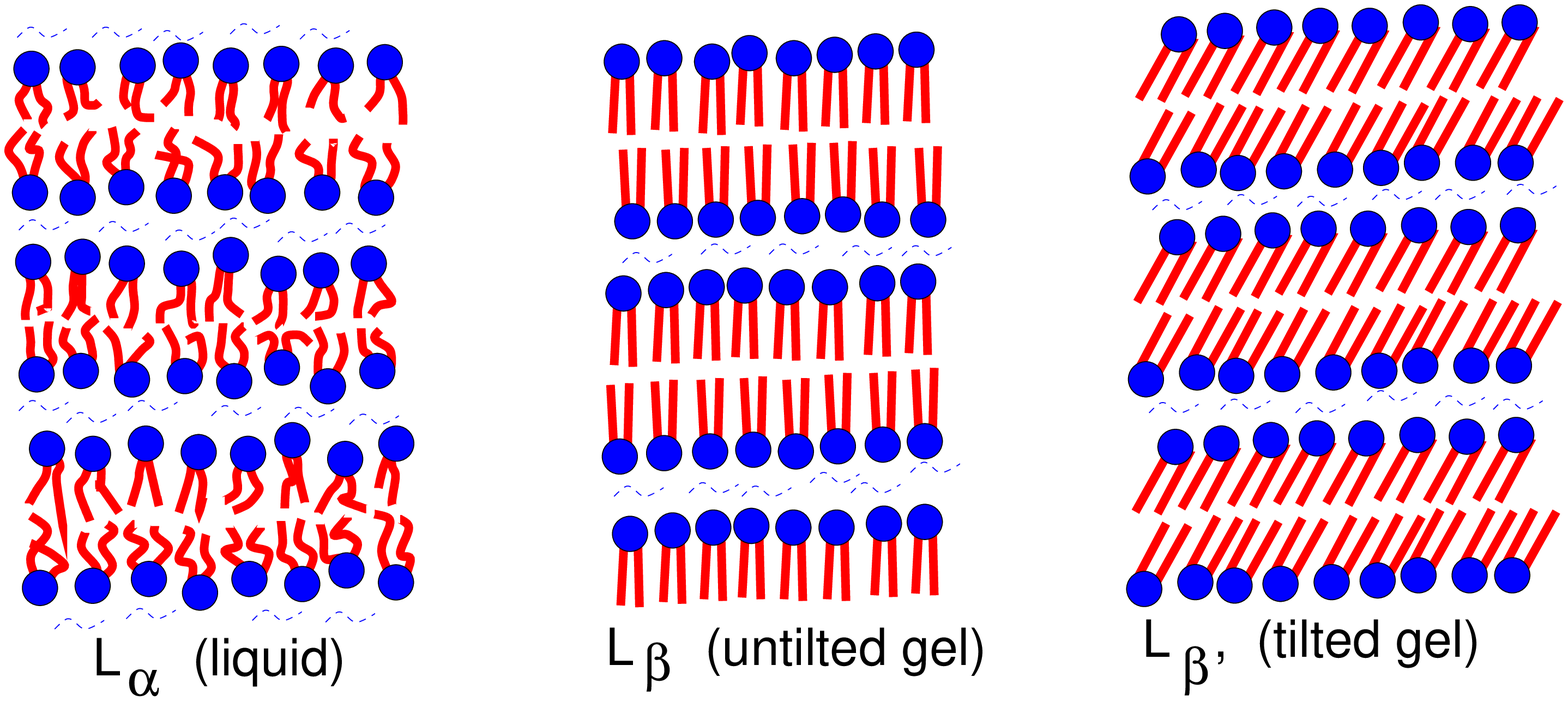}}
\vspace*{8pt}
\caption{
Some prominent membrane phases.
\label{fig:membrane_phases}
}
\end{figure}

Membranes also undergo internal phase transitions. Figure~\ref{fig:membrane_phases} 
shows schematically some characteristic phases of one-component membranes. The most 
prominent phase transition is the ``main'' transition, the transition from the liquid 
state into one of the gel states, where the layer thickness, the lipid mobility, 
and the conformational order of the chains, jump discontinuously as a function of 
temperature. In living organisms, most membranes are maintained in the liquid state. 
Nevertheless, the main transition presumably has some biological relevance,
since it is very close to the body temperature for some of the most common lipids 
(\eg, saturated phospholipids)~\cite{gennis}.

In sum, several aspects of membranes must be studied in order to understand 
their structure and function, in biological as well as in artificial systems:

\begin{itemize}
\item Self-assembling mechanisms
\item The inner structure and internal phase transitions, and the resulting 
  membrane properties (permeability, viscosity, stiffness)
\item The organization of membranes on a mesoscopic scale (vesicles, stacks, etc.)
\item The interaction of membranes with macromolecules.
\end{itemize}

These topics relate to the physics and chemistry of amphiphiles on many length and 
time scales, and several driving forces contribute to the unique properties of 
membranes: On the scale of atoms and small molecules, one has the forces which are 
responsible for the self-assembly: The hydrophobic effect, and the interactions 
between water and the hydrophilic head groups.  On the molecular scale, one has
the interplay of local chain conformations and chain packing, which is responsible 
for the main transition and determines the local structure and fluidity of the 
membrane. On the supramolecular scale, one has the forces which govern the 
mesoscopic structure and dynamics of membrane systems: The membrane elasticity, 
the thermal fluctuations, the hydrodynamic interactions with the surrounding fluid, 
the thermodynamic forces that drive phase separation in mixed membranes etc.

Studying all these aspects within one single computer simulation model is clearly 
impossible, and different models have been devised to study membranes at different 
levels of coarse-graining.  Among the oldest approaches of this kind are the 
Doniach~\cite{doniach} and the Pink model~\cite{pink1,pink2}, generalized Ising 
models, which have been used to investigate almost all aspects of membrane physics \
on the supramolecular scale~\cite{mouritsen,dammann}. With increasing 
computer power, numerous other simulation models have become treatable in the 
last decades, which range from atomistic models, molecular coarse-grained models, 
up to mesoscopic models that are based on continuum theories for membranes.

In the following, we shall focus on two important examples: Coarse-grained 
chain models designed to describe membranes on the molecular scale,
and random interface models that describe membranes on the mesoscopic scale.
We stress that this our overview is not complete, and there exist other
successful approaches, which are not included in our presentation. 
A relatively recent review can be found in Ref.~\refcite{fs_review}.

\subsection{Coarse-grained molecular models}  

In coarse-grained molecular models, molecules are still treated as individual 
particles, but their structure is highly simplified. Only the ingredients deemed 
essential for the material properties are kept. For amphiphiles, these are the 
amphiphilic character (for the self-assembly), and the molecular flexibility 
(for the internal phase transitions).

\subsubsection{Spring-bead models \label{sec:model1}}

The first coarse grained molecular models have been formulated on a lattice, 
relatively shortly after the introduction of the above-mentioned Pink 
model~\cite{pink1,pink2}. One particularly popular model is the Larson 
model~\cite{larson1,liverpool}: Water molecules ($w$) occupy single sites of 
a cubic lattice, and amphiphile molecules are represented by chains of ``tail'' 
($t$) and ``head'' ($h$) monomers. Only particles on neighbor lattice sites 
interact with each another. The lattice is entirely filled with $w$, $h$, or $t$ 
particles. If one further takes all hydrophilic particles, $h$ and $w$, to be 
identical, the interaction energy is determined by a single interaction parameter, 
which describes the relative repulsion between hydrophobic and hydrophilic particles.
The model reproduces self-assembly and exhibits many experimentally observed
phases, \ie, the lamellar phase, the hexagonal phase, the cubic micellar phase,
and even the bicontinuous gyroid phase~\cite{liverpool,larson2}.

Lattice models can be simulated efficiently by Monte Carlo methods. However, they 
have the obvious drawback of imposing an {\em a priori} anisotropy on space,
which restricts their versatility. For example, internal membrane phase transitions 
and tilted phases cannot be studied easily. In dynamical studies, one is 
restricted to using Monte Carlo dynamics, which is unrealistic.
Therefore, most more recent investigations rely on off-lattice models.

Off-lattice molecular approaches to study self-assembling amphiphilic systems have 
been applied for roughly 15 years. Presumably the first model was introduced by
B.~Smit {\em et al.}~\cite{smit1} in 1990. A schematic sketch is shown in 
Fig.~\ref{fig:solvents} a). As in the Larson model, the amphiphiles are represented 
by chains made of very simple $h$- or $t$-units, which are in this case spherical 
beads. ``Water'' molecules are represented by free beads. Beads in a chain are 
joined together by harmonic springs, where a cutoff is sometimes imposed in order 
to ensure that the beads cannot move arbitrarily far apart from one another. 
Non-bonded beads interact via simple short-range pairwise potentials, \eg, a 
truncated Lennard-Jones potential. The parameters of the potentials are chosen such
that $ht$ pairs and $hw$ pairs effectively repel each other. The Smit model 
reproduces self-assembly, micelle formation and membrane 
formation~\cite{smit2,smit3,palmer1,palmer2,goetz1}.

Most molecular off-lattice models are modifications of the Smit 
model~\cite{fs_review,shillcock1,stevens,kranenburg1}. They differ from one 
another in the specific choice of the interaction potentials between non-bonded 
beads, in the bond potentials between adjacent beads on the chain, in the chain 
architecture (number of head and tail beads, number of tails), and sometimes 
contain additional potentials such as bending potentials. Nowadays, Smit models 
are often studied with dissipative particle (DPD) dynamics, and the interaction 
potentials are the typical soft DPD potentials~\cite{shillcock1,kranenburg1}.
A ``water particle'' is then taken to represent a lump of several water 
molecules. An introduction into the DPD method is given in B. D\"unweg's 
lecture (this book). It works very well for membrane simulations as long as 
the time step is not too large~\cite{jakobsen}.

This approach is presented in more detail in B. Smit's seminar (this book).

\subsubsection{Implicit solvent and phantom solvent \label{sec:model2}}

The primary requirement for a lipid membrane model is to ensure that the lipids
maintain a bilayer structure. In the models discussed in the previous
section~\ref{sec:model1}, the bilayers are stabilized by the surrounding solvent 
particles. If the interactions between $w$, $h$, and $t$ particles are chosen 
suitably, the solvent particles and the amphiphiles arrange themselves such that 
the hydrophilic and hydrophobic particles are separated (microphase separation).

This ``explicit solvent'' approach is in some sense the most natural one and has 
several advantages: The bilayers are fully flexible, and they are surrounded
by a fluid. The fact that the fluid is composed of large beads as opposed to small 
water molecules is slightly problematic, since the solvent structure may introduce 
correlations, \eg, in simulations with periodic boundary conditions.
However, the correlations disappear if the membranes or their periodic images are 
separated by a large amount of solvent.

On the other hand, explicit solvent models also have a disadvantage: The simulation 
time depends linearly on the {\em total} number of beads. A large amount
of computing time is therefore spent on the uninteresting solvent particles. 
Therefore, effort has been spent on designing membrane models without explicit
solvent particles.

\begin{figure}[b] 
\centerline{
\parbox{1.2in}{ \includegraphics[width=1.2in]{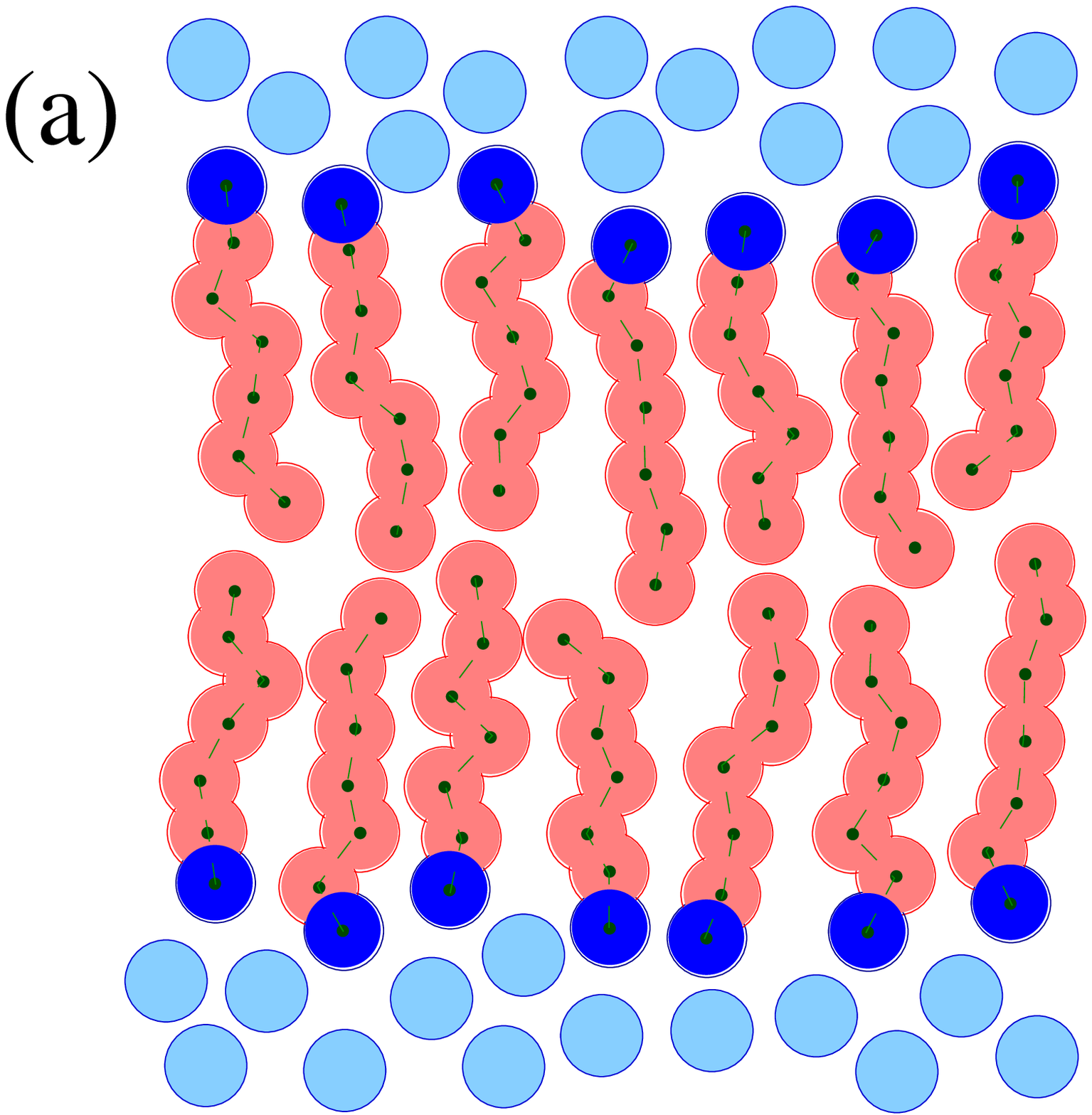} }
\parbox{1.2in}{ \includegraphics[width=1.2in]{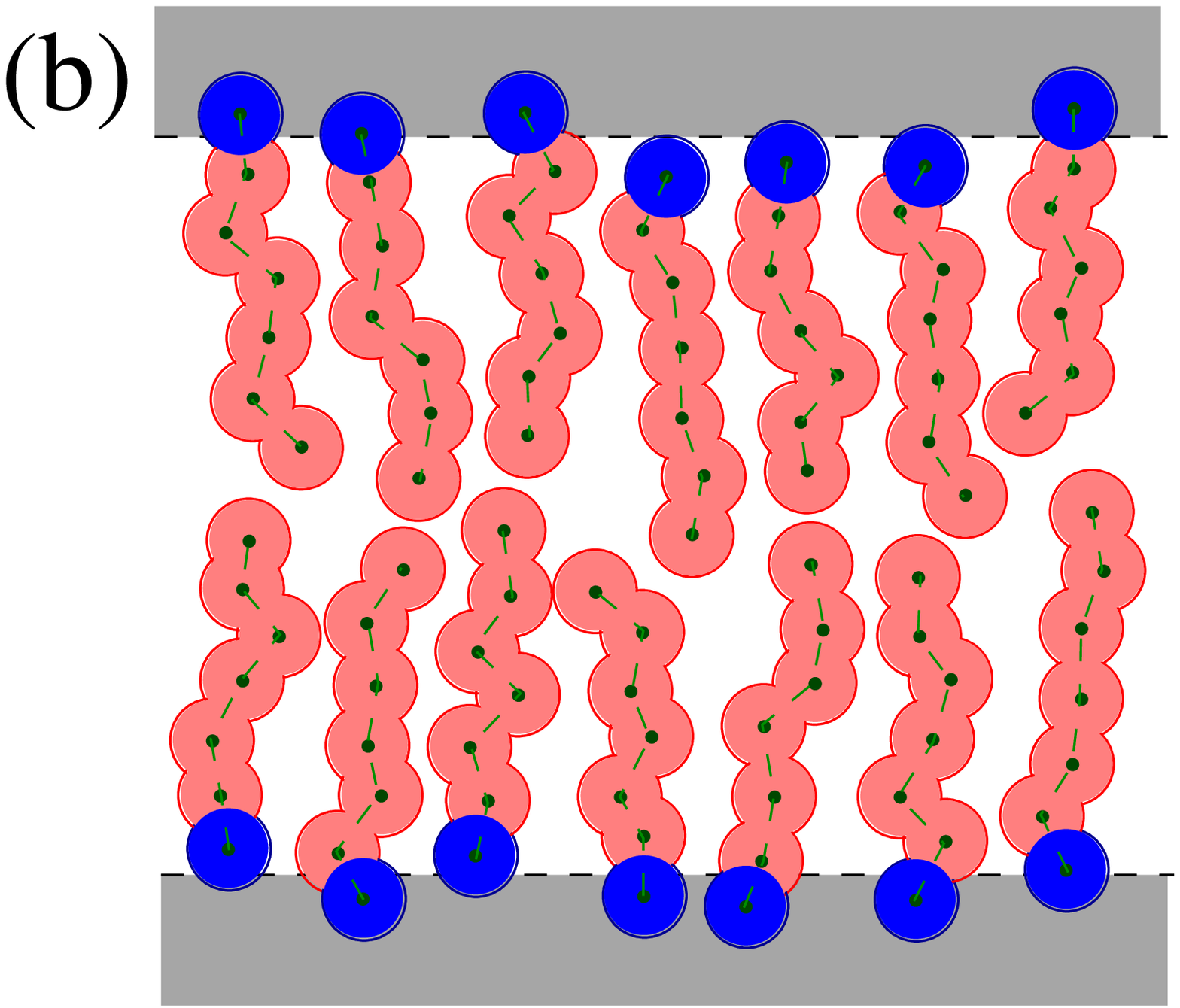} \\ \vfill }
\parbox{1.2in}{ \includegraphics[width=1.2in]{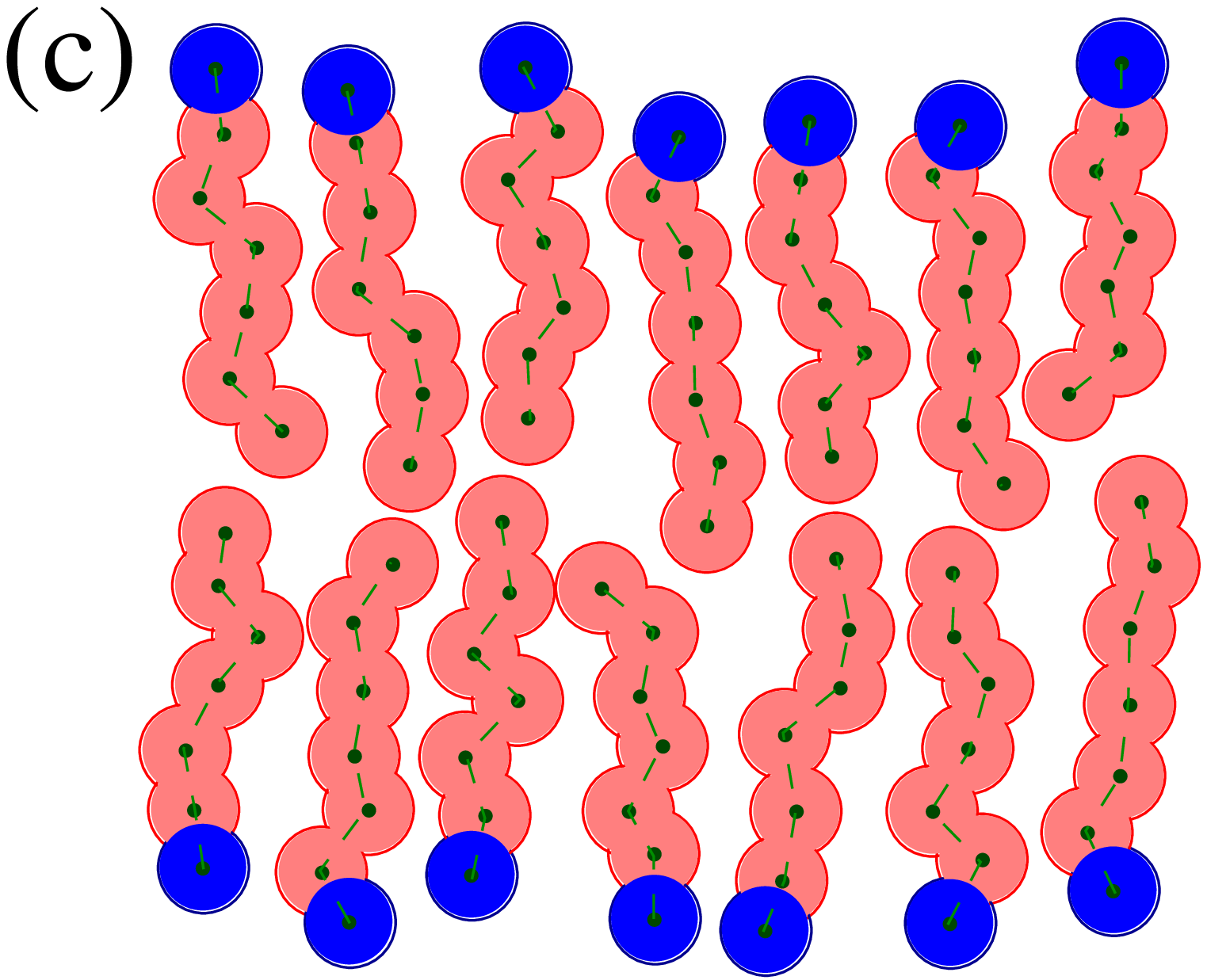} \\ \vfill}
\parbox{1.2in}{ \includegraphics[width=1.2in]{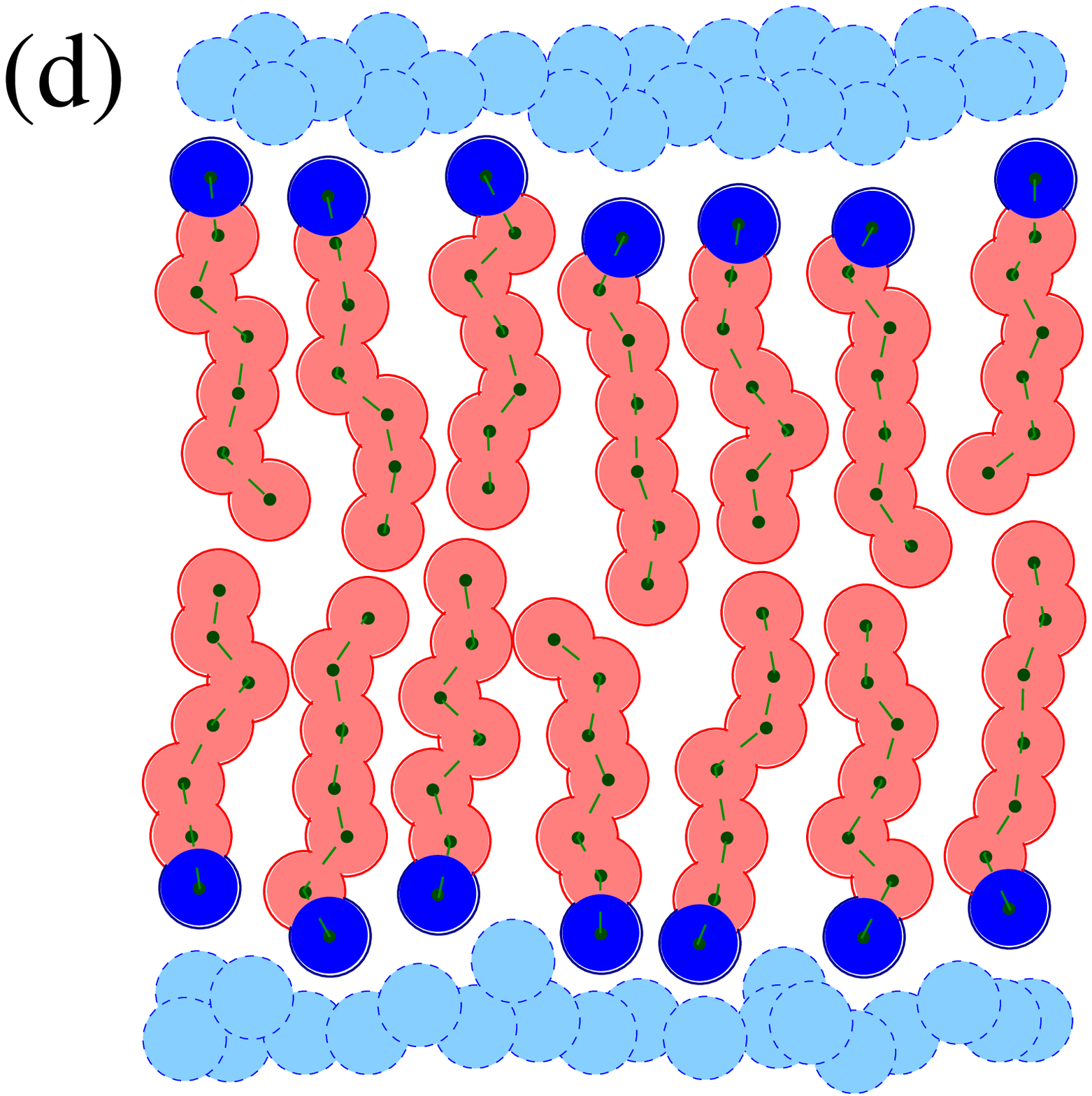} }
}
\vspace*{8pt}
\caption{
Schematic picture of bead-spring models for self-assembling
membranes.
(a) Explicit solvent model
(b) Sandwich model
(c) Solvent free model
(d) Phantom solvent model.
\label{fig:solvents}
}
\end{figure}

One early idea has been to force the amphiphiles into sheets by tethering the 
head groups to two-dimensional opposing surfaces, \eg, by a harmonic potential, 
or by rigid constraints~\cite{harries,baumgaertner1,baumgaertner2,sintes,olenz1}
(Fig.~\ref{fig:solvents} b)). This model, which we call ``sandwich model'', is
simple, extremely efficient, and the configurations are easy to analyze, since 
the bilayer is always well-defined. On the other hand, the bilayer has no 
flexibility, \ie, undulations and protrusions~\cite{goetz2} are supressed and 
important physics is lost.

A second approach is to eliminate the solvent, and represent its effect 
on the amphiphiles by appropriate effective interactions between monomers 
(Fig.~\ref{fig:solvents} c)). This way of modeling solvent is common in polymer 
simulations. Nevertheless, producing something as complex as membrane self-assembly 
is a non-trivial task. The first implicit solvent model was proposed only rather 
recently, in 2001, by Noguchi and Takasu~\cite{noguchi1}. They mimick the effect 
of the solvent by a multibody ``hydrophobic potential'', which depends on the 
local density of hydrophobic particles. From a physical point of view, the 
introduction of multibody potentials in an implicit solvent model is reasonable -- 
multibody potentials emerge automatically whenever degrees of freedom are 
integrated out systematically. Nevertheless, people usually tend to favor
pair potentials. Farago~\cite{farago} and Cooke {\em et al.}~\cite{cooke} have 
shown that it is possible to stabilize fluid bilayers with purely pairwise
interactions in a solvent-free model. The crucial requirement seems to be 
that the interactions between hydrophobic beads have a smooth attractive part
of relatively long range~\cite{cooke}.

The clear benefit of implicit solvent models is their high efficiency. They are 
very well suited to study large scale problems like vesicle 
formation~\cite{noguchi1,cooke}, vesicle fusion~\cite{noguchi2}, vesicles subject 
to external forces~\cite{noguchi3}, phase separation on vesicles~\cite{cooke} etc.
On the other hand, they also have limitations: They require tricky potentials; 
the pressure is essentially zero (since the membranes are surrounded by empty 
space); also, they cannot be used for dynamical studies where the hydrodynamical
coupling with the solvent becomes important.

A third approach was recently put forward by ourselves: the ``phantom solvent'' 
model~\cite{olenz1} (Fig.~\ref{fig:solvents}d)). We introduce explicit solvent 
particles, which do not interact with one another, only with the amphiphiles.
The amphiphiles perceive them as repulsive soft beads. In the bulk, they simply 
form an ideal gas. They have the simple physical interpretation that they probe
the free volume which is still accessible to the solvent, once the membrane has 
self-assembled. Thus the self-assembly is to a large extent driven entropically.

In Monte Carlo simulations, this model is just as efficient as solvent-free models, 
because Monte Carlo moves of phantom beads that are not in contact with a membrane 
are practically free of (computational) charge. It is robust, we do not need to 
tune the lipid potentials in order to achieve self-assembly. Pressure can be 
applied if necessary, and with a suitable dynamical model for the solvent
(\eg, dissipative particle dynamics), we can also study (hydro)dynamical phenomena. 
Compared with Smit-type models, the phantom solvent model has the advantage that 
the solvent is structureless and cannot induce artificial correlations.
On the other hand, the fact that it is compressible like a gas, may cause problems 
in certain dynamical studies.

To summarize, we have presented several possible ways to model the solvent 
environment of a self-assembled membrane, and discussed the advantages and 
disadvantages.  The different models are sketched schematically in
Fig.~\ref{fig:solvents}.

We shall now illustrate the potential of molecular coarse-grained models with two 
examples.

\subsubsection{First application example: The ripple phase}

\begin{figure}[t] 
\centerline{\includegraphics[width=1.3 in,angle=-90]{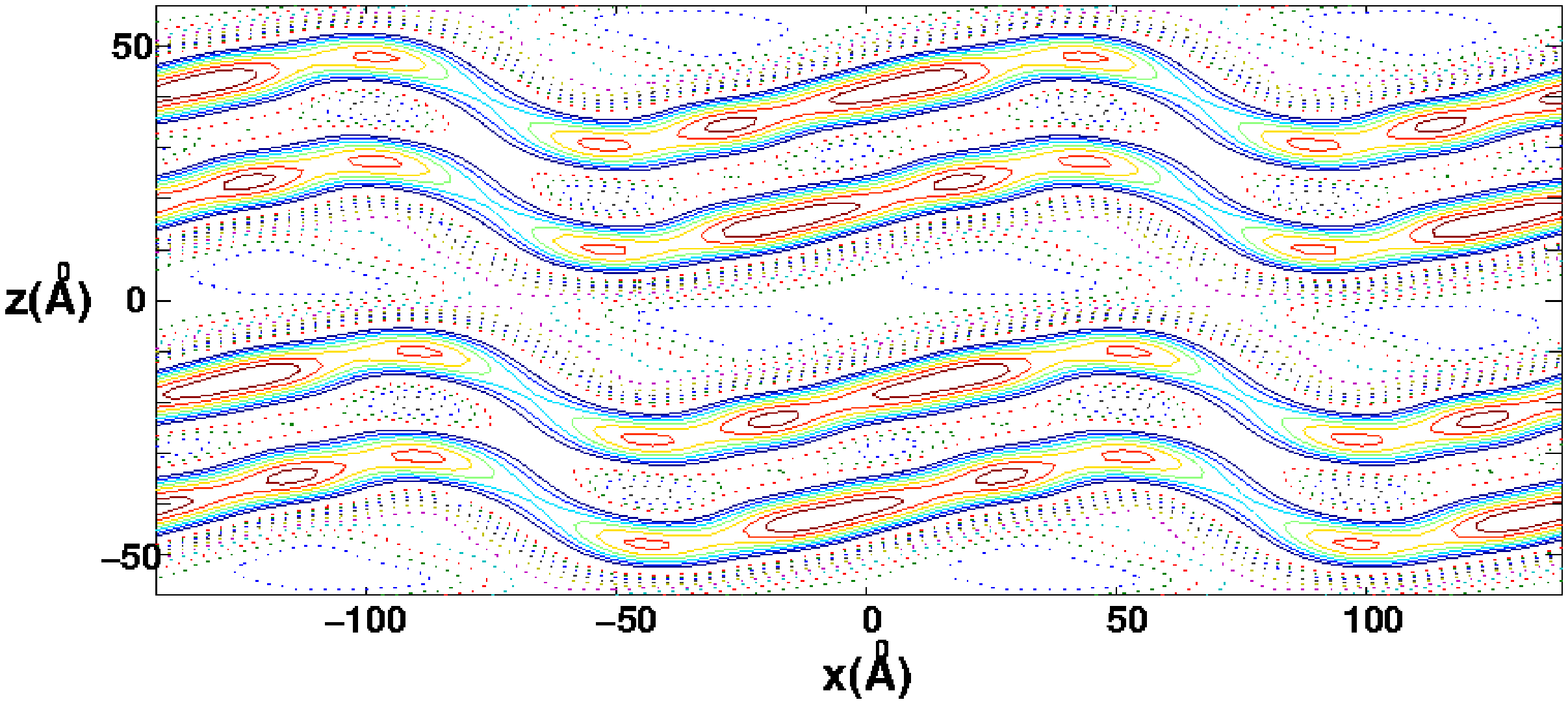}}
\centerline{\includegraphics[width=1.3 in,angle=-90]{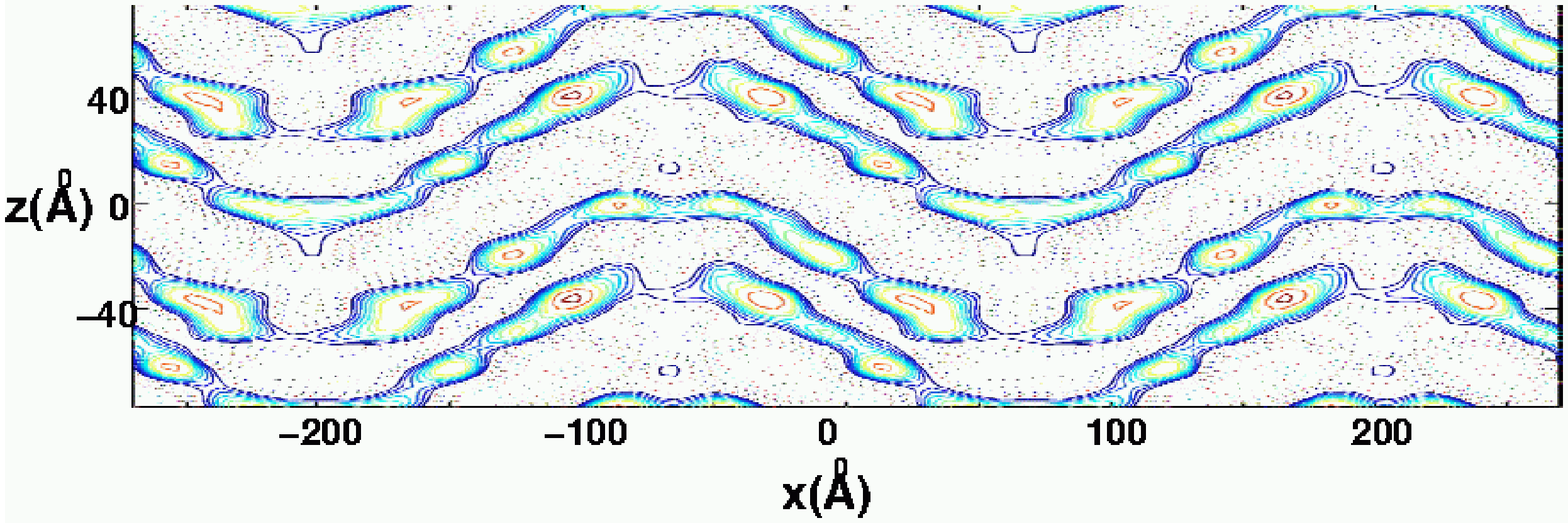}}
\vspace*{8pt}
\caption{
Electron density maps of (top) an asymmetric rippled state
(DMPC at 18.2 degrees) and (bottom) a symmetric rippled
state (DPPC at 39.2 ${}^0C$). Reprinted with permission from
Sengupta, Raghunathan, Katsaras~\protect\refcite{sengupta} (rescaled).
Copyright 2003 by the American Physical Society.
\label{fig:edms}
}
\end{figure}

Our first example shows how coarse-grained simulations can help to understand the 
inner structure of membranes. We have introduced earlier the liquid and gel 
phases in membranes (Fig.~\ref{fig:membrane_phases}). However, one rather mysterious 
phase was missing in our list: The ripple phase ($P_{\beta'}$). It emerges as an 
intermediate state between the tilted gel phase ($L_{\beta'}$) and the liquid phase 
($L_{\alpha}$), and according to freeze-etch micrographs, AFM pictures, and X-ray 
diffraction studies, it is characterized by periodic wavy membrane undulations.  
More precisely, one observes two different rippled structures: Figure~\ref{fig:edms} 
shows electron density maps of these two states, as calculated from X-ray data by 
Sengupta {\em et al}~\cite{sengupta}. The more commonly reported phase is the 
asymmetric ripple phase, which is characterized by sawtooth profile and has a 
wavelength of about 150 Angstrom. The symmetric rippled phase tends to form upon 
cooling from the liquid $L_{\alpha}$ phase and has a period twice as long. It is 
believed to be metastable. The molecular structure of both rippled states has 
remained unclear until very recently.

\begin{figure}[b] 
\centerline{\includegraphics[width=1.5in,angle=-90]{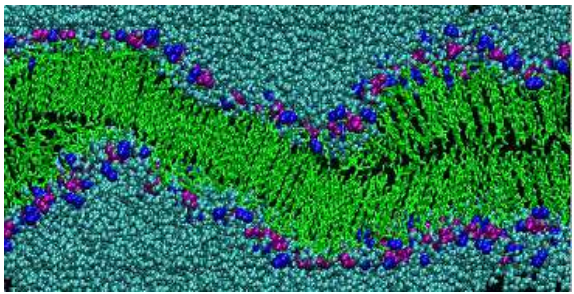}}
\vspace*{8pt}
\caption{
Asymmetric ripple in a Lecithin Bilayer, as observed
in an atomistic simulation. From de Vries {\em et al}~\protect\cite{vries}.
Copyright 2005 National Academy of Sciences, U.S.A.
\label{fig:lecithin_ripple}
}
\end{figure}

Atomistic simulations of de Vries {\em et al}, have finally shed light on the 
microscopic structure of the asymmetric ripple phase~\cite{vries}. These authors 
observed an asymmetric ripple in a Lecithin bilayer, which had a structure that
was very different from the numerous structures proposed earlier. 
Most strikingly, the rippled membrane is not a continuous bilayer. The ripple
is a membrane defect where a single layer spans the membrane, connecting
the two opposing sides (Fig.~\ref{fig:lecithin_ripple}).

The simulations of de Vries {\em et al} seem to clarify the structure of the 
asymmetric ripple phase, at least for the case of Lecithin. Still the question 
remains whether their result can be generalized to other lipid layers, or whether 
it is related to the specific properties of Lecithin head groups. This question 
can be addressed with idealized coarse-grained models.

Rippled phases had also been observed in simulations of coarse-grained models. 
Kranenburg {\em et al} have reported the existence of a rippled state in a Smit 
model with soft DPD interactions~\cite{kranenburg2,kranenburg3}. The structure 
of their ripple is different from that of Fig.~\ref{fig:lecithin_ripple}. 
Judging from this result, one is tempted to conclude that the structure of 
de Vries {\em et al} may not be generic.

However, we recently found ripples with a structure very similar to that 
of Fig.~\ref{fig:lecithin_ripple} in simulations of a simple bead-spring model by 
(Lenz and Schmid~\cite{olenz2,olenz3}). Our lipids are represented by linear chains 
with six tail beads (size $\sigma$) and one head bead (size $1.1 \sigma$), which are
connected by springs of size $0.7 \sigma$. Tail beads attract one another with a 
truncated and shifted Lennard-Jones potential. Head beads are slightly larger and 
purely repulsive. The solvent environment is modelled by a fluid of phantom beads 
of size $\sigma$. The system was studied with Monte Carlo simulations at constant 
pressure, using a simulation box of fluctuating size and shape in order to ensure 
that the pressure tensor is diagonal.

\begin{figure}[t] 
\centerline{\includegraphics[width=1.2in,angle=-90]{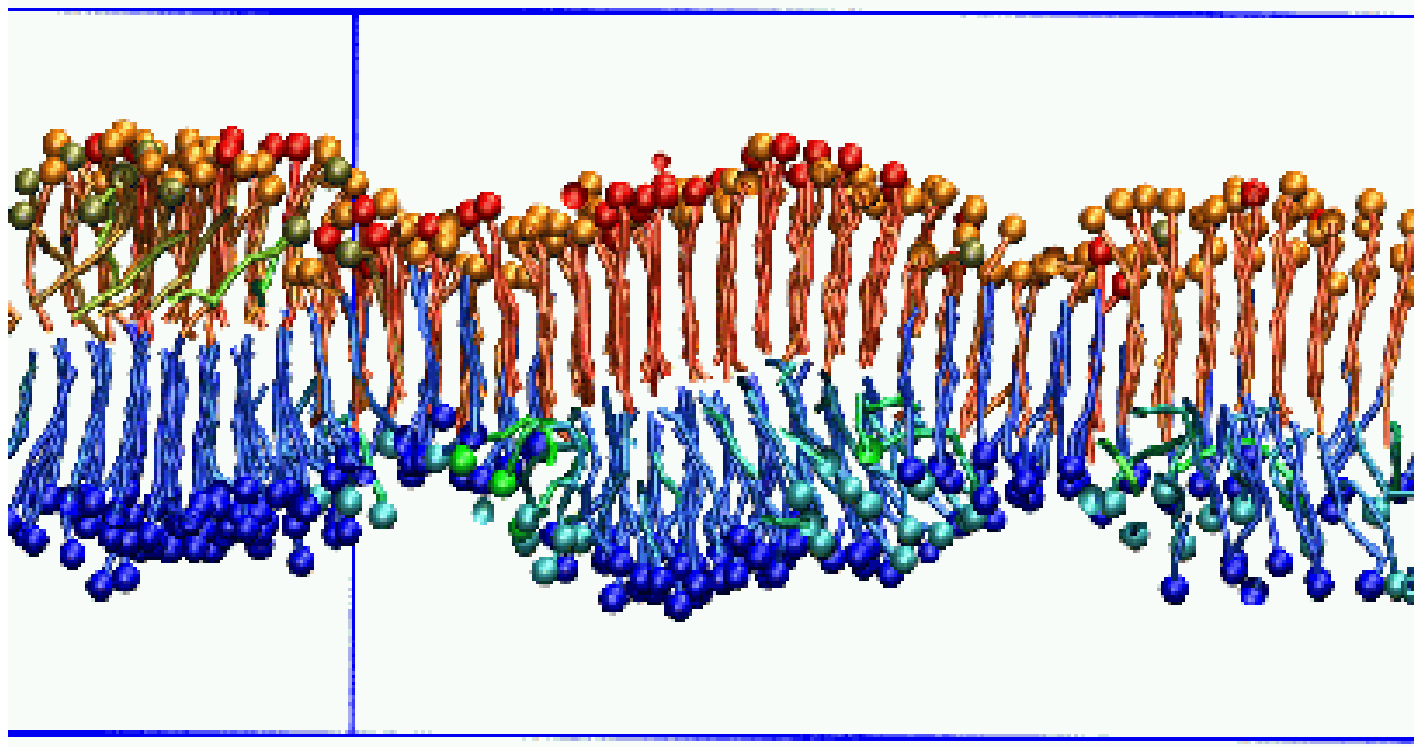}}
\centerline{\includegraphics[width=3.5in]{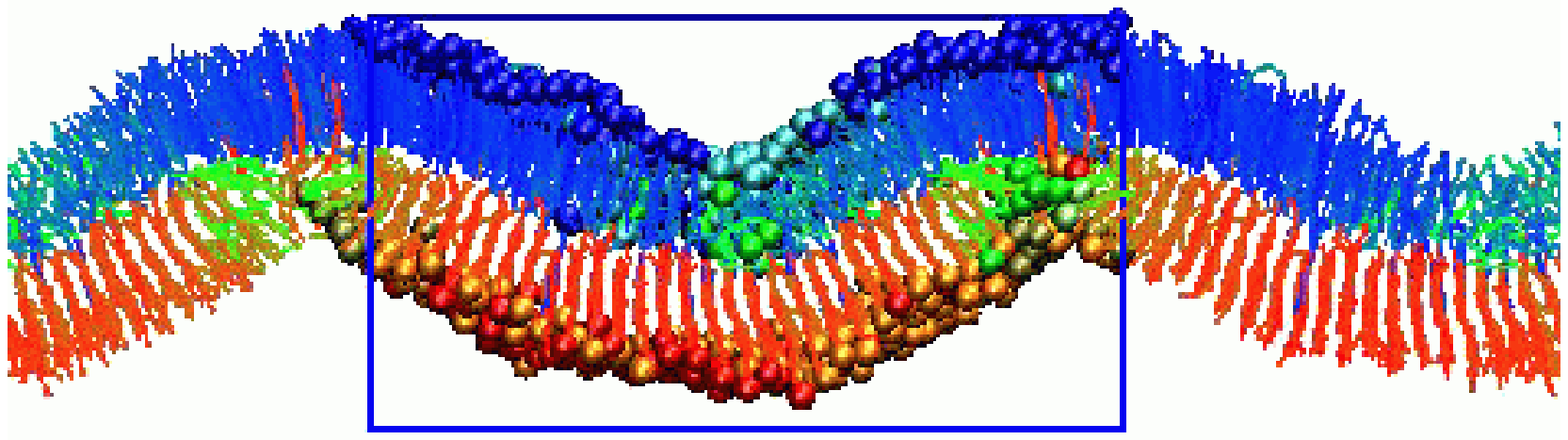}}
\vspace*{8pt}
\caption{
Ripples from simulations of a coarse-grained model.
(a) Asymmetric ripple
(b) Symmetric ripple.
\label{fig:ripples}
}
\end{figure}

Membranes were found to self-assemble spontaneously at sufficiently low 
temperatures, they exhibit a fluid $L_{\alpha}$-phase as well as a tilted gel 
$L_{\beta'}$-phase. The two phases are separated by a temperature region where 
ripples form spontaneously. In small systems, the ripples are always asymmetric. 
In larger systems, asymmetric ripples form upon heating from the $L_{\beta'}$-phase. 
Upon cooling from the $L_{\alpha}$-phase, a second, symmetric, type of ripple forms,
which has twice the period than the asymmetric ripple and some structural 
similarity, except that the bilayer remains continuous (see Fig.~\ref{fig:ripples}). 
The sideview of this second ripple has striking similarity with the density maps of
Fig.~\ref{fig:edms} b).

The coarse-grained simulations of Lenz thus not only demonstrate that the asymmetric 
ripple structure found by de Vries is generic, and can be observed in highly 
idealized membrane models as well. They also suggest a structural model for the yet
unresolved symmetric ripple.

\subsubsection{Second application example: Membrane stacks\label{sec:stacks}}

In our second example, we consider the fluctuations and defects in stacks of 
membranes. We shall show how simulations of coarse-grained membrane models 
can be used to test and verify mesoscopic models, and how they can bridge between 
coarse-graining levels. 

At the mesoscopic level, membranes are often represented by random interfaces 
(see also Section \ref{sec:random_interfaces}). One of the simplest theories for 
thermal fluctuations in membrane stacks is the ``discrete harmonic 
model''~\cite{lei}. It describes membranes without surface tension, and assumes 
that the fluctuations are governed by two factors: The bending stiffness $K_c$ 
of single membranes, and a penalty for compressing or swelling the stack,
which is characterized by a compressibility modulus $B$. The free energy is
given in harmonic approximation
\begin{equation}
\label{dh} 
{\cal F}_{\mbox{\tiny DH}} = \sum_n \int_A dx \: dy \:
\Big\{ \frac{K_c}{2} (\frac{\partial^2 h_n}{\partial x^2} +
\frac{\partial^2 h_n}{\partial y^2})^2 + \frac{B}{2} (h_n -
h_{n+1} + \bar{d})^2 \Big\},
\end{equation}
where $\bar{d}$ is the average distance between layers. The first part describes 
the elasticity of single membranes, and the second part accounts for the 
interactions between membranes.  Being quadratic, the free energy functional 
(\ref{dh}) is simple enough that statistical averages can be calculated exactly.

We have tested this theory in detail with coarse-grained molecular simulations 
(C. Loison {\em et al}~\cite{cloison1}) of a binary mixture of amphiphiles and 
solvent, within a Smit-type model originally introduced by Soddemann 
{\em et al}~\cite{soddemann1}. The elementary units are spheres with a hard core 
radius $\sigma$, which may have two types: ``hydrophilic'' or ``hydrophobic''. 
Beads of the same type attract each other at distances less than $1.5 \sigma$. 
``Amphiphiles'' are tetramers made of two hydrophilic and two hydrophobic beads, 
and ``solvent'' particles are single hydrophilic beads. We studied systems of up 
to 153600 beads with constant pressure molecular dynamics, using a simulation
box of fluctuating size and shape, and a Langevin thermostat to maintain constant
temperature. 

\begin{figure}[t]
\begin{center}
\centerline{
\parbox{1.5in}{ \includegraphics[width=1.3in]{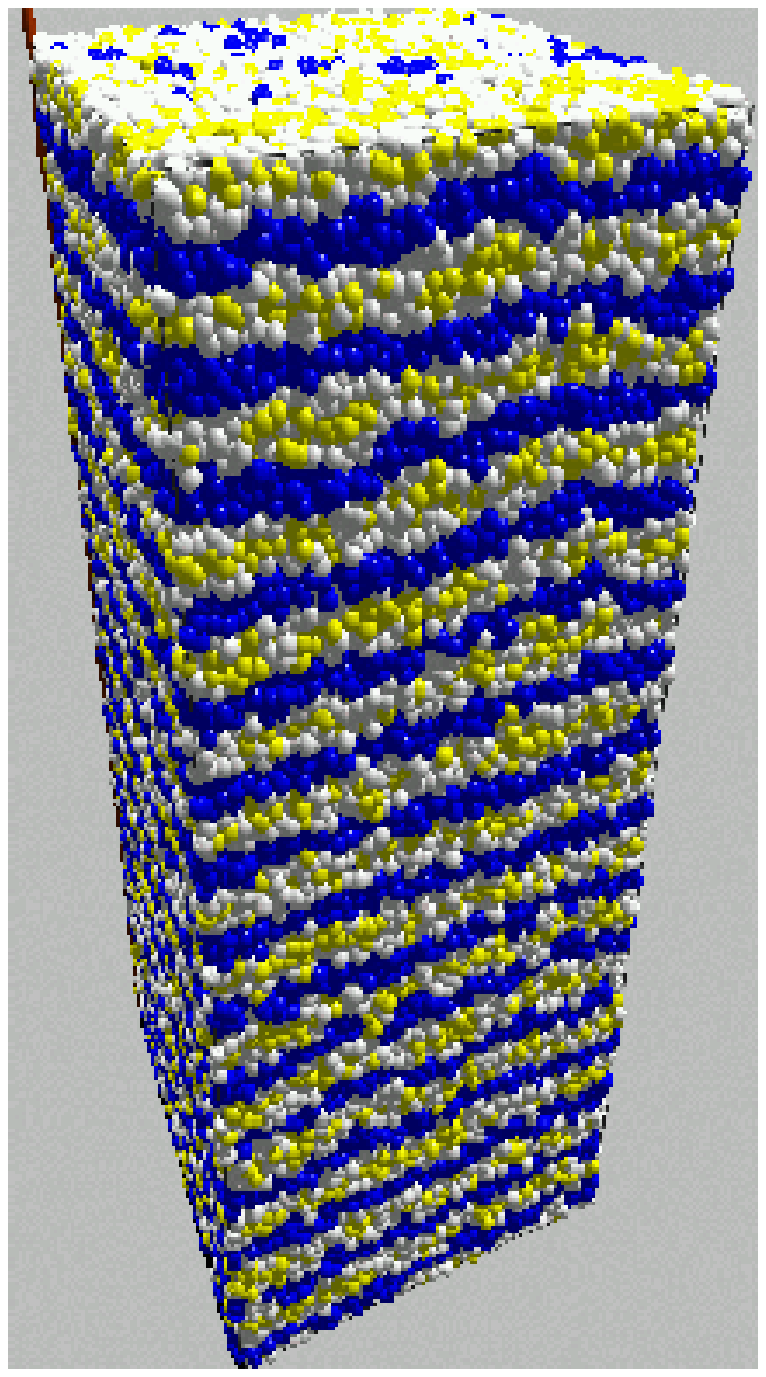}}
\hspace*{1cm}
\parbox{2in}{ \includegraphics[width=2in]{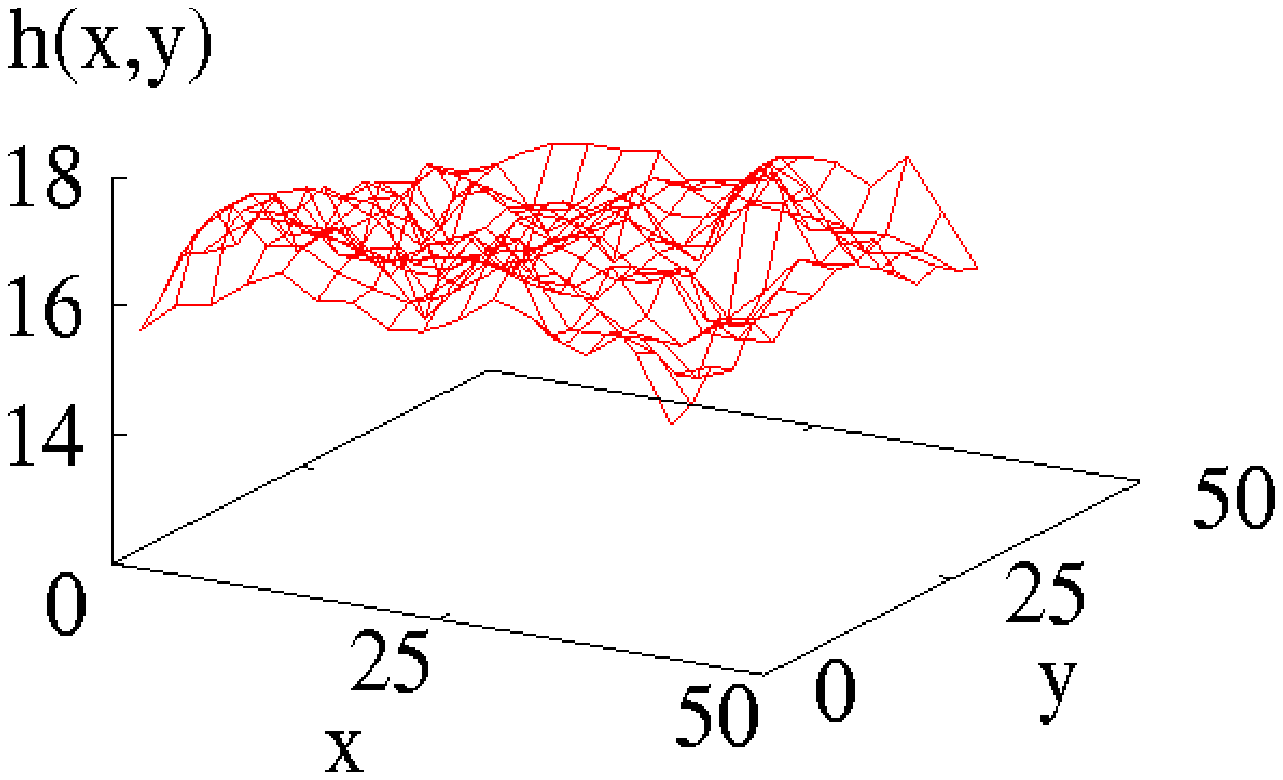}} 
}
\caption{
Left: Snapshot of a bilayer stack (30720 amphiphiles
and 30720 solvent beads). The ``hydrophobic'' tail
beads are dark, the ``hydrophilic'' head beads and
the solvent beads are light;
Right: Snapshot of a single membrane position $h_n(x,y)$. 
\label{fig:claire_snapshot}
}
\end{center}
\end{figure}

At suitable pressures and temperatures, the system assumes a fluid lamellar phase. 
Our configurations contained up to 15 stacked bilayers, which contained 20 volume 
percent solvent. A configuration snapshot is shown in 
Fig.~\ref{fig:claire_snapshot} (left).
In order to test the theory (\ref{dh}), one must first determine the local positions 
of the membranes in the stack. This was done as follows:
\begin{enumerate}
\item
  The simulation box was divided into $N_x N_y N_z$ cells.  ($N_x = N_y = 32$). 
  The size of the cells fluctuates because the dimensions of the box fluctuate.
\item
  In each cell, the relative density of tail beads $\rho_{\mbox{\tiny tail}}(x,y,z)$ 
  was calculated. It is defined as the ratio of the number of tail beads and the 
  total number of beads.
\item
  The hydrophobic space was defined as the space where the relative density of tail 
  beads is higher than a given threshold $\rho_0$. (The value of the threshold was 
  roughly 0.7, \ie, 80 \% of the maximum value of $\rho_{\mbox{\tiny tail}}$).
\item
  Cells belonging to the hydrophobic space are connected to clusters. Two 
  hydrophobic cells that share at least one vortex are attributed to the same 
  cluster.  Each cluster defines a membrane. 
\item
  For each membrane $n$ and each position $(x,y)$, the two heights 
  $h_n^{\mbox{\tiny min}}(x,y)$ and $h_n^{\mbox{\tiny max}}(x,y)$, where the density
  $\rho_{\mbox{\tiny tail}}(x,y,z)$ passes through the threshold $\rho_0$, are 
  determined. The mean position is defined as the average 
  $h_n(x,y) = (h_n^{\mbox{\tiny min}} + h_n^{\mbox{\tiny max}})/2. $
\end{enumerate}
The algorithm identifies membranes even if they have pores. At the
presence of other defects, such as necks or passages (connections
between membranes), additional steps must be taken, which are not
described here. A typical membrane conformation $h_n(x,y)$ is
shown in Fig.~\ref{fig:claire_snapshot} (right). 

\begin{figure}[b]
\begin{center}
\centerline{\includegraphics[width=1.3in,angle=-90]{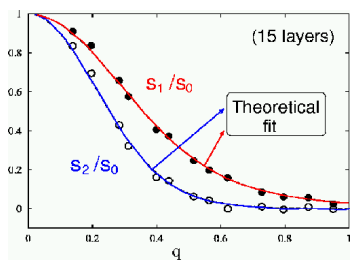}}
\caption{\label{fig:trans_struc}
Ratios of transmembrane structure factors $s_1/s_0$ and $s_2/s_0$
vs. in-plane wavevector $q$ in units of $\sigma^{-1}$.
The solid lines correspond to a theoretical fit to Eq.~(\ref{sn})
with one (common) fit parameter $K_c/B$.
After Ref.~\protect\refcite{cloison1}.
}
\end{center}
\end{figure}

The statistical distribution of $h_n(x,y)$ can be analyzed and compared with
theoretical predictions. For example, the transmembrane structure factor, which 
describes correlations between membrane positions in different membranes is
given by~\cite{cloison1}
\begin{equation}
\label{sn}
s_n(q) = \langle h_m({\bf q})^* h_{m+n}({\bf q}) \rangle
= s_0(q) \: \Big[
1 + \frac{X}{2} - \frac{1}{2} \sqrt{X (X + 4)} \Big]^n,
\end{equation}
where $h_n({\bf q})$ is the Fourier transform of $h_n(x,y)$ in the $(x,y)$-plane 
and $X = q^4 K_c/B$ is a dimensionless parameter.  The function $s_0(q)$ can also 
be calculated explicitly~\cite{cloison1}. The prediction (\ref{sn}) can be tested 
by simply plotting the ratio $s_n/s_0$ vs. $q$ for different $n$. The functional 
form of the curves should be given by the expression in the square brackets, with 
only one fit parameter $K_c/B$. Figure~\ref{fig:trans_struc} shows the simulation 
data. The agreement with the theory is very good over the whole range of wave 
vectors $q$.

Thus the molecular simulations confirm the validity of the mesoscopic model 
(\ref{dh}). Moreover, the analysis described above yields a value for the 
phenomenological parameter $K_c/B$. By analyzing other quantities, one can
also determine $K_c$ and $B$ separately. This gives the elastic parameters of 
membranes and their effective interactions, and establishes a link between 
the molecular and the mesoscopic description.

Many important characteristics of membranes not only depend on their elasticity, 
but also on their defects. For example, the permeability is crucially influenced 
by the number and properties of membrane pores. A number of atomistic and coarse 
grained simulation studies have therefore addressed pore 
formation~\cite{mueller,marrink,zahn,tolpekina,wang}, mostly in membranes under 
tension. In contrast, the membranes in the simulations of Loison {\em et al} 
have almost zero surface tension. This turns out to affect the characteristics 
of the pores quite dramatically~\cite{cloison2,cloison3}.

\begin{figure}[t]
\begin{center}
\centerline{\includegraphics[width=1.6in,angle=-90]{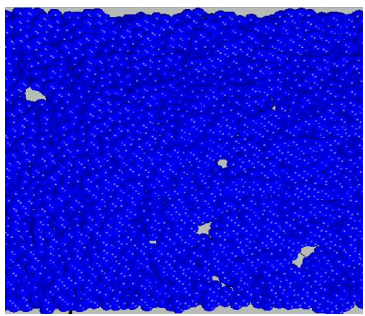}}
\caption{\label{fig:pore_snapshot} Snapshot of a single bilayer
(top view. Only hydrophobic beads are shown). }
\end{center}
\end{figure}

Figure~\ref{fig:pore_snapshot} shows a snapshot of a hydrophobic layer which 
contains a number of pores. The pores have nucleated spontaneously. They ``live'' 
for a while, grow and shrink without diffusing too much, until they finally 
disappear. Most pores close very quickly, but a few large ones stay open for a 
long time. A closer analysis shows that pores are hydrophilic, \ie, the amphiphiles 
rearrange themselves at the pore edge, such that solvent beads in the pore center 
are mainly exposed to head beads. The total number of pores is rather high in our 
system. This is because the amphiphiles are short and the membranes are thin.

The analysis algorithm presented above not only localizes membranes, but also 
identifies pores, their position and their shape. Therefore one can again test 
the appropriate mesoscopic theories for pore formation. The simplest Ansatz for 
the free energy of a pore with the area $A$ and the contour length $c$ has the 
form~\cite{lister}
\begin{equation}
\label{pore_energy}
E = E_0 + \lambda \: c - \gamma \: A,
\end{equation}
where $E_0$ is a core energy, $\lambda$ a material parameter called line tension, 
and $\gamma$ the surface tension. The second term describes the energy penalty at 
the pore rim. The last term accounts for the reduction of energy due to the release 
of surface tension in a stretched membrane. In our case, the surface tension is 
close to zero ($\gamma \approx 0$), because the simulation was conducted such that 
the pressure tensor is isotropic. Thus the last term vanishes. 

\begin{figure}[t]
\centerline{
\includegraphics[width=2.in]{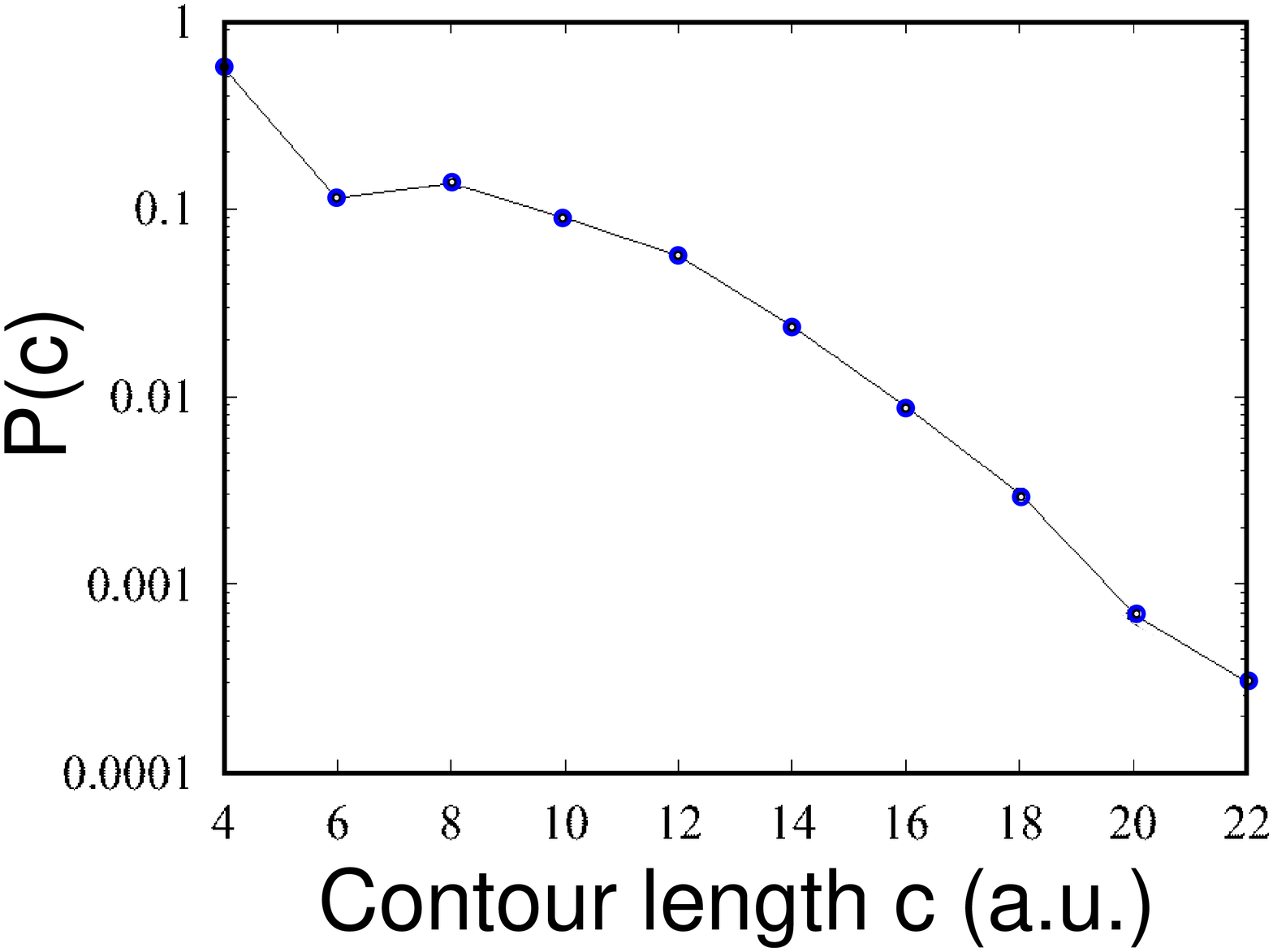}
\hspace*{1cm}
\includegraphics[width=2.in]{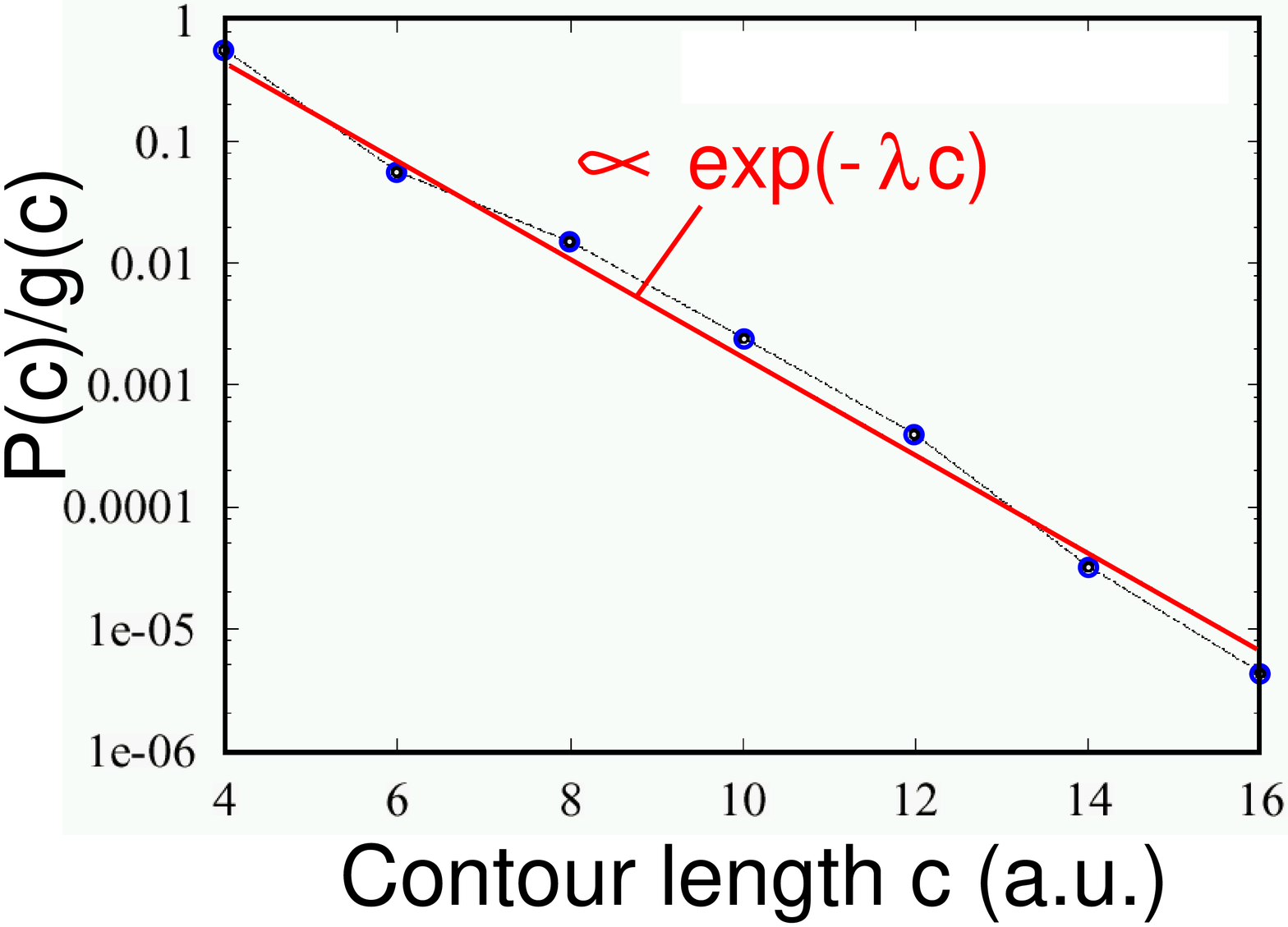}
}
\caption{
\label{fig:hist_contours}
Distribution of pore contours in a semi-logarithmic plot.
Left: Raw data,
Right: Divided by the degeneracy function $g(c)$.
After Ref.~\protect\refcite{cloison2}.
}
\end{figure}

In this simple free energy model, the pore shapes should be distributed according 
to a Boltzmann distribution, $P(c) \propto \exp(- \lambda c)$.
Figure~\ref{fig:hist_contours} (left) shows a histogram of contour lengths $P(c)$. 
The bare data do not reflect the expected exponential behavior. Something is 
missing. Indeed, a closer look reveals that the naive exponential Ansatz disregards 
an important effect: The ``free energy'' (\ref{pore_energy}) gives only {\em local} 
free energy contributions, \ie, those stemming from local interactions and local 
amphiphile rearrangements. In addition, one must also account for the {\em global} 
entropy of possible contour length conformations. Therefore, we have to evaluate the
``degeneracy'' of contour lengths $g(c)$, and test the relation
\begin{equation}
\label{pc_2}
P(c) \propto g(c) \exp(- \lambda c).
\end{equation}
Figure~\ref{fig:hist_contours} (right) demonstrates that this second Ansatz
describes the data very well. From the linear fit to the data, one can extract 
a value for the line tension $\lambda$.

\begin{figure}[t]
\begin{center}
\centerline{\includegraphics[width=1.3in,angle=-90]{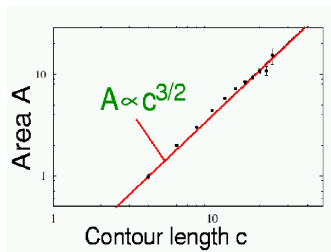}}
\caption{\label{fig:area_contour}
Pore area vs. contour length (arbitrary units).
After Ref.~\protect\refcite{cloison2}.
}
\end{center}
\end{figure}

The model (\ref{pore_energy}) makes a second important prediction: Since the free 
energy only depends on the contour length, pores with the same contour length are 
equivalent, and the shapes of these pores should be distributed like those of 
two-dimensional self-avoiding ring polymers. In other words, they are not round,
but have a fractal structure. From polymer theory, one knows that the size $R_g$ 
of a two dimensional self-avoiding polymer scales roughly like $R_g \propto N^{3/4}$ 
with the polymer length $N$. In our case, the ``polymer length'' is the contour 
length $c$. Thus the area $A$ of a pore should scale like
\begin{equation}
A \propto R_g^2 \propto ( C^{3/4}) ^2 = C^{3/2}.
\end{equation}
Figure~\ref{fig:area_contour} shows that this is indeed the case.

Other properties of pores have been investigated, \eg, dynamical properties,
pore lifetimes, pore correlations etc.~\cite{cloison2}. All results were in 
good agreement with the line tension model (\ref{pore_energy}). In sum, we 
find that the fluctuations and (pore) defects in membrane stacks can be described 
very well by a combination of two simple mesoscopic theories: An effective 
interface model for membrane undulations (Eq.~(\ref{dh})), and a line tension 
model for the pores (Eq.~(\ref{pore_energy})~\cite{cloison3}.

The simulations of Loison {\em et al} thus demonstrate nicely that simulations 
of coarse-grained molecular models can be used to test mesoscopic theories, 
and to bridge between the microscopic and the mesoscopic level.

\subsection{Mesoscopic membrane models \label{sec:random_interfaces}}  

When it comes to large scale structures and long time dynamics, simulations of 
molecular models become too cumbersome. It then proves useful to drop the notion 
of particles altogether and work directly with continuous mesoscopic theories. 
In general, continuum theories are mostly formulated in terms of continuous fields, 
which stand for local, coarse-grained averages of some appropriate microscopic 
quantities. In the case of membranes systems, it is often more adequate to keep 
the membranes as separate objects, despite their microscopic thickness, and treat 
them as interfaces in space. This leads to random-interface theories.

\subsubsection{Random interface models}

In random-interface models, membranes are represented by two-dimensional sheets 
in space, whose conformations are distributed according to an effective interface 
free energy functional. An example of such a functional has already been given 
in Eq.~(\ref{dh}). However, the free energy (\ref{dh}) can only be used to 
describe slightly fluctuating planar membranes in the $(xy)$-plane. In order to 
treat membranes of arbitrary shape (vesicles etc.), one must resort to the 
formalism of differential geometry~\cite{safran}.

One of the simplest approaches of this kind is the spontaneous curvature model. 
The elastic free energy of a membrane is given by~\cite{helfrich}
\begin{equation}
\label{interface}
{\cal F}_{\mbox{\tiny el}} = 
\int dA \big[
\frac{\kappa}{2} (2 H - H_0)^2 + \bar{\kappa} K),
\big]
\end{equation}
where $\int \!\! dS \cdot$ integrates over the surface of the membrane, and 
$H$ and $K$ are the mean and Gaussian curvature, which are related to the 
local curvature radii $R_1$ and $R_2$ via \mbox{$H = (1/R_1 + 1/R_2)/2$}, 
\mbox{$K = 1/(R_1 R_2)$}~\cite{weisstein}. The parameters $\kappa$, 
$\bar{\kappa}$ (the bending stiffnesses) and $H_0$ (the spontaneous curvature) 
depend on the specific material properties of the membrane. The total 
membrane surface $A = \int \!\! dS $ is taken to be constant.

In the case $H_0= 0$ and $\kappa >0, \bar{\kappa} > 0$, the free energy
(\ref{interface}) describes membranes that want to be flat, and imposes a 
bending penalty on all local curvatures. $H_0 \ne 0$ implies that the membrane 
has a favorite mean curvature. It is worth noting that the integral over the 
Gaussian curvature, $\int \!\! K \: dS$, depends solely on the topology of 
the interfaces. For membranes without rims, the Gauss-Bonnet theorem
states that
\begin{equation}
\int dS \: K = 2 \pi \chi_E,
\end{equation}
where the Euler characteristics $\chi_E = 2(c-g)$ counts the number of closed 
surfaces $c$ (including cavities) minus the number of handles $g$. 

Other free energy expressions have been proposed, which allow for an asymmetric 
distribution of lipids between both sides of the membrane, and/or for fluctuations 
of $A$~\cite{evans,seifert}. Here, we shall only discuss the spontaneous curvature 
model. Our membranes shall be self-avoiding, \ie, they have excluded volume 
interactions and cannot cross one another. 

Our task is to formulate a simulation model that mimicks the continuous space theory 
as closely as possible. More precisely, we first need a method that samples
random self-avoiding surfaces, and second a way to discretize the free energy
functional (\ref{interface}) on those surfaces.

\begin{figure}[t] 
\centerline{\includegraphics[width=2.3in]{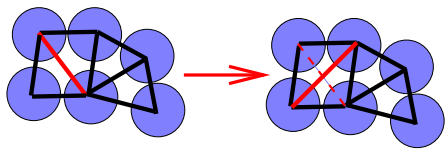}}
\vspace*{8pt}
\caption{
Connectivity update in the
dynamic triangulation algorithm
\label{fig:triangulation}
}
\end{figure}

A widely used method to generate self-avoiding surfaces is the dynamic triangulation 
algorithm~\cite{kazakov,billoire,kantor,ho,kroll,gompper_rev}. Surfaces are modeled 
by triangular networks of spherical beads, which are linked by tethers of some 
given maximum length $l_0$ (tethered-bead model). The tethers define the neighbor 
relations on the surface. In order to enforce self-avoidance, the beads are 
equipped with hard cores (diameter $\sigma$), and the maximum tether length is 
chosen in the range $\sigma < l_0 < \sqrt{3} \sigma$. The surfaces are sampled 
with Monte Carlo, with two basic types of moves: Moves that change the 
{\em position} of beads, and moves that change the {\em connectivity} of the 
network, \ie, redistribute the tethers between the chains. The latter is done 
by randomly cutting tethers and flipping them between two possible diagonals of 
neighboring triangles (Fig.~\ref{fig:triangulation}). Updates of bead positions 
are subject to the constraint that the maximum tether length must not exceed $l_0$.
In addition, connectivity updates have to comply with the requirement that at 
least three tethers are attached to one bead, and that two beads cannot be connected
to one another by more than one tether. Otherwise, attempted updates are accepted 
or rejected according to the standard Metropolis prescription. Of course,
additional, more sophisticated Monte Carlo moves can be introduced as well.

Next, the free energy functional (\ref{interface}) must be adapted for the 
triangulated surface. If no rims are present, we only need an expression for 
the first term. Since the integral over the Gaussian curvature $K$ is then
essentially given by the Euler-characteristic, it is much more convenient and 
accurate to evaluate the latter directly (\ie, count closed surfaces minus handles) 
than to attempt a discretization of $K$.

The question how to discretize the bending free energy in an optimal way is 
highly non-trivial, and several approximations have been proposed over the 
years~\cite{gompper_rev}. Here we give a relatively recent expression due to Kumar 
{\em et al} in 2001~\cite{kumar1}: The surface integral $\int dS$ is replaced by 
a sum over all triangles $t$ with area $A_t$. For each triangle $t$, one considers 
the three neighbor triangles $t'$ and determines the length $L_{tt'}$ of the common 
side as well as the angle $\theta_{tt'}$ between the two triangle normal vectors. 
The free energy is then approximated by
\begin{equation}
\label{interface2}
{\cal F}_{\mbox{\tiny el}} = 
\frac{\kappa}{2} \int dA (2 H - H_0)^2 
\approx
\frac{\kappa}{2} \sum_t A_t 
\big( \frac{\sum_{t'} L_{tt'} \theta_{tt'}}{2 A_t} - H_0 \big)^2.
\end{equation}

This completes the tethered-bead representation for random interfaces. 
We have presented one of the most simple variants. The models can be 
extended in various directions, \eg, to represent mixed membranes~\cite{kumar2}, 
membranes that undergo liquid/gel transitions~\cite{gompper1}, membranes with
fluctuating geometry~\cite{gompper2} or pores~\cite{shillcock2} etc. 
They have even been used to study vesicle dynamics, \eg, vesicles in shear
flow~\cite{noguchi4,noguchi5}. The dynamics can be made more realistic 
by using hybrid algorithms, where the beads are moved according to molecular 
dynamics or Brownian dynamics, and the tethers are flipped with Monte Carlo.
Unfortunately, the tether flips cannot easily be integrated in a pure molecular
dynamics simulation. This is a slight drawback of the tethered-bead approach.

An interesting alternative has recently been proposed by Noguchi 
{\em et al}~\cite{noguchi6}. He developed a specific type of potential, 
which forces (single) beads to self-assemble into a two-dimensional membrane 
without being connected by tethers. The membrane is coarse-grained as a curved 
surface as in tethered-bead models, but tethers are no longer necessary. 
This opens the way for a fully consistent treatment of dynamics in membrane 
systems with frequent topology changes. 

\subsubsection{Application example: passage of vesicles through pores}

As an example for an application of a tethered-bead model, we shall discuss 
a problem from the physics of vesicles. Linke, Lipowsky and Gruhn have recently 
investigated the question whether a vesicle can be forced to cross a narrow
pore by osmotic pressure gradients~\cite{linke1,linke2}. Vesicle passage through 
pores plays a key role in many strategies for drug delivery. On passing through 
the pore, the vesicle undergoes a conformational change, which is expensive and 
creates a barrier. On the other hand, the vesicle can reduce its energy, if it 
crosses into a region with lower osmotic pressure. Linke {\em et al} studied the
question, whether the barrier height can be reduced by a sufficient amount
that the pore is crossed spontaneously.

\begin{figure}[t] 
\centerline{\includegraphics[width=1.5in,angle=-90]{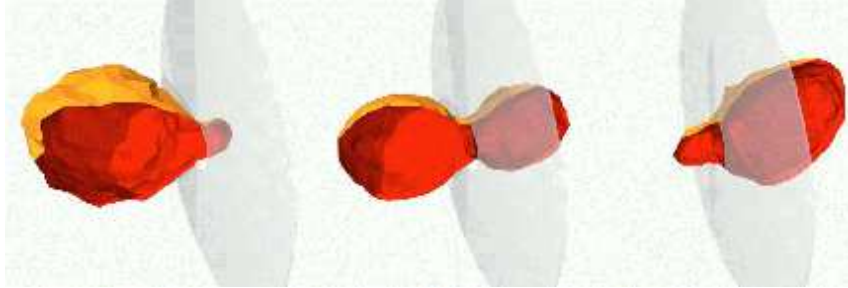}}
\vspace*{8pt}
\caption{
Osmotically driven vesicle translocation through a pore.
Configuration snapshots.
From G. T. Linke~\protect\cite{linke1}.
\label{fig:vesicle}
}
\end{figure}

To this end, they considered a vesicle filled with $N$ osmotically active particles.
The concentration of these particles outside the vesicle is given by $c_{1,2}$ on 
the two sides of the pore. For a vesicle in the process of crossing the pore
from side 1 to side 2, the vesicle area on both sides of the membrane is denoted 
$A_{1,2}$, and the volumes are $V_{1,2}$. Hence the osmotic contribution to the 
free energy is
\begin{equation}
\label{interface3}
{\cal F}_{\mbox{\tiny osm}} = 
(-N \ln ((V_1+V_2)/V_0) + c_1 V_1 + c_2 V_2),
\end{equation}
where $V_0$ is some reference volume. In the Monte Carlo simulations, a series
of different crossing stages was sampled (see Fig.~\ref{fig:vesicle}). This was 
done by introducing an appropriate reweighting factor, which kept the area $A_2$
within a certain, pre-determined range. Several simulations were done with
different windows for $A_2$. The procedure has similarity with umbrella sampling, 
except that the windows did not overlap. Nevertheless, the extrapolation of the 
histograms allowed to evaluate the free energy quite accurately, as a function 
of $A_2$.

\begin{figure}[htb] 
\centerline{\includegraphics[width=2.5in]{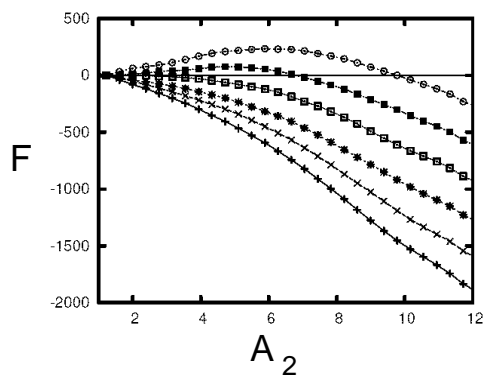}}
\vspace*{8pt}
\caption{
Free energy landscape on crossing the pore for different osmolarities $c_1$ 
starting from $c_1 = 2700$ (open circles) to $c_1 = 3100$ (plus) in steps
of 100. All units are arbitrary. The osmolarity $c_2$ is kept fixed. See text
for more explanation.
From G. Linke~{\em et al}\protect\cite{linke1,linke2}.
\label{fig:barrier}
}
\end{figure}

The resulting free energy landscape is shown in Fig.~\ref{fig:barrier}, for 
different osmolarities $c_1$ at the side where the vesicle starts. As expected, 
one finds a free energy barrier for low $c_1$. On increasing $c_1$, the height 
of the barrier decreases, until it finally disappears. Hence the vesicle
can be forced to cross the pore spontaneously, if $c_1$ is chosen sufficiently
high. In contrast, changing the bending rigidity has hardly any effect on 
the height of the potential barrier.

This application shows how mesoscopic models of membrane 
systems can be used to study the physical basis of a
potentially technologically relevant process. A comparable study 
with a particle-based model would have been much more expensive, 
and both the setup and the analysis would have been more 
difficult: One has no well-defined interface, one has only
indirect information on the bending stiffness etc.
The simplifications in the model clearly help not only
to get results more quickly, but also to understand
them more systematically.

\subsection{Summary}

In this section, we have introduced and discussed two important 
classes of coarse-grained models for membrane systems: 
Bead-spring models for simulations on the coarse-grained 
molecular level, and random interface models for the 
coarse-grained mesoscopic level. We have shown that such 
studies can give insight into generic properties of membranes 
and into basic physical mechanisms on different scales. 
We have also shown that it is possible to bridge between 
different levels, and that simulations of particle-based 
models can be used to test mesoscopic theories. 

\section{Liquid crystals and surfactant layers under shear\label{sec:shear}} 

So far, we have talked about equilibrium simulations. The special properties 
of complex fluids lead to particularly peculiar behavior at nonequilibrium. 
In this section, we will discuss coarse-grained simulations of complex fluids 
under shear. As in the previous section, we will move from the coarse-grained 
molecular level to the mesoscopic level. However, we will now focus on the 
simulation methods rather than on the construction of simulation models. 
As example systems, we will consider surfactant layers and liquid crystals.

\subsection{Introduction \label{sec:intro}}  

\subsubsection{Strain rate and shear stress}  

We begin with recalling some basic fluid mechanics~\cite{guyon}. In the continuum 
description, the fluid is thought to be divided into fluid elements, which are 
big enough that microscopic details are washed out, but small enough that each 
element can still be considered to be homogoneous. On a macroscopic scale, the 
fluid is described by spatially varying fields, \eg, the velocity or flow field 
$\vec{u}(\vec{r},t)$. The velocity gradient in the direction perpendicular 
to the flow is called the local strain rate.

One of the central quantities in fluid mechanics is the stress tensor $\sigma$. 
It describes the forces that the surrounding fluid exerts on the surfaces of a 
fluid element: For a surface with orientation $\vec{N}$ ($|\vec{N}|=1$), the force 
per area is $F_i/A = \sum_j \sigma_{ij} N_j$. The diagonal components of $\sigma$ 
give the longitudinal stress and correspond to normal forces. The off-diagonal 
components give the shear stress and correspond to tangential forces.

A fluid element with the velocity $\vec{u}$ is thus accelerated according to
\begin{equation}
\label{acceleration}
\rho \frac{d}{dt} u_j = \sum_i \partial_i \sigma_{ij},
\end{equation}
where $\rho$ is the local mass density. The stress tensor $\sigma_{ij}$ is often 
divided in three parts.
\begin{equation}
\label{total_stress}
\sigma_{ij} = - p \: \delta_{ij} + \sigma_{ij}^{\mbox{\tiny elastic}}
+ \sigma_{ij}'
\end{equation}
The first term describes the effect of the local pressure. It is always present, 
even in an equilibrium fluid at rest. The second term, 
$\sigma_{ij}^{\mbox{\tiny elastic}}$, accounts for elastic restoring forces,
if applicable. The last term, $\sigma_{ij}'$, gives the contribution of viscous 
forces and is commonly called the viscous stress tensor. It vanishes in the absence 
of velocity gradients, \ie, if $\partial_i v_j \equiv 0$. The concrete relation 
between the stress tensor and the local fields is called constitutive equation and
characterizes the specific fluid material. 

In standard hydrodynamics, one considers so-called Newtonian fluids, which have no 
elasticity ($\sigma_{ij}^{\mbox{\tiny elastic}} \equiv 0$) and a linear, 
instantaneous relation between the viscous stress tensor and the velocity gradient,
\begin{equation}
\label{stress_linear}
\sigma_{ij}' = \sum_{kl} L_{ijkl} \: \partial_k u_l.
\end{equation}
In the case of an incompressible, isotropic, Newtonian fluid, one has only one 
independent material parameter, the kinematic viscosity $\eta$, and the constitutive 
equation reads
\begin{equation}
\label{newton}
\sigma_{ij} = - p \: \delta_{ij} + 
\eta (\partial_i u_j + \partial_j u_i),
\end{equation}
where the value of the local pressure $p(\vec{r})$ is determined by the
requirement that the incompressibility condition 
\begin{equation}
\label{incompressibility}
\nabla \vec{u} \equiv 0.
\end{equation}
is always fulfilled. Equations~(\ref{acceleration}) in combination with 
(\ref{newton}) are the famous Navier-stokes equations (without external forces).

Complex fluids have internal degrees of freedom, and large characteristic length 
and time scales. Therefore, they become non-Newtonian already at low shear stress, 
and standard hydrodynamics does not apply. The relation between the stress tensor 
and the velocity gradient is commonly not linear, sometimes one has no unique 
relation at all. If the characteristic time scales are large, one encounters 
memory effects, \eg, the viscosity may decrease with time under shear 
stress~\cite{larson_book,kroeger,guyon}.

\begin{figure}[t] 
\centerline{
\includegraphics[width=2.in]{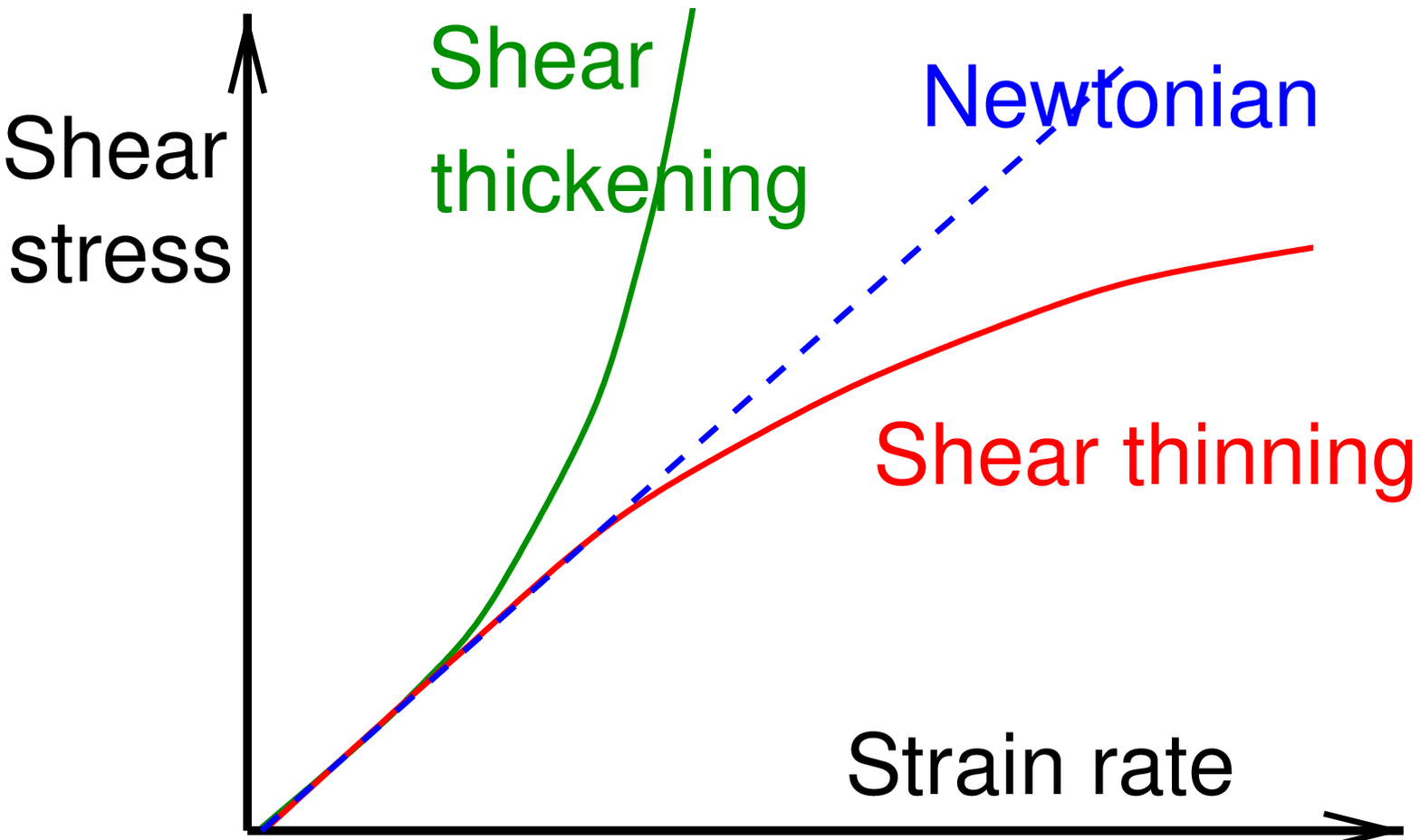}
\hspace*{1cm}
\includegraphics[width=1.7in]{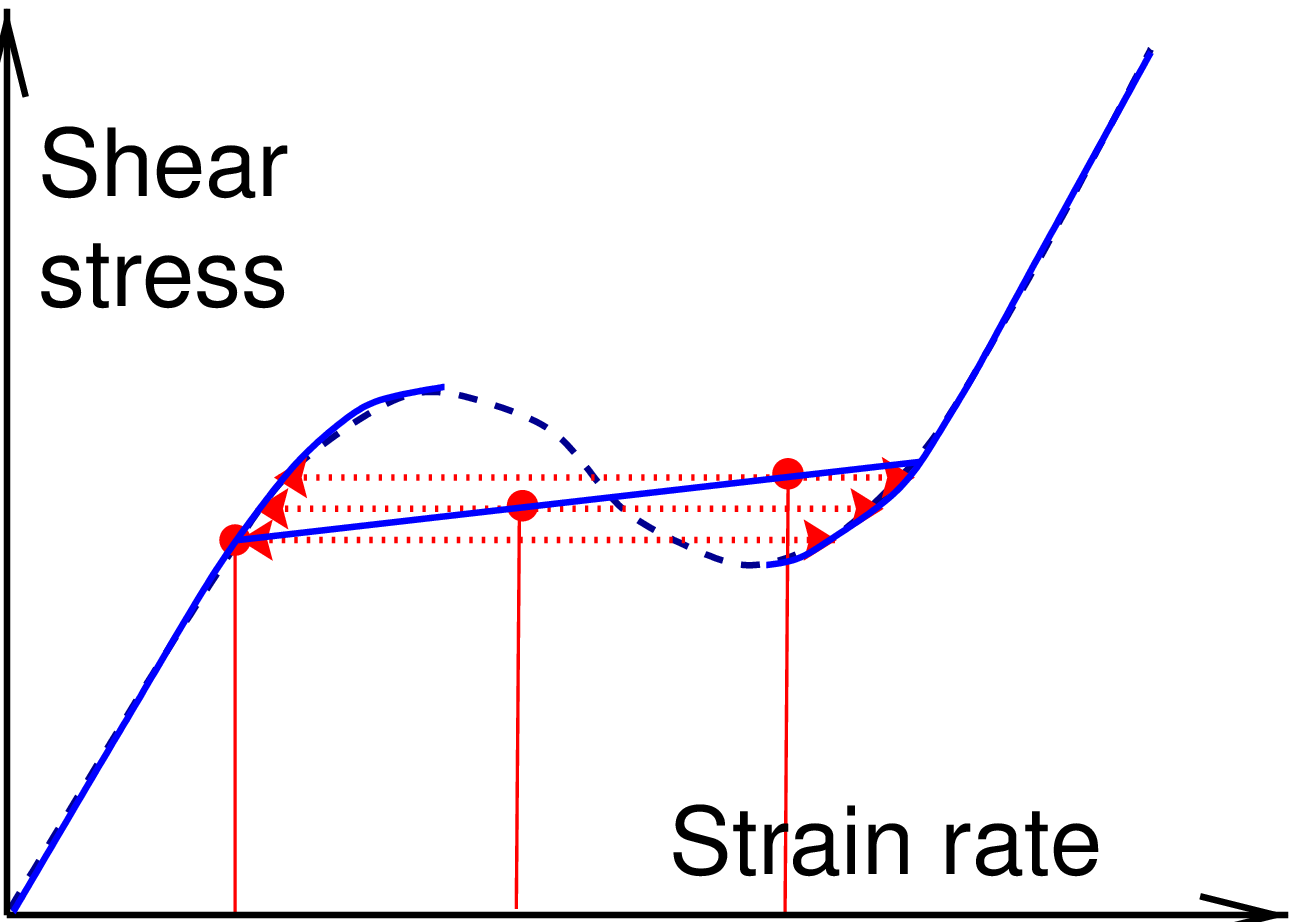}
}
\vspace*{8pt}
\caption{
Sketches of nonlinear stress-strain flow curves
Left: Shear thickening and shear thinning;
Right: Nonmonotonic stress-strain relation leading
   to phase separation. The horizontal dotted lines
   connect the two coexisting phases for three
   selected global strain rates.
\label{fig:stress-strain}
}
\end{figure}

These phenomena are often considered in planar Couette geometry: The flow 
$\vec{u}$ points in the $x$-direction, a velocity gradient in the $y$-direction 
is imposed, and the relation between the strain rate 
$\dot{\gamma} = \partial u_x/\partial y$ and the shear stress $\sigma_{xy}$ is 
investigated. For Newtonian fluids, the two are proportional. In complex fluids, 
the differential viscosity $\partial \sigma_{xy}/\partial \dot{\gamma}$
may decrease with increasing strain rate (shear thinning), or increase 
(shear thickening). In some materials, the stress-strain flow curve may even be 
nonmonotonic. Since such flow curves are mechanically unstable in macroscopic 
systems, this leads to phase separation, and the system becomes inhomogeneous 
(Fig.~\ref{fig:stress-strain})~\cite{bagley,mcleish,cates,roux,berret1,britton,fischer,lopez,olmsted1,porte,olmsted3,olmsted4}.
The separation of a fluid into two states with distinctly different strain rates 
is called shear banding. As in equilibrium phase transitions, the signature of 
the coexistence region is a plateau region in the resulting stress-strain flow 
curve of the inhomogeneous systems. However, one has two important differences to 
the equilibrium case: First, the location of the plateau cannot be determined by 
a Maxwell construction, but has to be calculated by solving the full dynamical 
equations. Second, the plateau is not necessarily horizontal. Even though mechanical 
stability requires that two coexisting phases must have the same shear stress, 
the ``plateau'' may have a slope or even be curved because different coexisting 
phases may be selected for different shear rates~\cite{schmitt,fielding} 
(see Fig.~\ref{fig:stress-strain}). 

\subsubsection{Nematic liquid crystals}  

\begin{figure}[b] 
\centerline{\includegraphics[width=5in]{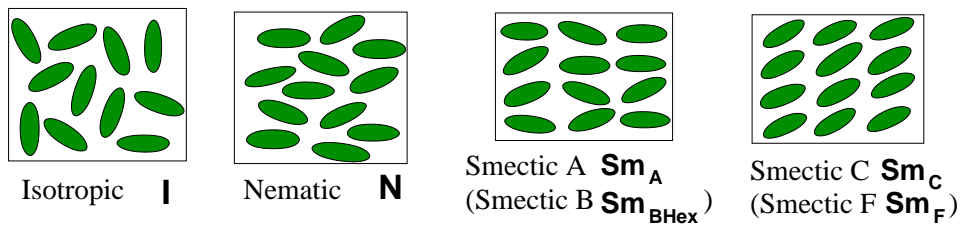}}
\vspace*{8pt}
\caption{
Examples of liquid crystalline phases. In the nematic
phase (N), the particles are oriented, but translationally
disordered. In the smectic phases, they form layers in
one direction, but remain fluid within the layers.
In the smectic B${}_{\mbox{\tiny Hex}}$ and smectic F phase, 
the structure within the layers is not entirely 
disordered, but has a type of order called hexatic.
See Refs.~\protect\refcite{degennes2,chandrasekhar} for explanation. 
The isotropic phase (I) is entirely disordered (regular
isotropic liquid).
\label{fig:liquid_crystals}
}
\end{figure}

Next we recapitulate some basic facts on liquid crystals. The building blocks of 
these materials are anisotropic particles, \eg, elongated molecules, associated 
structures such as wormlike micelles, or even rodlike virusses. A liquid crystal
phase is an intermediate state between isotropic liquid and crystalline solid,
which is anisotropic (particles are orientationally ordered), but in some sense 
still fluid (particles are spatially disordered in at least one direction). Some 
such structures are sketched in Fig.~\ref{fig:liquid_crystals}. The phase 
transitions are in some cases triggered by the temperature (thermotropic liquid 
crystals), and in other cases by the concentration of the anisotropic particles 
(lyotropic liquid crystals). Lamellar membrane stacks (see previous section) are 
examples of smectic liquid crystalline structures. In this section, we shall mainly
be concerned with nematics, where the particles are aligned along one common 
direction, but otherwise fluid. 

For symmetry reasons, the transition between the isotropic fluid and the nematic 
fluid must be first order. The nematic order is commonly characterized by the 
symmetric traceless order tensor
\begin{equation}
\label{order_tensor}
Q_{ij} = \langle \frac{1}{2} 
(3 d_i d_j - 1) \rangle,
\end{equation}
where $\vec{d}$ is a unit vector characterizing the orientation of a particle. In an 
isotropic fluid, ${\bf Q}$ is zero. Otherwise, the largest eigenvalue gives the 
nematic order parameter $S$, and the corresponding eigenvector is the direction 
of preferred alignment, the ``director'' $\vec{n}$. Note that only the direction 
of $\vec{n}$ matters, not the orientation, \ie, the vectors $\vec{n}$ and 
$-\vec{n}$ are equivalent.

\begin{figure}[b] 
\centerline{\includegraphics[width=3in]{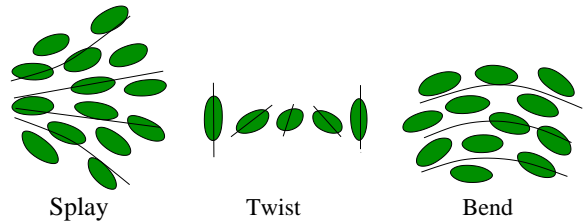}}
\vspace*{8pt}
\caption{
Elastic modes in nematic liquid crystals.
\label{fig:elastic_modes}
}
\end{figure}

Since the ordered nematic state is degenerate with respect to a continuous symmetry 
(the direction of the director), it exhibits elasticity: The free energy costs of 
spatial director variations must vanish if the wavelength of the modulations tends 
to infinity. At finite wavelength, the system responds to such distortions with 
elastic restoring forces. Symmetry arguments show that these depend only on three 
independent material constants~\cite{chaikin}, the Frank constants $K_{11}, K_{22}$, 
and $K_{33}$. The elastic free energy is given by~\cite{degennes2}
\begin{equation}
\label{elastic}
{\cal F}_{\mbox{\tiny elastic}}
= \int d \vec{r} \: 
\big\{
K_{11} (\nabla \vec{n})^2 +
K_{22} (\vec{n} \cdot (\nabla \times \vec{n}))^2 +
K_{33} (\vec{n} \times (\nabla \times \vec{n}))^2
\big\}.
\end{equation}
Figure~\ref{fig:elastic_modes} illustrates the corresponding fundamental distortions, 
the splay ($K_{11}$), twist ($K_{22}$), and bend mode ($K_{33}$).

How does shear affect a liquid crystalline fluid? In the isotropic phase and at 
low strain rate , the situation is still comparably simple. The velocity gradient 
imposes a background angular velocity 
\begin{equation}
\vec{\Omega} = \frac{1}{2}(\nabla \times \vec{u}) 
\end{equation}
on the fluid, which the particles pick up~\cite{ailawadi}. Since they rotate faster 
while perpendicular to the flow, they also acquire a very slight effective 
orientation at the angle $\pi/4$ relative to the flow direction. At higher strain 
rate, the order parameter increases, and the alignment angle decreases~\cite{yuan}. 
This sometimes even leads to a shear-induced transition into the nematic 
phase~\cite{see,olmsted5,berret2,berret3,berret4,berret5,mather}.

Deep in the nematic phase, the nematic fluid can be described by the flow field 
$\vec{u}(\vec{r},t)$ and the director field $\vec{n}(\vec{r},t)$ 
($|\vec{n}| \equiv 1$). We shall now assume incompressibility 
($\nabla \vec{u} = 0$) and a linear, instantaneous relation stress-strain relation 
as in Eq.~(\ref{stress_linear}). Compared to simple fluids, there are several 
complications.

First, nematic fluids are elastic, hence the stress tensor $\sigma_{ij}$ 
(\ref{total_stress}) has an elastic contribution
\begin{equation}
\label{stress_elastic}
\sigma_{ij}^{\mbox{\tiny elastic}}
= - \frac{\delta {\cal F}_{\mbox{\tiny elastic}} }
{\delta (\partial_i n_k) }
\: \partial_j n_k.
\end{equation}
The free energy ${\cal F}_{\mbox{\tiny elastic}}$ is given by Eq.~(\ref{elastic}).

Second, the fluid is locally anisotropic, therefore the constitutive equation for 
the viscous stress tensor contains more independent terms than in simple fluids.
The structure of $\sigma_{ij}'$ which is compatible with all symmetry requirements 
reads~\cite{chandrasekhar}
\begin{eqnarray}
\label{stress_lc_viscous}
\sigma_{ij}' &= &
\alpha_1 \: \sum_{kl} n_i n_j n_k n_l A_{kl}
+ \alpha_2 \: n_i N_j
+ \alpha_3 \: N_i n_j \\
&& 
+ \: \alpha_4 \: A_{ij}
+ \alpha_5 \sum_k n_i n_k A_{jk}
+ \alpha_6 \sum_k n_j n_k A_{ik},
\nonumber
\end{eqnarray}
where ${\bf A}$ is the symmetric flow deformation tensor,
\mbox{$ A_{ij} = \frac{1}{2}(\partial_i u_j + \partial_j u_i)$},
and the vector $\vec{N}$ gives the rate of change of the director relative to the 
angular velocity of the fluid, 
\mbox{$\vec{N}=\frac{d \vec{n}}{dt}-\vec{\Omega} \times \vec{n}$}.
The parameters $\alpha_i$ are the so-called Leslie coefficients. 
They are linked by one relation~\cite{chandrasekhar},
\begin{equation}
\alpha_2+\alpha_3 = \alpha_6 - \alpha_5
\end{equation}
hence one has five independent material constants.

Third, the flow field and the director field are coupled. Thus one must take
into account the dynamical evolution of the director field,
\begin{equation}
\label{director1}
\rho_1 \ddot{n}_j = g_j + \partial_i \pi_{ij},
\end{equation}
which is driven by the intrinsic body force $\vec{g}$, and the director stress
tensor $\pi_{ij}$. Here $\rho_1$ is the moment of inertia per unit volume.
As for the stress tensor, one can derive the general form of the constitutive 
equations for $\vec{g}$ and ${\bf \pi}$ with symmetry arguments. The individual
expressions can be found in the literature~\cite{chandrasekhar}.
Here, we only give the final total dynamical equation for $\vec{n}$,
\begin{equation}
\label{director2}
\rho_1 \ddot{n}_j = \lambda n_j
+ \sum_i \partial_i \big( \frac{\delta {\cal F}_{\mbox{\tiny el}} }
{\delta (\partial_i n_j)} \big)
- \frac{\partial {\cal F}_{\mbox{\tiny el}} }{\partial n_j}
- \gamma_1 N_j - \gamma_2 \sum_i n_i A_{ij},
\end{equation}
where $\gamma_1 = \alpha_3 - \alpha_2$, $\gamma_2 = \alpha_6 - \alpha_5$, and 
$\lambda$ is a Lagrange parameter which keeps $|\vec{n}|$ constant. 

To summarize, the dynamical equations of nematohydrodynamics for incompressible 
nematic fluids are given by Eqs.~(\ref{acceleration}) and (\ref{director2})
together with (\ref{total_stress}), (\ref{elastic}), (\ref{stress_elastic}), 
and (\ref{stress_lc_viscous}). These are the Leslie-Ericksen equations of
nemato-hydrodynamics~\cite{chandrasekhar}. The theory depends on five 
independent viscosity coefficients $\alpha_i$ and on three independent elastic 
constants $K_{ii}$. Not surprisingly, it predicts a rich spectrum of phenomena. 
For example, in the absence of director gradients, steady shear flow is found to 
have two opposing effects on the director: It rotates the molecules, and it aligns 
them. The relative strength of these two tendencies depends on the material 
parameters. As a result, one may either encounter stable flow alignment or a 
rotating ``tumbling'' state~\cite{larson3,berret6,zakharov,hess}.

The Ericksen-Leslie theory describes incompressible and fully ordered nematic 
fluids. The basic version presented here does not account for the possibility 
of disclination lines and other topological defects, and it does not include 
the coupling with a variable density and/or order parameter field. The situation 
is even more complicated in the vicinity of stable or metastable (hidden) 
thermodynamic phase transitions. Thermodynamic forces may contribute to a 
nonmonotonic stress-strain relationship, which eventually lead to mechanical 
phase separation~\cite{olmsted1}.

\begin{figure}[htb] 
\centerline{\includegraphics[width=2.5in]{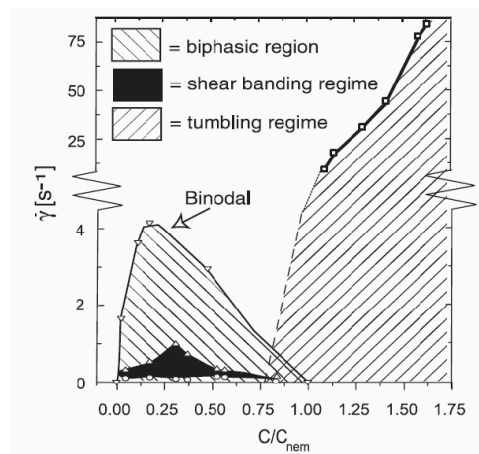}}
\vspace*{8pt}
\caption{
Nonequilibrium phase diagram of a lyotropic liquid crystal
under shear as a function of strain rate $\dot{\gamma}$
and particle density $C$. ($C_{\mbox{\tiny nem}}$ is the
density of the equilibrium isotropic/nematic transition).
The system is a mixture of rodlike virusses (fd virus) 
and polymers (dextran). From Ref.~\protect\refcite{lettinga}.
\label{fig:virus}
}
\end{figure}

Figure~\ref{fig:virus} shows an experimental example of a nonequilibrium phase 
diagram for a system of rodlike virusses~\cite{lettinga}.

\subsection{Simulating shear on the particle level: NEMD} 

In the linear response regime, \ie, in the low-shear limit, many rheological 
properties of materials can be determined from equilibrium simulations. For 
example, the viscosity coefficients introduced in the previous section can be 
related to equilibrium fluctuations with Green-Kubo relations~\cite{sarman1,tang}. 

However, the non-Newtonian character of most complex fluids and their resulting
unique properties manifest themselves only beyond the linear response regime. 
To study these, non-equilibrium simulations are necessary. We will now briefly 
review non-equilibrium molecular dynamics (NEMD) methods for the simulation of
molecular model systems under shear. Some of the methods are covered in
much more detail in Refs.~\refcite{allen_book,evans_book,sarman_rev}.

In NEMD simulations, one faces two challenges. First, one must mechanically impose 
shear. Second, most methods that enforce shear constantly pump energy into the 
system. Hence one must get rid of that heat, \ie, apply an appropriate thermostat.
We will address these two issues separately.

The most direct way of imposing shear is to confine the system between two rough 
walls, and either move one of them (for Couette flow), or apply a pressure 
gradient (for Poiseuille flow). Varnik and Binder have shown that this approach 
can be used to measure the shear viscosity in polymer melts~\cite{varnik}. 
It has the merit of being physical: Strain is enforced physically, and the heat 
can be removed in a physical way by coupling a thermostat to the walls. On the 
other hand, the system contains two surfaces, and depending on the material under 
consideration, one may encounter strong surface effects. This can cause problems.

\begin{figure}[b] 
\centerline{\includegraphics[width=2.5in]{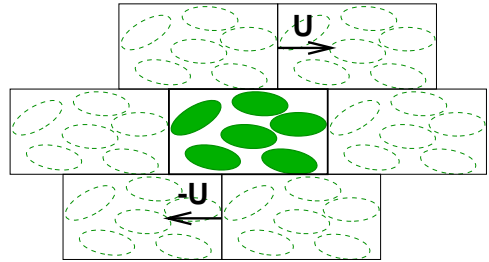}}
\vspace*{8pt}
\caption{
Lees-Edwards boundary conditions.
\label{fig:lees-edwards}
}
\end{figure}

An alternative straightforward way to generate planar Couette flow is to use 
moving periodic boundary conditions as illustrated in Fig.~\ref{fig:lees-edwards} 
(Lees-Edwards boundary conditions~\cite{allen_book,lees}): 
In order to enforce shear flow $\vec{u} = (u_x,0,0)$ with an average strain
rate $\dot{\gamma} = \partial u_x/\partial y$, one proceeds as follows:
One replicates the particles in the $x$ and $z$ direction like in
regular periodic boundary conditions,
\begin{equation}
\begin{array}{rcl}
  r_{x} &   \rightarrow & r_{x} \pm L_x  \\
  r_{y,z} & \rightarrow & r_{y,z} \\
  \vec{v} & \rightarrow & \vec{v}
\end{array}
\qquad 
\begin{array}{rcl}
  r_{x,y} & \rightarrow & r_{x,y} \\
  r_{z} & \rightarrow & r_{z} \pm L_z \\
  \vec{v} & \rightarrow & \vec{v}
\end{array}.
\end{equation}
In the $y$-direction, the replicated particles acquire an additional velocity 
$U = \dot{\gamma} L_y$ and an offset $U t$ in the $x$-direction, 
\begin{equation}
\begin{array}{rcl}
  r_{x} & \rightarrow & r_{x} \pm  (U t \: \mbox{mod} L_x)\\
  r_{y} & \rightarrow & r_{y} \pm L_y \\
  r_{z} & \rightarrow & r_{z} \\
  \vec{v} & \rightarrow & \vec{v} \pm U \vec{e}_{x} 
\end{array}
\end{equation}
Here $L_{x,y,z}$ are the dimensions of the simulation box, and $\vec{e}_x$ the
unit vector in the $x$-direction.

Lees-Edwards boundary conditions are perfectly sufficient to enforce planar Couette 
flow. In some applications, it has nevertheless proven useful to supplement them 
with fictitious bulk forces that favor a linear flow profile. One particularly
popular algorithm of this kind is the SLLOD algorithm~\cite{evans_book,sllod}. 
For the imposed flow $\vec{u}(\vec{r}) = \dot{\gamma} y \vec{e}_x$, 
the SLLOD equations of motion for particles $i$ read
\begin{eqnarray}
\frac{d \vec{r}_i}{dt} &=& \frac{ \vec{p}_i'}{m_i} 
+ \dot{\gamma} y_i \vec{e}_x \\
\frac{d \vec{p}_i'}{dt} &=& \vec{F}_i 
- \dot{\gamma} p_{yi}' \vec{e}_x,
\end{eqnarray}
where $\vec{p}_i' = m_i (\vec{v}_i - \vec{u}(\vec{r_i}))$ is the momentum of particle 
$i$ in a reference frame moving with the local flow velocity $\vec{u}(\vec{r}_i)$, 
and $\vec{F}_i$ is the regular force acting on the particle $i$. The equations
of motion can be integrated with standard techniques, or with specially devised
operator-split algorithms~\cite{zhang,pan}. It is also possible to formulate 
SLLOD variants for arbitrary steady flow~\cite{edwards}.
The SLLOD algorithm is very useful for the determination of viscosities in
complex fluids~\cite{baalss,sarman1,sarman2,sarman3,whirter1,whirter2}. 
It has the advantage that the system relaxes faster towards a steady state, 
and that the analysis of the data is simplified~\cite{evans_book}.

\begin{figure}[bt] 
\centerline{\includegraphics[width=2.in]{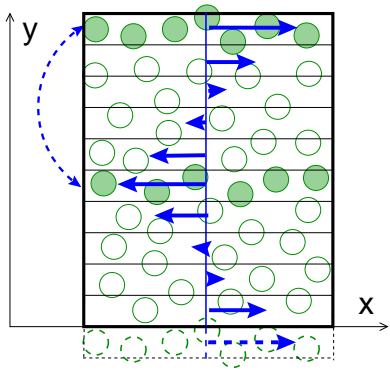}}
\vspace*{8pt}
\caption{
Schematic sketch of the M\"uller-Plathe method for imposing shear.
In periodical intervals, the particle in the middle slab that moves 
fastest in the $x$-direction and the particle in the top slab that 
moves fastest in the ($-x$)-direction exchange their momentum components 
$p_x$.  After Ref.~\protect\refcite{mplathe}.
\label{fig:mueller-plathe}
}
\end{figure}

A fourth method to enforce shear has recently been proposed by 
M\"uller-Plathe~\cite{mplathe}. One uses regular periodic boundary conditions, and
divides the system into slabs in the desired direction of flow gradient. 
In periodic intervals, the particle in the central slab ($y=0$) with the 
largest velocity component in the $x$ direction, and the particle in the topmost 
slab $(y=L_y/2)$ with the largest velocity component in the $(-x)$-direction is 
determined, and the momentum components $p_x$ of the two particles are swapped. 
This leads to a zigzag-shaped shear profile (Fig.~\ref{fig:mueller-plathe}).

The M\"uller-Plathe algorithm can be implemented very easily. Just one small
change turns a molecular dynamics program for thermal equilibrium  
into a NEMD program that produces shear flow. Furthermore, the algorithm conserves 
momentum and energy: Contrary to the algorithms discussed before, it does not 
pump heat into the system. Instead, the heat is constantly redistributed. It is 
drained out of the system at $y=0$ and $y=L_y/2$ by the momentum swaps, but it 
is also produced in the bulk through the energy dissipation associated with the 
shear flow. As a result, the zigzag shear profile is accompanied with a W-shaped 
temperature profile. Since this is not always wanted, the algorithm is sometimes 
used in combination with a thermostat.  A slight drawback of the algorithm is 
the fact that it produces intrinsically inhomogeneous profiles, due to the presence 
of the two special slabs at $y=0$ and $y=L_y/2$. This may cause problems
in small systems.

Having introduced these four methods to impose shear, we now turn to the problem 
of thermostatting. All shearing methods except for the M\"uller-Plathe method 
produce heat, therefore the thermostat is an essential part of the simulation 
algorithm. 

A thermostat is a piece of algorithm that manipulates the particle velocities 
such that the temperature is adjusted to its desired value. This is commonly 
done by adjusting the kinetic energy. In a flowing fluid with flow velocity
$\vec{u}(\vec{r})$, the kinetic energy has a flow contribution and a thermal 
contribution. Therefore, it is important to define the thermostat in a frame that 
moves with the fluid, \ie, to couple it to the ``peculiar'' velocities 
$\vec{v}_i' = \vec{v}_i - \vec{u}(\vec{r}_i)$, rather than to the absolute 
velocities $\vec{v}_i$. Unfortunately, most thermostats don't account for this
automatically, and one has to put in the flow profile by hand. The only exception 
is the dissipative particle dynamics (DPD) thermostat (see below). 

This raises the question how to determine the flow $\vec{u}(\vec{r})$. In a 
homogeneous system, at low strain rates, it can be acceptable to use the 
pre-imposed profile ($\vec{u}(\vec{r}) = \dot{\gamma} y \vec{e}_x$ in our
earlier examples). Thermostats that use the pre-imposed flow profile are 
called ``profile-biased thermostats''. At high strain rates, and in inhomogeneous 
systems, the use of profile-biased thermostats is dangerous and may twist the 
results~\cite{evans_book,whirter1}. In that case, the profiles $\vec{u}(\vec{r})$ 
must be calculated directly from the simulations, ``on the fly''. Such thermostats 
are called ``profile-unbiased thermostats''.

One can distinguish between deterministic and stochastic thermostatting methods. 
The simplest deterministic thermostat simply rescales the peculiar velocities of 
all particles after every time step such that the total thermal kinetic
energy remains constant. Among the more sophisticated deterministic thermostats, 
the ones that have been used most widely in NEMD simulations are the so-called 
Gaussian thermostats~\cite{evans_book}, and the Nos\'e Hoover 
thermostats~\cite{allen_book,frenkel_book}. 

The idea of Gaussian thermostats is very simple: For a system of particles $j$ 
subject to forces $\vec{f}_j$, one minimizes the quantity 
$C=\sum_j(\dot{\vec{p}}_j'- \vec{f}_j)^2/(2 m_j)$ with respect to 
$\dot{\vec{p}}_j$, with the constraint that either the time derivative of the 
total energy, or the time derivative of the total kinetic energy be zero (Gaussian 
isoenergetic and Gaussian isothermal thermostat, respectively). 
In the absence of constraints, the minimization procedure would yield Newton's 
law. The constraints modify the equations of motion in a way that can be 
considered in some sense minimal. One obtains
\begin{equation}
\dot{\vec{p}}_j' = \vec{f}_j - \alpha \vec{p}_j,
\end{equation}
where $\alpha$ is a Lagrange multiplier, which has to be chosen such that the 
desired constraint -- constant energy or constant kinetic energy -- is fulfilled.

The Nos\'e Hoover thermostats are well-known from equilibrium molecular dynamics 
and shall not be discussed in detail here. We only note that in NEMD simulations, 
one uses not only the standard isothermal Nos\'e-Hoover thermostats,
but also isoenergetic variants.

Deterministic thermostats have the slightly disturbing feature that they change 
the dynamical equations in a rather unphysical way. It is sometimes hard to say 
how that might affect the system. At equilibrium, Nos\'e-Hoover thermostats 
rest on a well-established theoretical basis: They can be derived from extended 
Hamiltonians, they produce the correct thermal averages for static quantities etc. 
Outside of equilibrium, the situation is much more vague. For the case of
steady states, some results on the equivalence of thermostats are fortunately
available. Evans and Sarman have shown that steady state averages and time 
correlation functions are identical for Gaussian isothermal and isoenergetic 
thermostats and for Nos\'e-Hoover thermostats~\cite{sarman_rev,evans1}. Ruelle 
has recently proved that the Gaussian isothermal and the Gaussian isoenergetic 
thermostat are equivalent in an infinite system which is ergodic under space
translations~\cite{ruelle}. 

An alternative approach to thermostatting are the stochastic methods, such as 
the Langevin thermostat~\cite{kolb} and the DPD thermostat~\cite{soddemann2}.
They maintain the desired temperature by introducing friction and stochastic 
forces. This approach has the virtue of being physically motivated. The 
dissipative and stochastic terms stand for microscopic degrees of
freedom, which have been integrated out in the coarse-grained model, but can 
still absorb and carry heat. From the theory of coarse-graining, it is well-known
that integrating out degrees of freedom effectively turns deterministic
equations of motion into stochastic equations of 
motion~\cite{zwanzig,ottinger,gorban}. 

The Langevin thermostat is the simplest stochastic dynamical model.
One modifies the equation of motion for the peculiar momenta by introducing a 
dissipative friction $\zeta$ and a random force $\vec{\eta}_j$,
\begin{equation}
\dot{\vec{p}}_j' = \vec{f}_j - \zeta \vec{v}_j' + \vec{\eta}_j,
\end{equation}
where $\vec{\eta}_j$ fulfills
\begin{eqnarray}
\langle \vec{\eta}_k \rangle &=& 0\\
\langle \eta_{i\alpha}(t) \eta_{j \beta}(t') \rangle
&=&
2 \zeta k_B T \: \delta_{ij} \: \delta_{\alpha \beta} \:
\delta(t-t'),
\end{eqnarray}
\ie, the noise is delta-correlated and Gaussian distributed with mean zero and 
variance $2 \zeta k_B T$ (fluctuation-dissipation theorem). In practice, this can 
be done by adding $(- \zeta v_{j \alpha}')$ and a random number to each 
force component $f_{i \alpha}$ in each time step $\Delta t$. The random number 
must be distributed with mean zero and variance $\sqrt{2 k_B T \zeta/\Delta t}$. 
Because of the central limit theorem, the distribution does not necessarily have 
to be Gaussian, if the time step is sufficiently small~\cite{duenweg1}.

The DPD thermostat is an application of the dissipative particle dynamics 
algorithm~\cite{soddemann2}. The idea is similar to that of the Langevin algorithm, 
but instead of coupling to absolute velocities, the thermostat couples to velocity 
differences between neighbor particles. Therefore, the algorithm is Galilean 
invariant, and accounts automatically for the difference between flow
velocity and thermal velocity. We refer to B. D\"unweg's lecture (this book) for 
an introduction into the DPD method (see also \refcite{frenkel_book}).

This completes our brief introduction into NEMD methods for systems under shear.
We close with a general comment: The strain rate $\dot{\gamma}$ has the dimension 
of inverse time. It introduces a time scale $1/\dot{\gamma}$ in the system, 
which diverges in the limit $\dot{\gamma} \to 0$. Therefore, systems at low 
strain rates converge very slowly towards their final steady state, and the 
study of systems at low strain rates is very time-consuming. In fact, even in 
coarse-grained molecular models, simulated strain rates are usually much higher 
than experimentally accessible strain rates.

We now give two examples of recent large-scale NEMD studies of inhomogeneous 
complex systems under shear.

\subsubsection{First application example: Nematic-Isotropic Interfaces 
under shear} 

Our first example is a study of the behavior of a Nematic-Isotropic interface 
under shear (Germano and Schmid~\cite{germano1,germano2}). Interfaces play a 
central role for shear banding, and understanding their structure should provide 
a key to the understanding of phase separation under shear. Our study is a first 
step in this direction.  We will discuss the questions whether the structure of 
an equilibrium interface is affected by shear, whether the phase transition is 
shifted, and whether shear banding can be observed.

The model system was a fluid of soft repulsive ellipsoids with the aspect ratio 
15:1, in a simulation box of size (150:300:150) (in units of the ellipsoid 
diameter $\sigma$). The density was chosen in the coexistence region between 
the nematic and isotropic phase, such that the system phase separates at 
equilibrium.  The initial configuration was a relaxed equilibrium configuration 
2.2with a nematic slab and an isotropic slab, separated by two interfaces. The latter 
align the director of the nematic phase such that it is parallel to the 
interface~\cite{allen1,mcdonald,akino}.

\begin{figure}[hbt] 
\centerline{
\parbox{2.2in}{
\includegraphics[width=2.in]{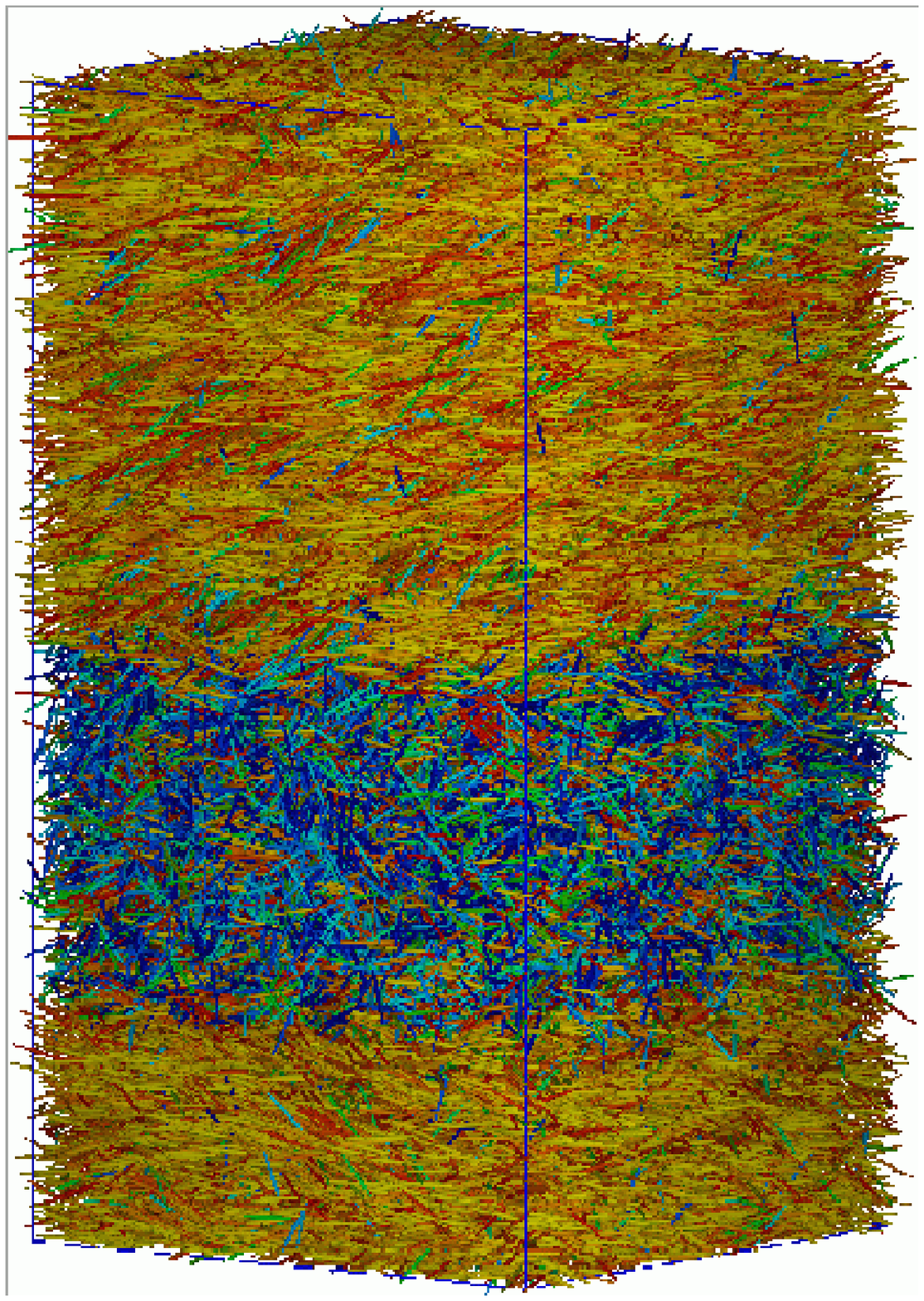}
}
\hspace*{1cm}
\parbox{1.5in}{
\includegraphics[width=1.2in]{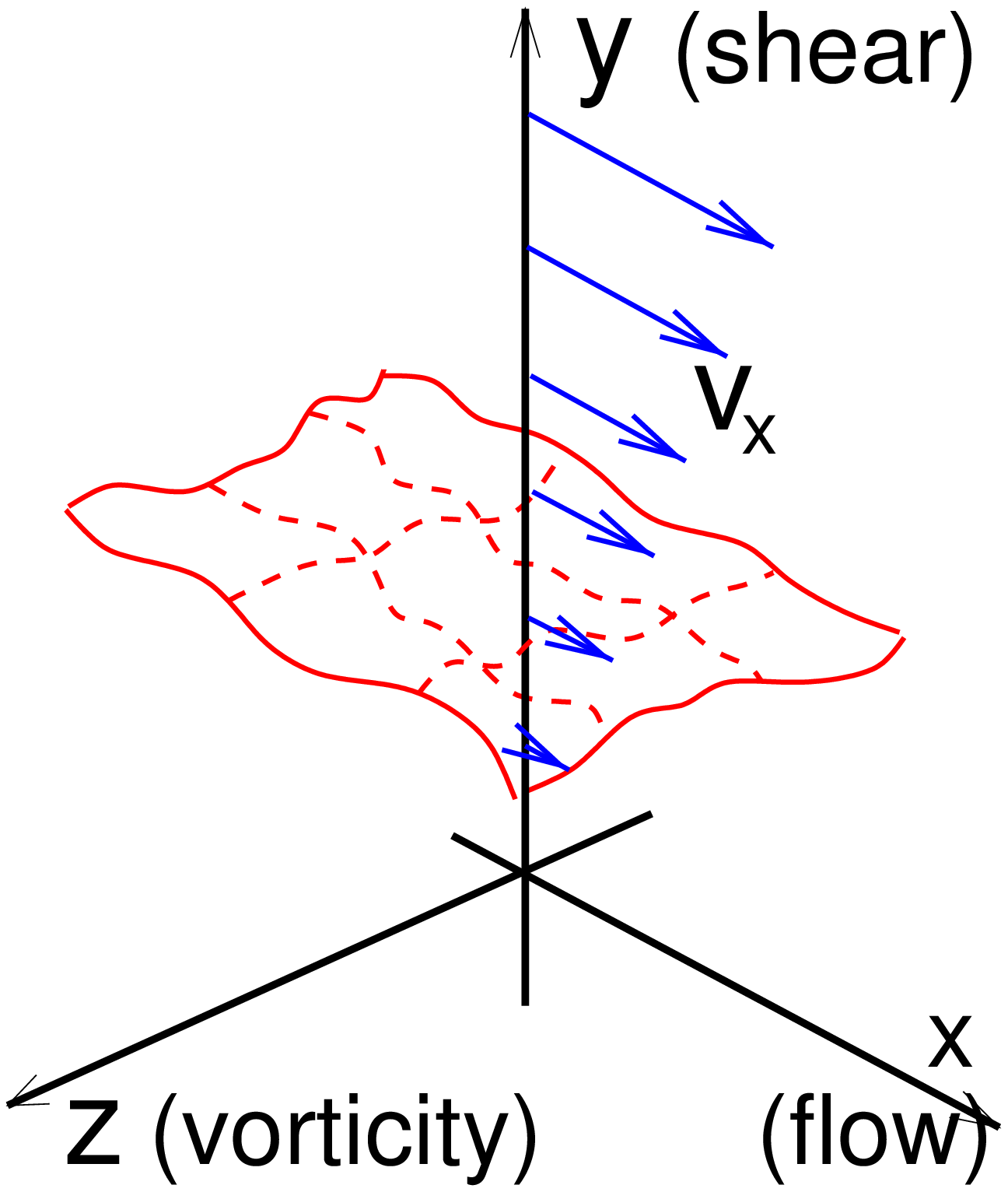}
}
}
\vspace*{8pt}
\caption{
Snapshot of a sheared nematic-isotropic interface (115200 particles) 
at strain rate $\dot{\gamma} = 0.001 \tau^{-1}$.
Also shown is the coordinate system used throughout
this section. 
From Ref.~\protect\refcite{germano1}.
\label{fig:interface}
}
\end{figure}

\clearpage

Shear was imposed with Lees-Edwards boundary conditions in combination with a 
profile unbiased Nos\'e Hoover thermostat, in the direction normal to the interface.
The two coexisting phases thus share the same shear stress. Two setups are
possible within this ``common stress'' geometry. In the flow-aligned setup, 
the director points in the direction of flow $\vec{u}$ (the $x$ direction); 
in the log-rolling setup, it points in the direction of ``vorticity''
$\nabla \times \vec{u}$ (the $z$ direction). Only the flow-aligned setup
turned out to be stable in our system. We considered mainly the strain rate 
$\dot{\gamma} = 0.001 \tau^{-1}$. Here $\tau$ is the natural time unit, 
$\tau = \sigma \sqrt{m/k_B T}$, with the particle mass $m$, the particle 
diameter $\sigma$, and the temperature $T$. This strain rate is small enough 
that the interface is still stable. A configuration snapshot is shown in
Fig.~\ref{fig:interface}.

\begin{figure}[tb] 
\centerline{\includegraphics[width=1.3in,angle=-90]{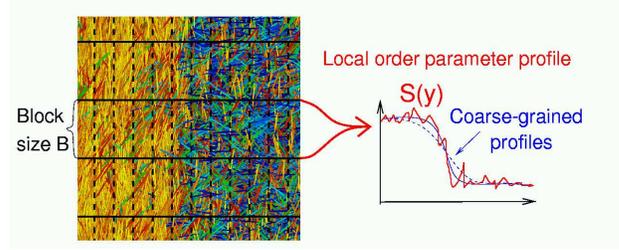}}
\vspace*{8pt}
\caption{
Illustration of block analysis to obtain local profiles.
\label{fig:binning}
}
\end{figure}

To analyze the system, it was split into columns of size $B \times L_y \times B$. 
The columns were further split into bins, which contained enough particles that 
a local order parameter could be determined. In this manner, one obtains local 
order parameter profiles for each column in each configuration. From the profile 
$S(y)$, we determined the local positions of the two interfaces as follows: 
We computed at least two coarse-grained profiles $\bar{S}(y;\delta_i)$ by 
convoluting the profile $S(y)$ with a symmetrical box-like coarse-graining 
function of width $\delta_i$. The coarse-graining width must be chosen such that 
the coarse-grained profiles still reflect the interfaces, but short ranged 
fluctuations are averaged out. The intersection of two averaged profiles locates 
a ``dividing surface'', where the negative order parameter excess on the nematic 
side balances the positive order parameter excess on the paranematic side. 
We define this to be the ``local position'' of the interface. Due to fluctuations,
it depends slightly on the particular choice of the $\delta_i$. Nevertheless, 
the procedure gives remarkably unambiguous values even for strongly fluctuating 
profiles~\cite{akino}.  The procedure is illustrated in 
Fig.~\ref{fig:binning}.

Once the positions $h_{NI}, h_{IN}$ of the two interfaces have been determined, 
we calculate profiles for all quantities of interest and shift them by the amount 
$h_{NI}$ or $h_{IN}$, respectively. This allows to perform averages over 
{\em local} profiles. Furthermore, the interface positions $h_{IN}$ and $h_{NI}$ 
themselves can be used to analyze the fluctuation spectrum of the interface 
positions.

\begin{figure}[hbt] 
\centerline{\includegraphics[width=1.8in]{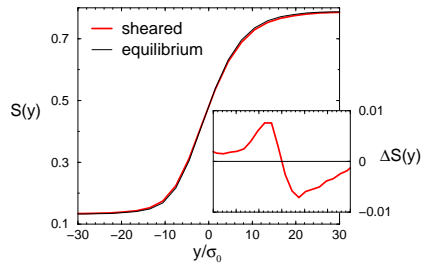}}
\vspace*{-8pt}
\caption{
Interfacial order parameter profile at strain rate $\dot{\gamma} = 0.001 \tau^{-1}$
(flow aligned setup). Inset shows the difference between profiles with and without
shear.
From Ref.~\protect\refcite{germano2}.
\label{fig:order}
}
\end{figure}

\begin{figure}[t] 
\centerline{
\includegraphics[width=1.7in]{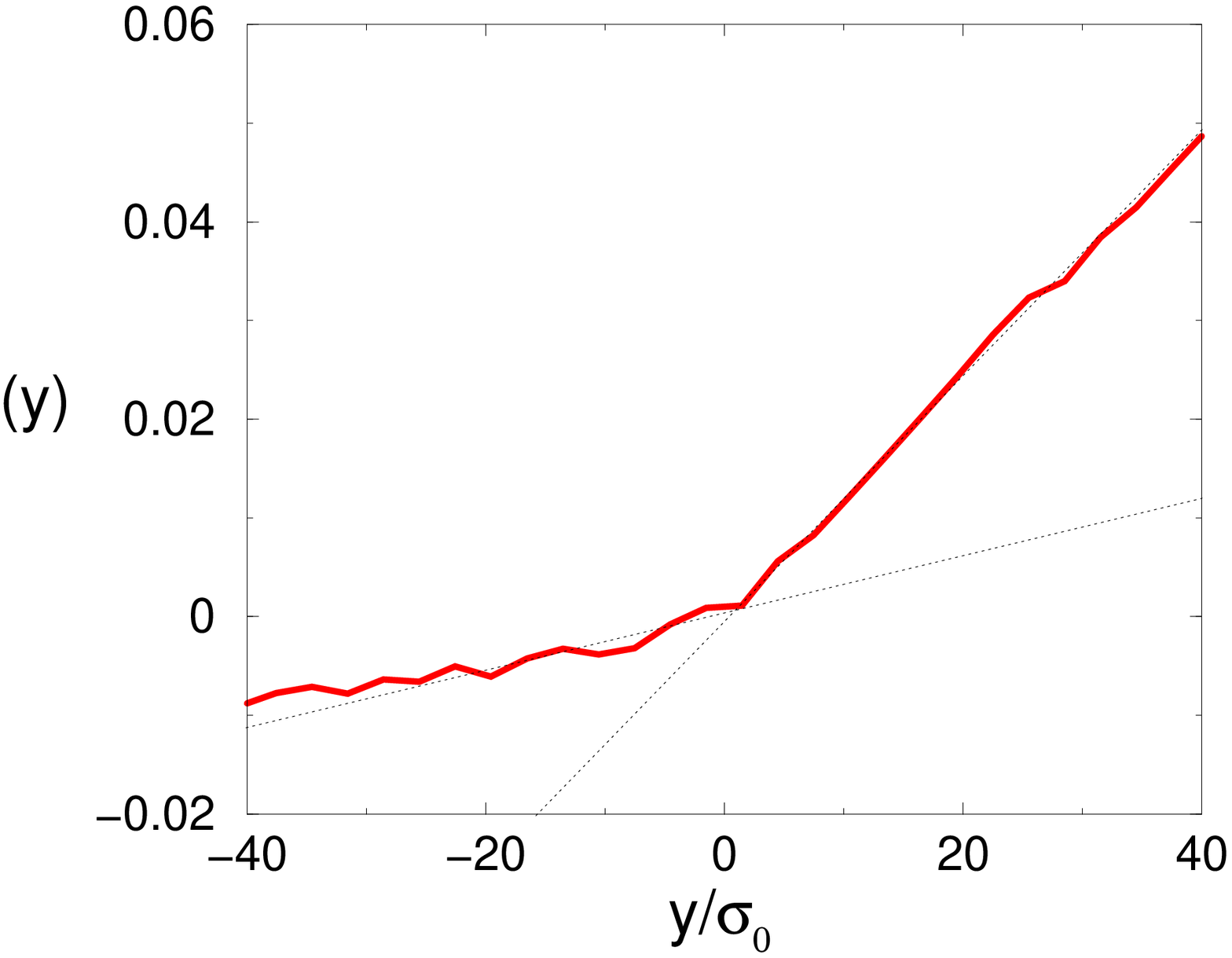}
\hspace*{1cm}
\includegraphics[width=1.7in]{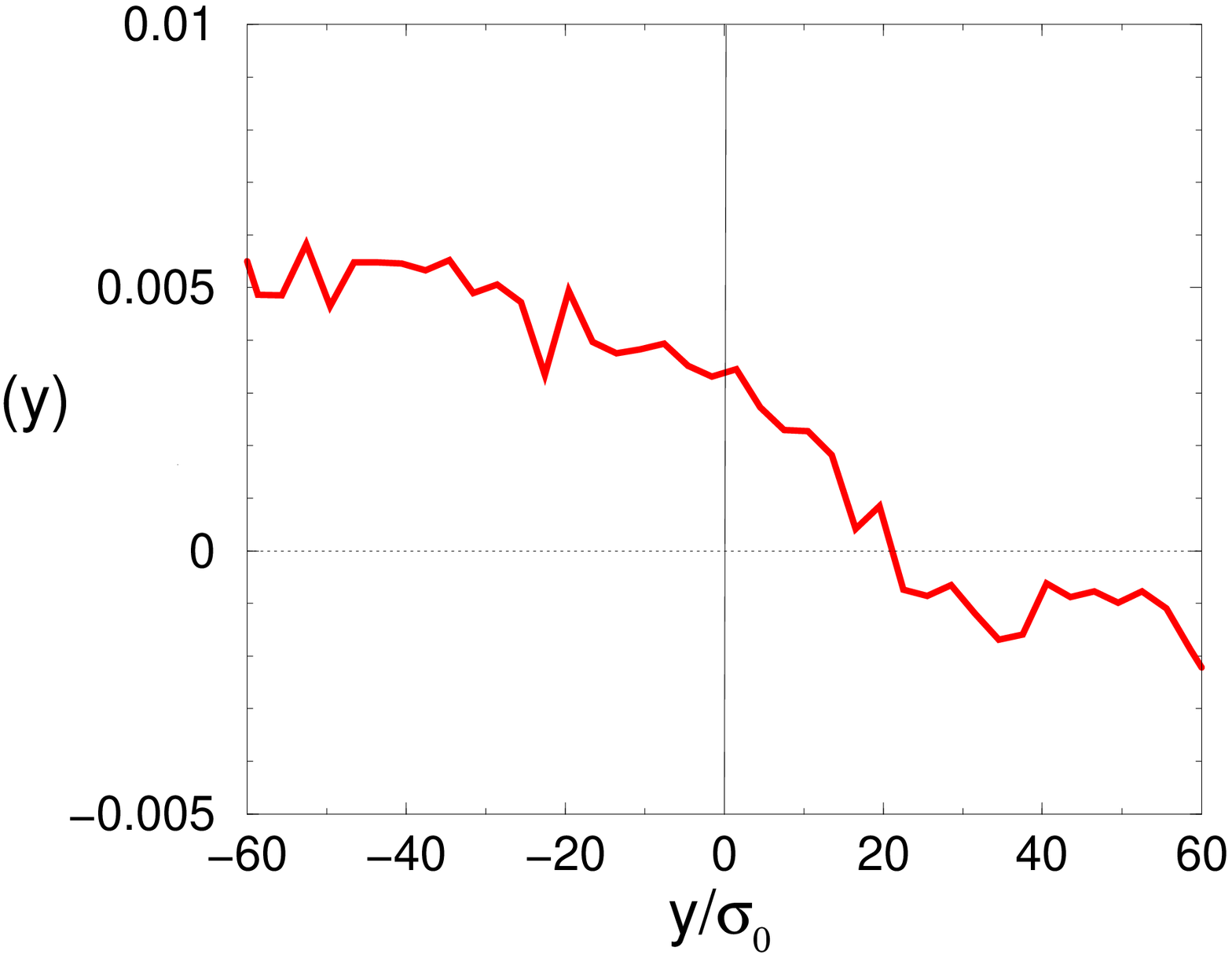}
}
\vspace*{-8pt}
\caption{
Velocity profile in the flow direction (left), and in the
vorticity direction (right), at strain rate
$\dot{\gamma} = 0.001 \tau^{-1}$.
From Ref.~\protect\refcite{germano2}.
\label{fig:velocity}
}
\end{figure}

Unexpectedly, the shear turned out to have almost no influence at all on the 
interfacial structure. Under shear, the interfaces are slightly broadened, 
indicating that the interfacial tension might be slightly reduced 
(Fig.~\ref{fig:order}). But the effect is barely noticeable. Likewise, the 
density profile and the capillary wave spectrum are almost undistinguishable
from those of the equilibrium interface, within the error. The velocity
profiles, however, are much more interesting. The flow profile exhibits a 
distinct kink at the interface position (Fig.~\ref{fig:velocity}). Hence we
clearly observe shear banding -- the total strain is distributed nonuniformly
between the two phases. At the same shear stress, the nematic phase supports 
a higher strain rate than the isotropic phase. Second and surprisingly, 
the interface apparently also induces a streaming velocity gradient in the 
vorticity direction (the $z$-direction). As a consequence, vorticity flow is
generated in opposite directions in the isotropic and in the nematic phase. 
This flow is symmetry breaking, and as yet, we have no explanation for it. 
We hope that future theoretical and simulation studies will clarify 
this phenomenon.

\begin{figure}[hb] 
\centerline{\includegraphics[width=1.7in,angle=-90]{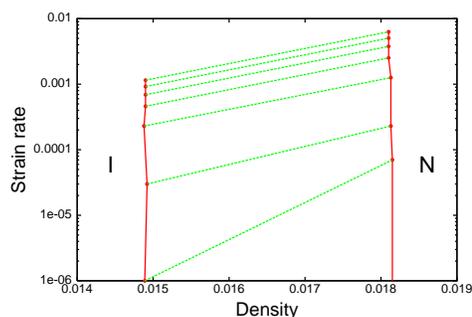}}
\vspace*{-8pt}
\caption{
Phase diagram of the flow-aligned system. The dashed
tie lines connect coexisting states. I denotes the isotropic
region, and N the nematic region.
From Ref.~\protect\refcite{germano2}.
\label{fig:shear_phases}
}
\end{figure}

At high strain rates, the interface gets destroyed. Figure~\ref{fig:shear_phases} 
shows a nonequilibrium phase diagram, which has been obtained from interface 
simulations of small systems. The local strain rate is always higher in the
nematic phase than in the isotropic phase. It is remarkable that the coexistence
region does not close up. Instead, the interface disappears abruptly at average 
strain rates above $\dot{\gamma} \approx 0.006/\tau$. 

\subsubsection{Second application example: Shear-induced phenomena
in surfactant layers}  

As a second example, we discuss the effect of shear on lamellar stacks of 
surfactants. This has been studied in coarse-grained molecular simulations 
by Soddemann, Guo, Kremer {\em et al}~\cite{guo,soddemann3}. The model system was 
very similar to that used in Section~\ref{sec:stacks}, except that it does not 
contain solvent particles~\cite{soddemann1}. Shear was induced with the 
M\"uller-Plathe algorithm~\cite{mplathe} in combination with a DPD 
thermostat~\cite{soddemann1}.

The first series of simulations by Guo {\em et al}~\cite{guo} addressed the 
question, whether the surfactant layers can be forced to reorient under shear. 
Three different orientations of the layers with respect to the shear geometry 
were considered: In the transverse orientation, the layers are normal to 
the flow direction, in the parallel orientation, they are normal to the direction
of the velocity gradient, and in the perpendicular orientation, they lie in the 
plane of the flow and the velocity gradient. The transverse orientation is 
unstable under shear flow. Experimentally, both a transition from the
transverse to the parallel orientation and from the transverse to the 
perpendicular orientation have been observed~\cite{gupta,koppi}.

\begin{figure}[bt] 
\centerline{\includegraphics[width=4in]{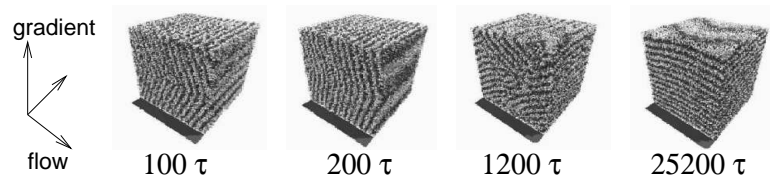}}
\vspace*{-8pt}
\caption{
Reorientation of surfactant layers under shear from an
initial transverse state at low strain rate, $\dot{\gamma} = 0.001/\tau$.
Reprinted with permission from Ref.~\protect\refcite{guo}.
Copyright 2002 by the American Physical Society.
\label{fig:guo1}
}
\end{figure}

\begin{figure}[bt] 
\centerline{\includegraphics[width=4.2in]{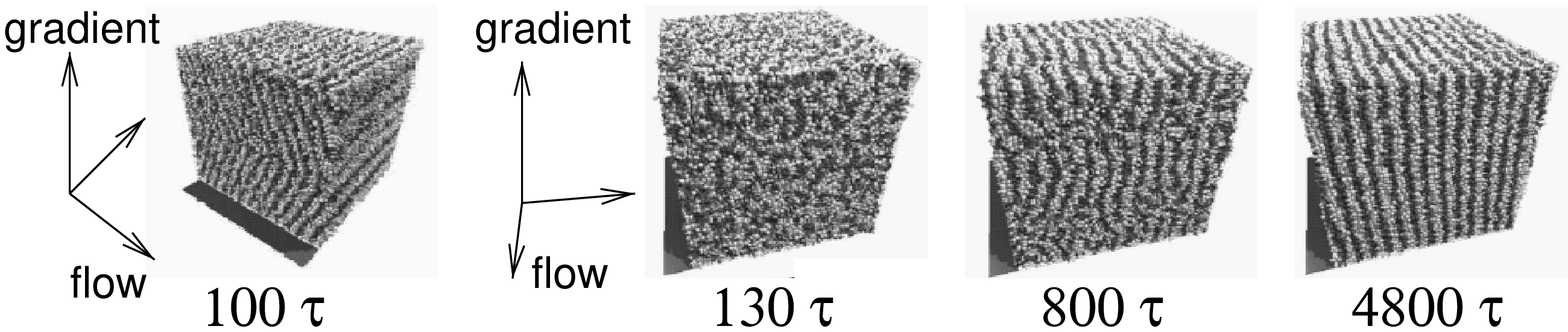}}
\vspace*{-8pt}
\caption{
Reorientation of surfactant layers under shear from an initial transverse state
at high strain rate, $\dot{\gamma} = 0.03/\tau$.
Reprinted with permission from Ref. \protect\refcite{guo}.
Copyright 2002 by the American Physical Society.
\label{fig:guo2}
}
\end{figure}

Guo {\em et al} have studied this by simulations of a system of dimer amphiphiles. 
Figures~\ref{fig:guo1} and \ref{fig:guo2} show a series of snapshots for different 
times at two different strain rates. Both systems were set up in the transverse 
state. At low strain rate, the shear generates defects in the transverse lamellae. 
Mediated by these defects, the lamellae gradually reorient into the parallel state. 
At high strain rate, the lamellae are first completely destroyed and then 
reorganize in the perpendicular state. 

\begin{figure}[bt] 
\centerline{
\parbox{1.5in}{
\includegraphics[width=1.5in]{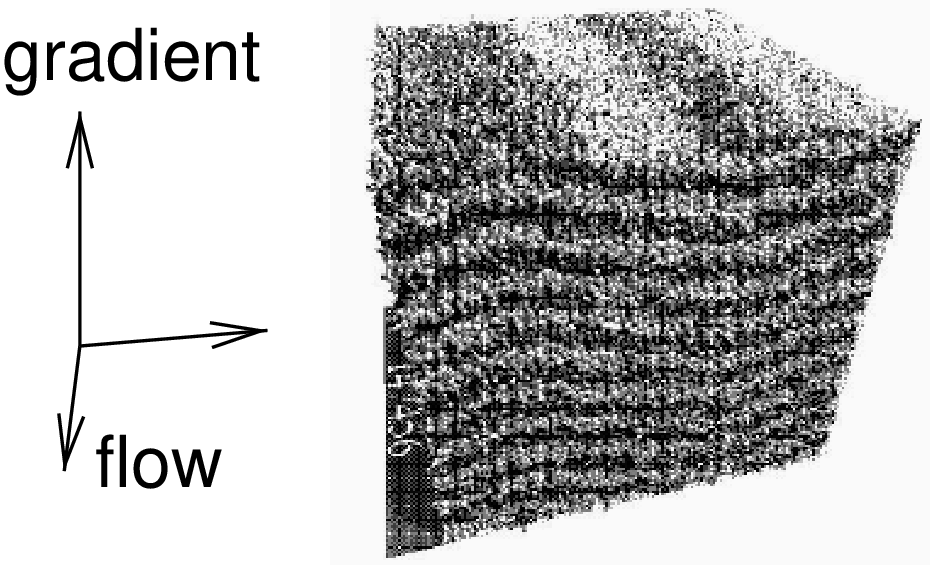}
}
\hspace*{1cm}
\parbox{2in}{
\includegraphics[width=1.5in]{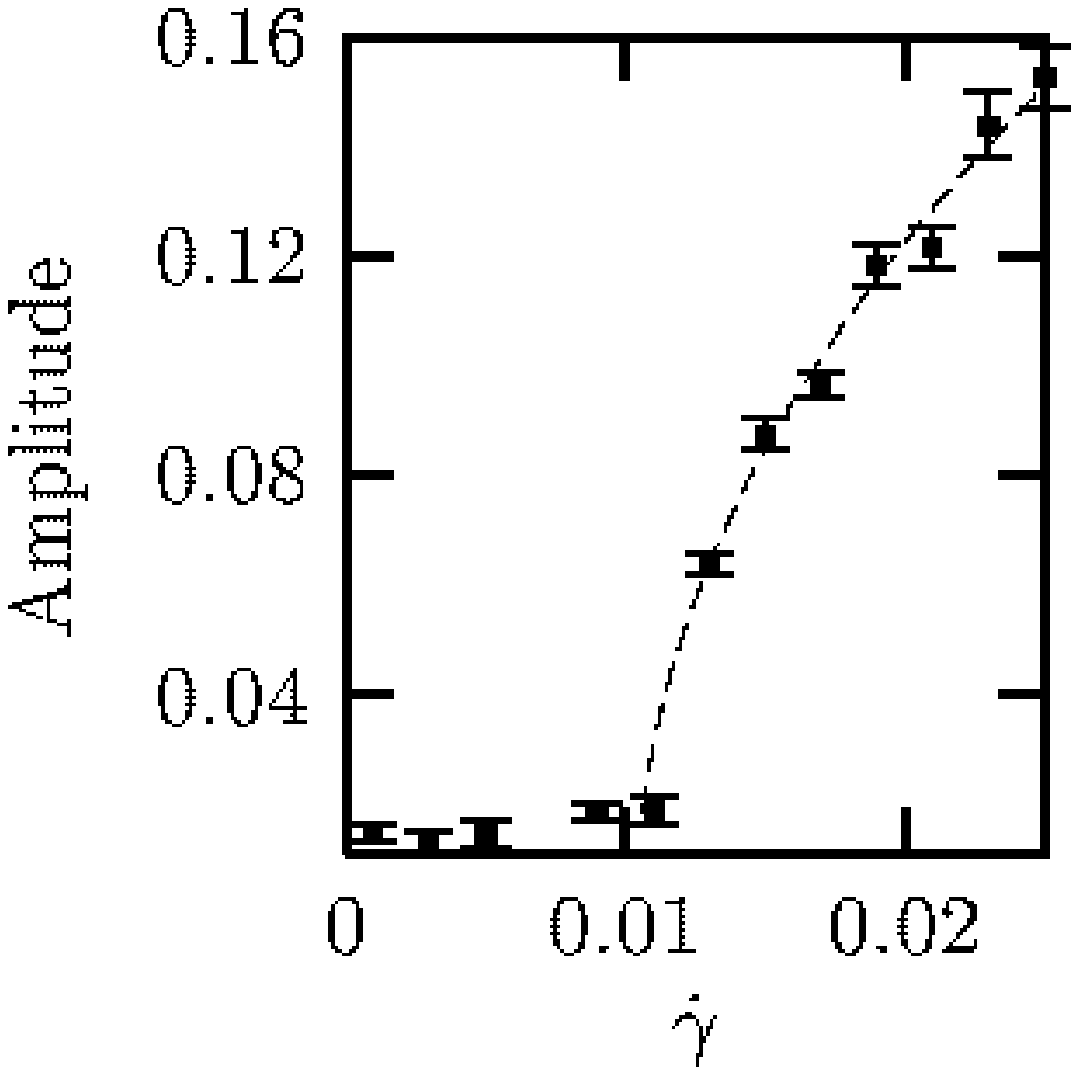}
}
}
\vspace*{-0.5cm}
\caption{
Shear induced undulations.
Left: configuration snapshot at strain rate $\dot{\gamma} = 0.015/\tau$.
Right: Undulation amplitude as a function of strain rate.
Undulations set in at a strain rate of $\dot{\gamma} \approx 0.01$.
From Ref. \protect\refcite{soddemann3}.
\label{fig:thoso}
}
\end{figure}

The final state is not necessarily the most favorable steady state. In fact, 
Guo {\em et al} found that the strain energy dissipation was always smallest in
the perpendicular state, regardless of the strain rate. When starting from the
transverse state at low strain rates, the parallel state formed nevertheless
for kinetic reasons. At intermediate strain rates, shear may induce undulations 
in the parallel state. This has been predicted by Auernhammer 
{\em et al}~\cite{auernhammer1,auernhammer2} and studied by 
Soddemann {\em et al} by computer simulation of a system of layered 
tetramers~\cite{soddemann3}. The phenomenon results from the coupling between 
the layer normal, the tilt of the molecules, and the shear flow. It is triggered
by the fact that shear flow induces tilt. The tilted surfactant layer dilates 
and uses more area, which eventually leads to an undulation instability. This 
could indeed be observed in the simulations.  Figure~\ref{fig:thoso} shows a 
snapshot of an undulated configuration, and a plot of the undulation amplitude
as a function of strain rate. The undulations set in at a well-defined strain-rate. 

Our two examples show that coarse-grained molecular simulations of 
complex, inhomogeneous fluids under shear are now becoming possible. A wealth
of new, intriguing physics can be expected from this field in the future.

\subsection{Simulations at the Mesoscopic Level} 

We will only touch very briefly on the wide field of mesoscopic simulations for 
complex fluids under shear.

The challenge of mesoscopic simulations is to find and/or formulate the 
appropriate mesoscopic model, and then put it on the computer. For liquid 
crystals, we have already discussed one candidate, the Leslie-Ericksen 
theory. However, this is by far not the only available mesoscopic theory 
for liquid crystals. A popular alternative for the description of 
lyotropic liquid crystals is the Doi theory, which is based on a Smoluchowski 
equation for the distribution function for rods. Numerous variants and extensions 
of the Doi model have been proposed. A relatively recent review on both
the Leslie-Ericksen and the Doi theory can be found in Ref.~\refcite{rey}.

The phenomenological equations can be solved with traditional methods of 
computational fluid dynamics. An alternative approach which is particularly 
well suited to simulate flows in complex and fluctuating geometries is the 
Lattice Boltzmann method, which has been discussed by B. D\"unweg in his lecture. 
Yeomans and coworkers have recently proposed a Lattice Boltzmann algorithm 
for liquid-crystal hydrodynamics, which allows to simulate liquid crystal 
flow~\cite{yeomans1,yeomans2}. The starting point is the 
Beris-Edwards theory~\cite{beris}, a dynamical model for liquid crystals 
that consists of coupled dynamical equations for the velocity profiles and the 
order tensor ${\bf Q}$ (Eq.~(\ref{order_tensor}). The Lattice-Boltzmann 
scheme works with the usual discrete velocities $i$ and distributions
$f_i(\vec{r}_n)$ for the fluid flow field on the lattice site $\vec{r}_n$.  
In addition, a second distribution $G_i(\vec{r}_n)$ is introduced, which 
describes the flow of the order tensor field. The time evolution includes
free streaming and collision steps. The exact equations are quite complicated 
and shall not be given here, they can be found in Refs.~\refcite{yeomans2}.

\begin{figure}[bt] 
\centerline{\includegraphics[width=0.8in,angle=-90]{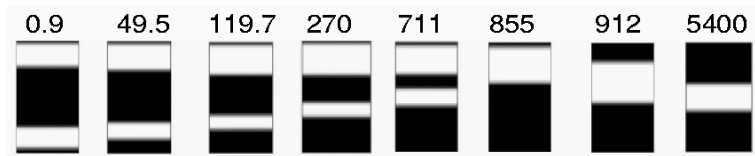}}
\vspace*{8pt}
\caption{
Snapshots from a Lattice-Boltzmann simulation of a sheared
hybrid aligned nematic cell. White regions are ordered,
and black regions are disordered. The numbers give time in milliseconds.
From Ref. \protect\refcite{yeomans3}.
\label{fig:yeomans}
}
\end{figure}

As an application example of this method, we discuss a recent simulation by 
Marenduzzo {\em et al} of a sheared hybrid aligned nematic (HAN) 
cell~\cite{yeomans3,yeomans4}. The surfaces in a HAN cell impose conflicting 
orientations on the director of an adjacent nematic fluid, \ie, one surface 
orients (``anchors'') the fluid in a parallel way (``planar''), and the other 
in a perpendicular way (``homeotropic''). In the system under consideration, 
the parameters were chosen such that the fluid is in the isotropic phase, but 
very close to the nematic phase. The fluid is at rest at time $t>0$. 
At times $t > 0$, the surfaces are moved relative to each other, and drag the 
fluid with them (no-slip boundary conditions). Figure ~\ref{fig:yeomans} shows 
snapshots of the cell for different times. The evolution of the structure inside 
the cell is quite complex. First, two ordered bands form close to the surface, 
due to the fact that the flow field still builds up and strain rates are quite 
high in the vicinity of the surfaces. The director in these bands is flow-aligned, 
and in the lower band, its direction competes with the anchoring energy of the 
homeotropic surface. As a result, the lower band unbinds and crosses the system 
to join the other band. Finally, the top band also unbinds and moves towards the 
center of the cell. This simulation shows nicely how a complex dynamical process, 
which results from an interplay of shear flow and elastic deformations, can be 
studied with a mesoscopic simulation method. We note that the time scale of this 
simulation is seconds, \ie, inaccessible for molecular simulations.

\subsection{Summary} 

Under shear, complex fluids exhibit a wealth of new, interesting, and
practically relevant phenomena. To study these, various simulation approaches 
for different levels of coarse-graining have been developed. In this section,
we have presented and discussed several variants of non-equilibrium molecular 
dynamics (NEMD), and illustrated their use with examples of large scale 
NEMD simulations of inhomogeneous liquid crystals and surfactant layers.
Furthermore, we have briefly touched on mesoscopic simulation methods for
liquid crystals under shear. 

\section{Conclusions}     

Due to the rapidly improving performance of modern computers, complex fluids 
can be studied on a much higher level today than just ten years ago. 
A large number of coarse-grained models and methods have been designed, which
allow to investigate different aspects of these materials, at equilibrium as well 
as far from equilibrium. The new computational possibilities have vitally
contributed to the current boost of soft matter science.

In the two central sections of this lecture, we have given an introduction 
into two important chapters of computational soft matter: 
In Section~\ref{sec:membranes}, we have given a rough overview over 
idealized coarse-grained simulation models for membranes, and hopefully 
conveyed an idea of the potential of such models. In Section~\ref{sec:shear}, 
we have discussed simulation methods for complex fluids under shear. 
Generally, nonequilibrium studies of complex fluids are still much
more scarce than equilibrium studies. Much remains to be done, 
both from the point of view of method development and applications,
and many exciting developments can be expected for the near future.

\section*{Acknowledgments}
\addcontentsline{toc}{section}{Acknowledgements}

I would like to thank Olaf Lenz, Claire Loison, Michel Mareschal, Kurt Kremer,
and Guido Germano, for enjoyable collaborations that have lead to some of 
the results discussed in this paper. I am grateful to Alex de Vries,
Siewert-Jan Marrink, Gunnar Linke, and Thomas Gruhn, for allowing to use their
configuration snapshots. This work was supported by the John von Neumann
institute for computing, and by the Deutsche Forschungsgemeinschaft.


\begin{thebibliography}{999}

\itemsep=0.cm
\parsep=0.cm

\bibitem{daoud}
  M. Daoud and C. E. Williams, Eds. (1995) {\it Soft Matter Physics}.
  Springer-Verlag, Berlin

\bibitem{larson_book}
  R. G. Larson (1999) {\it The Structure and Rheology of Complex Fluids}.
  Oxford University Press, New York

\bibitem{israelachvili}
  J. N. Israelachvili (1992) {\it Intermolecular and Surface Forces}.
  Academic Press, London

\bibitem{kroeger}
  M. Kr\"oger (2003) 
  Simple models for complex nonequilibrium fluids.
  {\it Physics reports} {\bf 390}, p. 453 

\bibitem{flory}
  P. J. Flory (1969)
  {\it Statistical Mechanics of Chain Molecules}. 
  Interscience Publishers, New York

\bibitem{degennes1} P.-G. de Gennes (1979) 
  {\it Scaling Concepts in Polymer Physics}. 
  Cornell University Press, Ithaca

\bibitem{doi1} M. Doi (1992)
  {\it Introduction to Polymer Physics}.
  Clarendon Press, Oxford

\bibitem{doi2} M. Doi and S. Edwards (1986) 
  {\it The Theory of Polymer Dynamics}.
  Clarendon Press, Oxford

\bibitem{russel} W. B. Russel, D. A. Saville, and W. R. Schowalter (1989)
  {\it Colloidal Dispersions}. 
  Cambridge University Press, Cambridge

\bibitem{hunter} R. J. Hunter (1989)
  {\it Foundations of Colloid Science}.
  Clarendon Press, Oxford

\bibitem{borowko} M. Borowko, Ed. (2000) 
  {\it Computational Methods in Surface and Colloid Science}.
  Marcel Dekker Inc., New York

\bibitem{safran} S. A. Safran (1994)
  {\it Statistical Thermodynamics of Surfaces, Interfaces, and Membranes}. 
  Addison-Wesley, Reading, MA

\bibitem{schick} G. Gompper and M. Schick (1994) 
  {\it Self-Assembling Amphiphilic Systems}, in 
  {\it Phase Transitions and Critical Phenomena}, 
  C. Domb and J. Lebowitz Eds.,
  Academic Press, London, 1994 {\bf 16}, p.~1.

\bibitem{degennes2} P. G. de Gennes and J. Prost (1995) 
  {\it The Physics of Liquid Crystals}.
  Oxford University Press, Oxford

\bibitem{chandrasekhar} S. Chandrasekhar (1992)
  {\it Liquid Crystals}.
  Cambridge University Press, Cambridge

\bibitem{chaikin}
  P. M. Chaikin and T. C. Lubensky (1995)
  {\it Principles of Condensed Matter Physics}.
  Cambridge University Press, Cambridge

\bibitem{mplathe1} F. M\"uller-Plathe (2002)
  Coarse-graining in polymer simulation: From the atomistic to the mesoscopic scale and back.
  {\it ChemPhysChem} {\bf 3}, p. 754

\bibitem{nielsen} S. O. Nielsen, C. F. Lopez, G. Srinivas, and M. L. Klein (2004)
  Coarse grain models and the computer simulation of soft materials.
  {\it J. Phys.: Cond. Matter} {\bf 16}, p. R481 

\bibitem{gennis}
  R.~B. Gennis (1989)
  {\em Biomembranes}.
  Springer Verlag, New York

\bibitem{lipowsky}
  R. Lipowsky and E. Sackmann Eds. (1995)
  {\it Structure and Dynamics of Membranes -- from Cells to Vesicles}. 
  Vol. 1 of {\it Handbook of Biological Physics}. 
  Elsevier, Amsterdam

\bibitem{mouritsen}
  O. G. Mouritsen (2005) 
  {\it Life -- As a Matter of Fat}.
  Springer, Berlin Heidelberg

\bibitem{schubert}
  K. V. Schubert and E. W. Kaler (1996)
  Nonionic microemulsions.
 {\it Ber. Bunseng. Phys. Chemie} {\bf 100}, p. 190

\bibitem{doniach}
  S. Doniach (1978) 
  Thermodynamic fluctuations in phospholipid bilayers.
  {\it J. Chem. Phys.} {\bf 68}, p. 4912 

\bibitem{pink1}
  D. A. Pink, T. J. Green, and D. Chapman (1980) 
  Raman-scattering in bilayers of saturated phosphatidylcholines - experiment and theory.
  {\it Biochemistry} {\bf 19}, p. 349 

\bibitem{pink2}
  A. Caill\'e, D. A. Pink, F. de Verteuil, and M. Zuckermann (1980)
  Theoretical models for quasi-two-dimensional mesomorphic monolayers and membrane bilayers.
  {\it Can. J. Physique} {\bf 58}, p. 581 

\bibitem{dammann}
  B. Dammann, H. C. Fogedby, J. H. Ipsen, C. Jeppesen, K. Jorgensen,
  O. G. Mouritsen, J. Risbo, M. C. Sabra, M. M. Sperooto, and M. J.  Zuckermann (1995)
  in {\it Handbook of nonmedical applications of liposomes} Vol. 1,
  Ed. D. D. Lasic, Y. Barenholz, CRC press, p. 85.

\bibitem{fs_review}
  F. Schmid (2000) 
  Systems involving surfactants.
  In {\it Computational methods in surface and colloid science}, 
  {\em Surfactant Science Series} Vol. 89, Ed. M. Borowko,
  Marcel Dekker Inc., New York, p. 631

\bibitem{larson1}
  R. G. Larson, L. E. Scriven, and H. T. Davis (1985)
  Monte-Carlo simulation of model amphiphilic oil-water systems.
  {\it J. Chem. Phys.} {\bf 83}, p. 2411 

\bibitem{liverpool}
  T. B. Liverpool (1996)
  Larson models of Amphiphiles in Complex Fluids.
  In {\it Ann. Rev. Comp. Phys. IV}, 
  Ed. D. Stauffer,
  World Scientific, Singapore, p. 317.

\bibitem{larson2}
  R. G. Larson (1996)
  Monte Carlo simulations of the phase behavior of surfactant solutions.
 {\it J. Physique II} {\bf 6}, p. 1441

\bibitem{smit1}
  B. Smit, A. G. Schlijper, L. A. M. Rupert, and N. M. van Os (1990) 
  Effects of chain-length of surfactants on the interfacial tension -
  molecular-dynamics simulations and experiments.
  {\it J. Phys. Chem.} {\bf 94}, p. 6933 

\bibitem{smit2}
  B. Smit, K. Esselink, P. A. J. Hilbers, N. M. van Os, L. A. M. Rupert, 
  and I. Szleifer (1993) 
  Computer simulations of surfactant self-assembly.
  {\it Langmuir} {\bf 9}, p. 9 

\bibitem{smit3}
  S. Karaborni, K. Esselink, P. A. J. Hilbers, B. Smit, J. Karth\"auser,
  N. M. van Os, and R. Zana, 
  Simulating the self-assembly of gemini (dimeric) surfactants.
  {\it Science} {\bf 266}, p. 254 

\bibitem{palmer1}
  B. J. Palmer and J. Liu (1996)
  Simulations of micelle self-assembly in surfactant solutions.
  {\it Langmuir} {\bf 12}, p. 746 

\bibitem{palmer2}
  B. J. Palmer and J. Liu (1996)
  Effects of solute-surfactant interactions on micelle formation in surfactant solutions 
  {\it Langmuir} {\bf 12}, p. 6015 

\bibitem{goetz1}
  R. Goetz and R. Lipowsky (1998)
  Computer simulations of bilayer membranes: Self-assembly and interfacial tension.
  {\it J. Chem. Phys.} {\bf 108}, p. 7397 

\bibitem{shillcock1}
  J. C. Shillcock and R. Lipowsky (2002)
  Equilibrium structure and lateral stress distribution of amphiphilic bilayers from dissipative
  particle dynamics simulations.
  {\it J. Chem. Phys.} {\bf 117}, p. 5048 

\bibitem{stevens}
  M. J. Stevens (2004)
  Coarse-grained simulations of lipid bilayers .
  {\it J. Chem. Phys.} {\bf 121}, p. 11942

\bibitem{kranenburg1}
  M. Kranenburg, J. P. Nicolas, and B. Smit (2004)
  Comparison of mesoscopic phospholipid-water models.
  {\it Phys. Chem. Chem. Phys.} {\bf 6}, p. 4142 

\bibitem{jakobsen}
  A. F. Jakobsen, O. G. Mouritsen, and G. Besold (2005)
  Artifacts in dynamical simulations of coarse-grained model lipid bilayers.
  {\it J. Chem. Phys.} {\bf 122}, p. 204901 

\bibitem{harries}
  D. Harries and A. Ben-Shaul (1997)
  Conformational chain statistics in a model lipid bilayer: Comparison between mean field and
  Monte Carlo calculations.
  {\it J. Chem. Phys.} {\bf 106}, p. 1609 

\bibitem{baumgaertner1}
  A. Baumg\"artner (1995)
  Asymmetric partitioning of a polymer into a curved membrane.
  {\it J. Chem. Phys.} {\bf 103}, p. 10669 

\bibitem{baumgaertner2}
  A. Baumg\"artner (1996)
  Insertion and hairpin formation of membrane proteins: A Monte Carlo study.
  {\it Biophys. J.} {\bf 71}, p. 1248 

\bibitem{sintes}
  T. Sintes and A. Baumg\"artner (1997)
  Short-range attractions between two colloids in a lipid monolayer.
  {\it Biophys. J.} {\bf 73}, p. 2251 

\bibitem{olenz1}
  O. Lenz and F. Schmid (2004)
  A simple computer model for liquid lipid bilayers.
  {\it J. Mol. Liquids} {\bf 117}, p. 147 

\bibitem{goetz2}
  R. Goetz, G. Gompper, and R. Lipowsky, 
  Mobility and elasticity of self-assembled membranes.
  {\it Phys. Rev. Lett.} {\bf 82}, p. 221 

\bibitem{noguchi1}
  H. Noguchi and M. Takasu (2001)
  Self-assembly of amphiphiles into vesicles: A Brownian dynamics simulation.
  {\it Phys. Rev. E} {\bf 64}, p. 041913 

\bibitem{farago}
  O. Farago (2003) 
  "Water-free" computer model for fluid bilayer membranes.
  {\it J. Chem. Phys.} {\bf 119}, p. 596

\bibitem{cooke}
  I. R. Cooke, K. Kremer, and M. Deserno (2005) 
  Tunable generic model for fluid bilayer membranes.
  {\it Phys. Rev. E} {\bf 72}, p. 011506 

\bibitem{noguchi2}
  H. Noguchi and M. Takasu (2001)
  Fusion pathways of vesicles: A Brownian dynamics simulation.
  {\it J. Chem. Phys.} {\bf 115}, p. 9547 

\bibitem{noguchi3}
  H. Noguchi and M. Takasu (2002)
  Structural changes of pulled vesicles: A Brownian dynamics simulation.
  {\it Phys. Rev. E} {\bf 65}, p. 051907 

\bibitem{sengupta}
  K. Sengupta, V. A. Raghunathan, and J. Katsaras (2003),
  Structure of the ripple phase of phospholipid multibilayers.
  {\it Phys. Rev. E} {\bf 68}, p. 031710 

\bibitem{vries}
  A. H. de Vries, S. Yefimov, A. E. Mark, and S. J. Marrink (2005)
  Molecular structure of the lecithin ripple phase. 
  {\it PNAS} {\bf 102}, p. 5392 

\bibitem{kranenburg2}
  M. Kranenburg, C. Laforge, and B. Smit (2004)
  Mesoscopic simulations of phase transitions in lipid bilayers.
  {\it Phys. Chem. Chem. Phys.} {\bf 6}, p. 4531 

\bibitem{kranenburg3}
  M. Kranenburg and B. Smit (2005)
  Phase behavior of model lipid bilayers.
  {\it J. Phys. Chem. B} {\bf 109}, p. 6553 

\bibitem{olenz2}
  O. Lenz and F. Schmid (2007)
  Structure of symmetric and asymmetric `ripple' phases in lipid bilayers.
  {\it Phys. Rev. Lett.} {\bf 98}, p. 058104

\bibitem{olenz3}
  F. Schmid, D. D\"uchs, O. Lenz, and B. West (2007)
  A generic model for lipid monolayers, bilayers, and membranes.
  {\it Comp. Phys. Comm.} {\bf 177}, p. 168

\bibitem{lei}
  N. Lei, C. R. Safinya, and R. F. Bruinsma (1995)
  Discrete harmonic model for stacked membranes - theory and experiment.
  {\it J. Phys. II} {\bf 5}, p. 1155 

\bibitem{cloison1}
  C. Loison, M. Mareschal, K. Kremer, and F. Schmid (2003)
  Thermal fluctuations in a lamellar phase of a binary
  amphiphile-solvent mixture: A molecular dynamics study.
  {\it J. Chem. Phys.} {\bf 119}, p. 13138 

\bibitem{soddemann1}
  T. Soddemann, B. D\"unweg, and K. Kremer (2001)
  A generic computer model for amphiphilic systems.
  {\it Eur. Phys. J. E} {\bf 6}, p. 409 

\bibitem{mueller}
  M. M\"uller and M. Schick (1996)
  New mechanism of membrane fusion.  
  {\it J. Chem. Phys.} {\bf 116}, p. 2342 

\bibitem{marrink}
  S.-J. Marrink, F. J\"ahning, and H. Berendsen (1996)
  Proton transport across transient single-file water pores in a lipid membrane studied by
  molecular dynamics simulations.
  {\it Biophys. J.} {\bf 71}, p. 632 

\bibitem{zahn}
  D. Zahn and J. Brickmann (2002)
  Molecular Dynamics Study of Water Pores in a Phospholipid Bilayer.
  {\it Chem. Phys. Lett.} {\bf 352}, p. 441 

\bibitem{tolpekina}
  T. V. Tolpekina, W. K. den Otter, and W. J. Briels (2004)
  Simulations of stable pores in membranes: System size dependence and line tension.
  {\it J. Chem. Phys.} {\bf 121}, p. 8014 

\bibitem{wang}
  Z. J. Wang and D. Frenkel (2005)
  Pore nucleation in mechanically stretched bilayer membranes.
  {\it J. Chem. Phys.} {\bf 123}, p. 154701 

\bibitem{cloison2}
  C. Loison, M. Mareschal, and F. Schmid (2004)
  Pores in bilayer membranes of amphiphilic molecules:
  Coarse-Grained Molecular Dynamics Simulations Compared with Simple
  Mesoscopic Models.
  {\it J. Chem. Phys.} {\bf 121}, p. 1890 

\bibitem{cloison3}
  C. Loison, M. Mareschal, and F. Schmid (2005)
  Fluctuations and defects in lamellar stacks of amphiphilic bilayers.
  {\it Comp. Phys. Comm.} {\bf 169}, p. 99 

\bibitem{lister}
  J. D. Lister (1975)
  Stability of lipid bilayers and red blood-cell membranes.
  {\it Physics Lett. A} {\bf A 53}, p. 193 

\bibitem{weisstein}
  E. W. Weisstein (2003)
  {\it CRC Concise encyclopaedia of mathematics}. 
  Chapman \& Hall CRC, http://mathworld.wolfram.com

\bibitem{helfrich}
  W. Helfrich (1973)
  Elastic properties of lipid bilayers - theory and possible experiments.
  {\it Z. Naturforschung C} {\bf 28}, p. 693 

\bibitem{evans}
  E. Evans (1974) 
  Bending resistance and chemically induced moments in membrane bilayers. 
  {\it Biophys. J.} {\bf 14}, p. 923

\bibitem{seifert}
  U. Seifert (1997)
  Configurations of fluid membranes and vesicles.
  {\it Adv. Phys.} {\bf 46}, p. 13

\bibitem{kazakov}
  V. A. Kazakov, I. K. Kostov, and A. A. Migdal (1985)
  Critical properties of randomly triangulated planar random surfaces.
  {\it Phys. Lett. B} {\bf 157}, p. 295

\bibitem{billoire}
  A. Billoire and F. David (1986)
  Scaling properties of randomly triangulated planar random surfaces - a numerical study.
 {\it Nucl. Phys. B} {\bf 275}, p. 617

\bibitem{kantor}
  Y. Kantor, M. Kardar, and D. R. Nelson (1986) 
  Statistical mechanics of tethered surfaces.
  {\it Phys. Rev. Lett.} {\bf 57}, p. 791 

\bibitem{ho} 
  J.-S Ho, A. Baumg\"artner (1990)
  Simulations of fluid self-avoiding membranes.
  {\it Europhys. Lett.} {\bf 12}, p. 295 

\bibitem{kroll}
  D. M. Kroll and G. Gompper (1992)
  The conformations of fluid membranes - Monte-Carlo simulations.
  {\it Science} {\bf 255}, p. 968 

\bibitem{gompper_rev}
  G. Gompper and D. M. Kroll (1997)
  Network models of fluid, hexatic and polymerized membranes.
  {\it J. Phys.: Cond. Matter.} {\bf 9}, p. 8795 

\bibitem{kumar1}
  P. B. S. Kumar, G. Gompper, and R. Lipowsky (2001)
  Budding dynamics of multicomponent membranes.
  {\it Phys. Rev. Lett.} {\bf 86}, p. 3911 

\bibitem{kumar2}
  P. B. S. Kumar and M. Rao (1998)
  Shape instabilities in the dynamics of a two-component fluid membrane.
  {\it Phys. Rev. Lett.} {\bf 80}, p. 2489 

\bibitem{gompper1}
  G. Gompper and D. M. Kroll (1997)
  Freezing flexible vesicles.
  {\it Phys. Rev. Lett.} {\bf 78}, p. 2859 

\bibitem{gompper2}
  G. Gompper and D. M. Kroll (1998)
  Membranes with fluctuating topology: Monte Carlo simulations.
  {\it Phys. Rev. Lett.} {\bf 81}, p. 2284 

\bibitem{shillcock2}
  J. C. Shillcock and D. H. Boal (1996) 
  Entropy-driven instability and rupture of fluid membranes.
  {\it Biophys. J.} {\bf 71}, p. 317
 
\bibitem{noguchi4}
  H. Noguchi and G. Gompper (2005)
  Fluid vesicles with viscous membranes in shear flow.
  {\it Phys. Rev. Lett.} {\bf 93}, p. 258102

\bibitem{noguchi5}
  H. Noguchi and G. Gompper (2005)
  Shape transitions of fluid vesicles and red blood cells in capillary flows.
  {\it PNAS} {\bf 102}, p. 14159

\bibitem{noguchi6}
  H. Noguchi and G. Gompper (2006)
  Meshless membrane model based on the moving least-squares method.
 {\it Phys. Rev. E} {\bf 73}, p. 021903 

\bibitem{linke1}
  G. T. Linke (2005) Dissertation Universit\"at Potsdam. \\
  URN: urn:nbn:de:kobv:517-opus-5835 \\
  URL: http://opus.kobv.de/ubp/volltexte/2005/583/.   

\bibitem{linke2}
  G. T. Linke, R. Lipowsky, and T. Gruhn (2006)
  Osmotically induced passage of vesicles through narrow pores,
  {\it Europhys. Lett.} {\bf 74}, p. 916 

\bibitem{guyon} E. Guyon, J.-P. Hulin, L. Petit, and C. D. Mitescu (2001)
  {\it Physical Hydrodynamics}.
  Oxford University Press, Oxford

\bibitem{bagley}
  E. B. Bagley, I. M. Cabott, and D. C. West (1958)
  Discontinuity in the flow curve of polyethylene.
  {\it J. Appl. Phys.} {\bf 29}, p. 109 

\bibitem{mcleish}
  T. C. B. McLeish and R. C. Ball (1986)
  A molecular approach to the spurt effect in polymer melt flow.
  {\it J .Polym. Sci.} {\bf 24}, p. 1735 

\bibitem{cates}
  M. E. Cates, T. C. B. McLeish, and G. Marrucci (1993)
  The rheology of entangled polymers at very high shear rates.
  {\it Europhys. Lett.} {\bf 21}, p. 451 

\bibitem{roux}
  D. C. Roux, J.-F. Berret, G. Porte, E. Peuvrel-Disdier, and P. Lindner (1995)
  Shear-induced orientations and textures of nematic wormlike micelles.
  {\it Macromolecules} {\bf 28}, p. 1681 

\bibitem{berret1}
  J.-F. Berret, G. Porte, and J.-P. Decruppe (1996)
  Inhomogeneous shear rows of wormlike micelles: A master dynamic phase diagram.
  {\it Phys. Rev. E} {\bf 55}, p. 1668 

\bibitem{britton}
  M. M. Britton and P. T. Callaghan (1999)
  Shear banding instability in wormlike micellar solutions 
  {\it Eur. Phys. J. B} {\bf 7}, p. 237 

\bibitem{fischer}
  E. Fischer and P. T. Callaghan (2001)
  Shear banding and the isotropic-to-nematic transition in wormlike micelles.
  {\it Phys. Rev. E} {\bf 64}, p. 011501 

\bibitem{lopez}
  M. R. Lopez-Gonzalex, W. M. Holmes, P. T. Callaghan, and P. J. Photinos (2004)
  Shear banding fluctuations and nematic order in wormlike micelles.
  {\it Phys. Rev. Lett.} {\bf 93}, p. 268302 

\bibitem{olmsted1}
  P. D. Olmsted and P. M. Goldbart (1992)
  Isotropic-nematic transition in shear flow: State selection, coexistence, phase transitions,
  and critical behavior.
  {\it Phys. Rev. A} {\bf 46}, p. 4966 

\bibitem{porte}
  G. Porte, J.-F. Berret, and J. L. Harden (1997)
  Inhomogeneous flows of complex fluids: Mechanical instability versus non-equilibrium phase
  transition.
  {\it J. de Physique II} {\bf 7}, p. 459 

\bibitem{schmitt}
  V. Schmitt, C. M. Marques, and F. Lequeux  (1995)
  Shear-induced phase separation of complex fluids - the role of flow-concentration coupling.
  {\it Phys. Rev. E} {\bf 52}, p. 4009 

\bibitem{olmsted2}
  P. D. Olmsted and C. Y. D. Lu (1997) 
  Coexistence and phase separation in sheared complex fluids.
  {\it Phys. Rev. E} {\bf 56}, p. R55 

\bibitem{lettinga}
  M. P. Lettinga and J. K. G. Dhont (2004)
  Non-equilibrium phase behaviour of rod-like viruses under shear flow.
  {\it J. Phys.: Cond. Matter} {\bf 16}, p. S3929 

\bibitem{olmsted3}
  P. D. Olmsted and C. Y. D. Lu (1999)
  Phase separation of rigid-rod suspensions in shear flow. 
  {\it Phys. Rev. E} {\bf 60}, p. 4397 

\bibitem{olmsted4}
  P. D. Olmsted (1999)
  Two-state shear diagrams for complex fluids in shear flow.
  {\it Europhys. Lett.} {\bf 48}, p. 339 

\bibitem{fielding}
  S. M. Fielding and P. D. Olmsted (2003)
  Flow phase diagrams for concentration-coupled shear banding.
  {\it Europ. Phys. J. E} {\bf 11}, p. 65 

\bibitem{ailawadi}
  N. K. Ailawadi, B. J. Berne, and D. Forster (1971) 
  Hydrodynamics and collective angular-momentum fluctuations in molecular fluids.
  {\it Phys. Rev. A} {\bf 3}, p. 1462 

\bibitem{yuan}
  X.~F. Yuan, M.~P. Allen (1997)
  Non-linear responses of the hard-spheroid fluid under shear flow.
  {\it Physica A} {\bf 240}, p. 145

\bibitem{see}
  H. See, M. Doi, and R. Larson (1990)
  The effect of steady flow fields on the isotropic-nematic phase transition of rigid rod-like
  polymers.
  {\it J. Chem. Phys.} {\bf 92}, p. 792 

\bibitem{olmsted5}
  P. D. Olmsted and P. Goldbart (1990)
  Theory of the nonequilibrium phase transition for nematic liquid crystals under shear flow.
  {\it Phys. Rev. A} {\bf 41}, p. 4578 

\bibitem{berret2}
  J. F. Berret, D. C. Roux, and G. Porte (1994)
  Isotropic-to-nematic transition in wormlike micelles under shear.
  {\it J. de Physique II} {\bf 4}, p. 1261 

\bibitem{berret3}
  J. F. Berret, D. C. Roux, G. Porte, and P. Lindner (1994)
  Shear-induced isotropic-to-nematic phase transition in equilibrium polymers.
  {\it Europhys. Lett.} {\bf 25}, p. 521 

\bibitem{berret4}
  E. Cappelaere, J.-F. Berret, J. P. Decruppe, R. Cressely, and P. Lindner (1997)
  Rheology, birefringence, and small-angle neutron scattering in a charged micellar system:
  Evidence of a shear-induced phase transition. 
 {\it Phys. Rev. E} {\bf 56}, p. 1869 

\bibitem{berret5}
  J.-F. Berret, D. C. Roux, and P. Lindner (1998)
  Structure and rheology of concentrated wormlike micelles at the shear-induced
  isotropic-to-nematic transition.  
  {\it Eur. Phys. J. B} {\bf 5}, p. 67 

\bibitem{mather}
  P. T. Mather, A .Romo-Uribe, C. D. Han, and S. S. Kim (1997)
  Rheo-optical evidence of a flow-induced isotropic-nematic transition in a thermotropic
  liquid-crystalline polymer.
  {\it Macromolecules} {\bf 30}, p. 7977 

\bibitem{larson3}
  R. G. Larson and D. W. Mead (1993)
  The Ericksen number and Deborah number cascades in sheared polymeric nematics.
  {\it Liquid Crystals} {\bf 15}, p. 151

\bibitem{berret6}
  J. F. Berret, D. C. Roux, G. Porte, and P. Lindner (1995)
  Tumbling behavior of nematic worm-like micelles under shear-flow 
  {\it Europhys. Lett.} {\bf 32}, p. 137

\bibitem{zakharov}
  A. V. Zakharov, A. A. Vakulenko, and J. Thoen (2003)
  Tumbling instability in a shearing nematic liquid crystal: 
  Analysis of broadband dielectric results and theoretical treatment.
  {\it J. Chem. Phys.} {\bf 118}, p. 4253 

\bibitem{hess} 
  S. Hess, M. Kr\"oger (2004)
  Regular and chaotic orientational and rheological behaviour of liquid crystals.
  {\it J. Phys.: Cond. Matter} {\bf 16}, p. S3835 

\bibitem{sarman1}
  S. Sarman and D. J. Evans (1993)
  Statistical mechanics of viscous flow in nematic fluids.
  {\it J. Chem. Phys.} {\bf 99}, p. 9021 

\bibitem{tang}
  S. Tang, G. T. Evans, C. P. Mason, and M. P. Allen (1995)
  Shear viscosity for fluids of hard ellipsoids -  
  A kinetic-theory and molecular-dynamics study.
  {\it J. Chem. Phys.} {\bf 102}, p. 3794 

\bibitem{allen_book}
  M. P. Allen and D. J. Tildesley (1989)
  {\it Computer Simulation of Liquids}.
  Oxford University Press, New York

\bibitem{evans_book}
  D. J. Evans and T. P. Morriss (1990)
  {\it Statistical Mechanics of Nonequilibrium Fluids}.
  Academic Press, San Diego

\bibitem{sarman_rev}
  S. S. Sarman, D. J. Evans, and P. T. Cummings (1992)
  Recent developments in non-Newtonian molecular dynamics.
  {\it Physics Reports} {\bf 305}, p. 1 

\bibitem{varnik}
  F. Varnik and K. Binder (2002)
  Shear viscosity of a supercooled polymer melt via 
  nonequilibrium molecular dynamics simulations.
  {\it J. Chem. Phys.} {\bf 117}, p. 6336 

\bibitem{lees}
  A. W. Lees and S. F. Edwards (1972)
  Computer study of transport processes under extreme conditions.
  {\it J. Phys. C} {\bf 5}, p. 1921 

\bibitem{sllod}
  D. J. Evans and G. P. Morriss (1984) 
  Nonlinear-response theory for steady planar couette-flow.
  {\it Phys. Rev. A} {\bf 30}, p. 1528 

\bibitem{edwards}
  B. J. Edwards, C. Baig, and D. J. Keffer (2005)
  An examination of the validity of nonequilibrium molecular-dynamics simulation algorithms 
  for arbitrary steady-state flows.
  {\it J. Chem. Phys.} {\bf 123}, p. 114106 

\bibitem{zhang}
  F. Zhang, D. J. Searles, D. J. Evans, J. S. D Hansen, and D. J. Isbister (1999)
  Kinetic energy conserving integrators for Gaussian thermostatted SLLOD.
  {\it J. Chem. Phys.} {\bf 111}, p. 18 

\bibitem{pan}
  G. A. Pan, J. F. Ely, C. McCabe, and D. J. Isbister (2005)
  Operator splitting algorithm for isokinetic SLLOD molecular dynamics.
  {\it J. Chem. Phys} {\bf 122}, p. 094114 

\bibitem{baalss}
  D. Baalss and S. Hess (1986) 
  Nonequilibrium molecular-dynamics studies on the anisotropic viscosity of 
  perfectly aligned nematic liquid crystals.
  {\it Phys. Rev. Lett.} {\bf 57}, p. 86 

\bibitem{sarman2}
  S. Sarman (1995)
  Nonequilibrium molecular dynamics of liquid-crystal shear-flow.
  {\it J. Chem. Phys.} {\bf 103}, p. 10378 

\bibitem{sarman3}
  S. Sarman (1997)
  Shear flow simulations of biaxial nematic liquid crystals.
  {\it J. Chem. Phys.} {\bf 107}, p. 3144 

\bibitem{whirter1}
  J. Liam McWhirter and G. N. Patey (2002) 
  Nonequilibrium molecular dynamics simulations of a simple dipolar fluid under shear flow.
  {\it J. Chem. Phys. 117}, p. 2747 

\bibitem{whirter2}
  J. Liam McWhirter and G. N. Patey (2002)
  Molecular dynamics simulations of a ferroelectric nematic liquid under shear flow. 
  {\it J. Chem. Phys. 117}, p. 8551 

\bibitem{mplathe}
  F. M\"uller-Plathe (1999)
  Reversing the perturbation in nonequilibrium molecular dynamics: An easy way to calculate the
  shear viscosity of fluids.
  {\it Phys. Rev. E} {\bf 59}, p. 4894
  
\bibitem{frenkel_book}
  D. Frenkel and B. Smit (2002) 
  {\it Understanding Molecular Simulations}.
  Academic Press, San Diego

\bibitem{evans1}
  D. J. Evans and S. Sarman (1993)
  Equivalence of thermostatted nonlinear responses.
  {\it Phys. Rev. E} {\bf 48}, p. 65

\bibitem{ruelle}
  D. Ruelle (2000)
  A Remark on the Equivalence of Isokinetic and Isoenergetic Thermostats 
  in the Thermodynamic Limit.
  {\it J. Stat. Phys.} {\bf 100}, p. 757

\bibitem{kolb}
  A. Kolb and B. D\"unweg (1999)
  Optimized constant pressure stochastic dynamics. 
  {\it J. Chem. Phys.} {\bf 111}, p. 4453

\bibitem{soddemann2}
  T. Soddemann, B. D\"unweg, and K. Kremer (2003)
  Dissipative particle dynamics: A useful thermostat for equilibrium and 
  nonequilibrium molecular dynamics simulations.
  {\it Phys. Rev. E} {\bf 68}, p. 046702 
  
\bibitem{zwanzig}
  R. Zwanzig (1961)
  Memory effects in irreversible thermodynamics.
  {\it Phys. Rev.} {\bf 124}, p. 983 

\bibitem{ottinger}
  H. C. \"Ottinger (1998)
  General projection operator formalism for the dynamics and thermodynamics of complex fluids.
  {\it Phys. Rev. E} {\bf 57}, p. 1416 

\bibitem{gorban}
  A. N. Gorban, I. V. Karlin, H. C. \"Ottinger, and L. L. Tatarinova (2001)
  Ehrenfests argument extended to a formalism of nonequilibrium thermodynamics.
  {\it Phys. Rev. E} {\bf 63}, p. 066124 

\bibitem{duenweg1}
  B. D\"unweg and W. Paul (1991) 
  Brownian dynamics simulations without Gaussian random numbers.
  {\it Int. J. Mod. Phys. C} {\bf 2}, p. 817 

\bibitem{germano1}
  G. Germano and F. Schmid (2003)
  Simulation of nematic-isotropic phase coexistence in liquid crystals
  under shear.
  In {\it Publication Series of the John von Neumann Institute for Computing},
  Vol. 20, p. 311 

\bibitem{germano2}
  G. Germano and F. Schmid (2005) 
  Nematic-isotropic interfaces under shear: A molecular dynamics simulation.
  {\it J. Chem. Phys.} {\bf 123}, p. 214703 

\bibitem{allen1}
  M. P. Allen (2000)
  Molecular simulation and theory of the isotropic-nematic interface.
  {\it J. Chem. Phys.} {\bf 112}, p. 5447 

\bibitem{mcdonald}
  A. J. McDonald, M. P. Allen, and F. Schmid (2001)
  Surface tension of the isotropic-nematic interface.
  {\it Phys. Rev. E} {\bf 63}, p. 10701R 

\bibitem{akino}
  N. Akino, F. Schmid, and M. P. Allen (2001)
  Molecular-dynamics study of the nematic-isotropic interface.
  {\it Phys. Rev. E} {\bf 63}, p. 041706 

\bibitem{guo}
  H. Guo, K. Kremer, and T. Soddemann (2002)
  Nonequilibrium molecular dynamics simulation of shear-induced alignment 
  of amphiphilic model systems.
  {\it Phys. Rev. E} {\bf 66}, p. 061503 

\bibitem{soddemann3}
  T. Soddemann, G. K. Auernhammer, H. Guo, B. D\"unweg, and K. Kremer (2004)
  Shear-induced undulation of smectic-A: Molecular dynamics simulations vs. analytical theory.
  {\it Eur. Phys. J. E} {\bf 13}, p. 141 

\bibitem{gupta}
  V. K. Gupta, R. Krishnamoorti, J. A. Kornfield, and S. D. Smith (1995)
  Evolution of microstructure during shear alignment in a polystyrene-polyisoprene
  lamellar diblock copolymer.
  {\it Macromolecules} {\bf 28}, p. 4464 

\bibitem{koppi}
  K. A. Koppi, M. Tirrell, F. S. Bates, K. Almdal, and R. H. Colby (1992)
  Lamellae orientation in dynamically sheared diblock copolymer melts.
  {\it J. Physique II} {\bf 2}, p. 1941 

\bibitem{auernhammer1}
  G. K. Auernhammer, H. R. Brand, and H. Pleiner (2000)
  The undulation instability in layered systems under shear flow - a simple model.
  {\it Rheol. Acta} {\bf 39}, p. 215 

\bibitem{auernhammer2}
  G. K. Auernhammer, H. R. Brand, and H. Pleiner (2002)
  Shear-induced instabilities in layered liquids.
  {\it Phys. Rev. E} {\bf 66}, p. 061707

\bibitem{rey}
  A. D. Rey and M. M. Denn (2002)
  Dynamical phenomena in liquid-crystalline materials.
  {\it Annu. Rev. Fluid Mech.} {\bf 34}, p. 233 

\bibitem{yeomans1}
  C. Denniston, E. Orlandini, and J. M. Yeomans (2000)
  Simulations of liquid crystal hydrodynamics
  in the isotropic and nematic phases.
  {\it Europhys. Lett.} {\bf 52}, p. 481 

\bibitem{yeomans2}
  C. Denniston, D. Marenduzzo, E. Orlandini, and J. M. Yeomans (2004)
  Lattice Boltzmann algorithm for three-dimensional liquid-crystal hydrodynamics.
  {\it Phil. Trans. Royal Soc. London A} {\bf 362}, p. 1745 

\bibitem{beris}
  A .N. Beris, B. J. Edwards, and M. Grmela (1990)
  Generalized constitutive equation for polymeric liquid crystals. 
  1. Model formulation using the Hamiltonian (Poisson bracket) formulation.
  {\it J. Non-Newton. Fluid Mechanics} {\bf 35}, p. 51 

\bibitem{yeomans3}
  D. Marenduzzo, E. Orlandini, and J. M. Yeomans (2003)
  Rheology of distorted nematic liquid crystals.
  {\it Europhys. Lett.} {\bf 64}, p. 406 

\bibitem{yeomans4}
  D. Marenduzzo, E. Orlandini, and J. M. Yeomans (2004)
  Interplay between shear flow and elastic deformations in liquid crystals.
  {\it J. Chem. Phys.} {\bf 121}, p. 582

\end{thebibliography}
\end{document}